 \documentclass[11pt]{article}

\usepackage{amsmath,amsfonts,amssymb}

 \usepackage[numbers,sort&compress]{natbib}

\usepackage{hyperref}
\hypersetup{
   colorlinks   =  true,
    linkcolor    = blue,
    citecolor    = red,
     urlcolor	=blue,
	urlbordercolor = {1 1 0}
}

\newcommand\be{\begin{equation}}
\newcommand\ee{\end{equation}}
\newcommand\bea{\begin{eqnarray}}
\newcommand\eea{\end{eqnarray}}

\usepackage{lmodern}
\usepackage[english]{babel}
\usepackage[utf8]{inputenc}
\usepackage[T1]{fontenc}
\usepackage{doi}
\renewcommand{\doitext}{\noexpand\textsc{doi:}}
\usepackage{pgf,tikz,pgfplots}
\usepackage{tikz-cd}
\usetikzlibrary{babel}
\pgfplotsset{compat=1.15}
\usetikzlibrary{matrix,arrows}
\usepackage{graphicx}
\usepackage{adjustbox}
\usepackage{subcaption,booktabs}
\usepackage{amsmath}
\usepackage{amsthm}
\usepackage{physics}
\usepackage{amssymb}
\usepackage{stmaryrd}
\usepackage{bm}
\usepackage{framed}
\usepackage{lscape}
\usepackage{makecell}
\usepackage{mathrsfs}
\providecommand{\abs}[1]{\lvert#1\rvert} %Valor absoluto
\providecommand{\norm}[1]{\lVert#1\rVert} %Norma
\newcommand{\set}[1]{\left\{ #1 \right\}} %Conjunto

\newcommand{\R}{\mathbb{R}} %Real numbers
\newcommand{\C}{\mathbb{C}} %Complex numbers
\newcommand{\Z}{\mathbb{Z}} %Integer numbers
\newcommand{\SO}{\mathrm{SO}}

\newcommand{\cH}{\mathcal{H}} %Caligraphic H

\newcommand{\E}{\mathbf{E}} %Euclidean
\newcommand{\AdS}{\mathbf{A}\mathbf{d}\mathbf{S}} %Anti de Sitter
\newcommand{\dS}{\mathbf{d}\mathbf{S}} %de Sitter
\newcommand{\M}{\mathbf{M}} %Minkowski
\newcommand{\NH}{\mathbf{N}\mathbf{H}} %Newton-Hooke
\newcommand{\G}{\mathbf{G}} %Galilean
\usepackage{svg}

\renewcommand{\G}{\mathbf{G}}

\newcommand{\g}{\mathfrak{g}}
\newcommand{\h}{\mathfrak{h}}
\newcommand{\p}{\mathfrak{p}}

\newcommand{\gl}{\mathfrak{gl}}
\renewcommand{\sl}{\mathfrak{s}\mathfrak{l}}
\newcommand{\so}{\mathfrak{s}\mathfrak{o}}

\newcommand{\cL}{\mathcal{L}}

\renewcommand{\sp}{\mathfrak{s}\mathfrak{p}}

\renewcommand{\k}{\bm{\kappa}}
\newcommand{\diag}{\mathrm{diag}}

\newcommand{\cv}{\mathfrak{X}} %Conjunto de campos vectoriales
\newcommand{\X}{\mathbf{X}} %Campo vectorial/Sistema X
\newcommand{\Y}{\mathbf{Y}} %Campo vectorial/Sistema Y
\newcommand{\x}{\mathbf{x}} %upla de coordenadas x
\newcommand{\cD}{\mathcal{D}}

\newcommand{\cF}{\mathcal{F}}
\renewcommand{\H}{\mathbf{H}}

\newcommand{\Cos}{\,\mathrm{C}}
\newcommand{\Sin}{\,\mathrm{S}}
\newcommand{\Tan}{\,\mathrm{T}}

\renewcommand{\S}{\mathbf{S}}
\newcommand{\pr}{\mathrm{pr}}

\newcommand{\e}{\mathrm{e}}

\newcommand{\Iq}{\mathbb{I}}

\newcommand{\cR}{\mathcal{R}}

\newcommand{\T}{\mathrm{T}}

\theoremstyle{definition}
\newtheorem{defi}{Definition}[section] %Definición
\newtheorem{problem}{Problem}

\theoremstyle{plain}
\newtheorem{thm}[defi]{Theorem} %Teorema
\newtheorem{lema}[defi]{Lemma} %Lema
\newtheorem{prop}[defi]{Proposition} %Proposición
\newtheorem{cor}[defi]{Corollary} %Corolario

\theoremstyle{remark}
\newtheorem*{remark}{Remark}

\numberwithin{equation}{section} %La numeración de las ecuaciones va con las secciones

\begin{document}
\
  \vskip0.5cm

  \begin{center}

 \noindent
 {\Large \bf
Contact Lie systems on  Riemannian and Lorentzian spaces: \\[6pt]   from scaling symmetries   to curvature-dependent reductions}

   \end{center}

\medskip

\begin{center}

{\sc  Rutwig Campoamor-Stursberg$^{1,2}$, Oscar Carballal$^{1,2}$\\[2pt] and Francisco J.~Herranz$^3$}

\end{center}

\medskip

 \noindent
$^1$ Instituto de Matem\'atica Interdisciplinar, Universidad Complutense de Madrid, E-28040 Madrid,  Spain

\noindent
$^2$ Departamento de \'Algebra, Geometr\'{\i}a y Topolog\'{\i}a,  Facultad de Ciencias
Matem\'aticas, Universidad Complutense de Madrid, Plaza de Ciencias 3, E-28040 Madrid, Spain

\noindent
{$^3$ Departamento de F\'isica, Universidad de Burgos,
E-09001 Burgos, Spain}

 \medskip

\noindent  E-mail: {\small
 \href{mailto:rutwig@ucm.es}{\texttt{rutwig@ucm.es}}, \href{mailto:oscarbal@ucm.es}{\texttt{oscarbal@ucm.es}}, \href{mailto:fjherranz@ubu.es}{\texttt{fjherranz@ubu.es}}
}

\medskip

\begin{abstract}
\noindent
We propose an adaptation of the notion of scaling symmetries for the case of Lie--Hamilton systems, allowing their subsequent reduction to contact Lie systems. As an illustration of the procedure, time-dependent frequency oscillators and time-dependent thermodynamic systems are analyzed from this point of view. The formalism provides a novel method for constructing contact Lie systems on the three-dimensional sphere, derived from recently established Lie--Hamilton systems arising from the fundamental four-dimensional representation of the symplectic Lie algebra $\sp(4, \R)$. It is shown that these systems are a particular case of a larger hierarchy of contact Lie systems on a special class of three-dimensional homogeneous spaces, namely the Cayley--Klein spaces. These include Riemannian spaces (sphere, hyperbolic and Euclidean spaces), pseudo-Riemannian spaces (anti-de Sitter, de Sitter and Minkowski spacetimes), as well as Newtonian or non-relativistic spacetimes. Under certain topological conditions, some of these systems retrieve well-known two-dimensional  Lie--Hamilton systems through a curvature-dependent reduction.
\end{abstract}
\medskip
\medskip

\noindent
\textbf{Keywords}: Cayley--Klein geometries, Lie systems;  Lie--Hamilton systems; nonlinear differential equations; contact structures

\noindent
\textbf{MSC 2020 codes}: 37J55, 34A26 (Primary),  17B66, 34C14, 70G45 (Secondary)

\smallskip

\medskip

 \newpage

\tableofcontents

\newpage

%%%%%%%%%%%%%%%%%%%%%%%%%%%%%%%%%%%%%%%%%%%%%%
\section{Introduction}
Contact geometry has been intimately related to the theory of differential equations since its origins, when S. Lie, in his seminal paper \cite{Lie1888}, introduced a type of symmetries of differential equations more general than point symmetries, the so-called {\it contact symmetries} (see \cite{Stephani1990,AbrahamShrauner1995,CampoamorStursberg2016} and references therein). In this context, its influence on the study of partial differential equations (PDEs) is vast. This is the case, for instance, of the Monge-Amp{\`e}re equations \cite{Kushner2006}, where cohomological techniques can be successfully applied \cite{Kruglikov1998} by means of Lie algebras that are compatible with a certain contact form \cite{Khakimdjanov2004,Ancochea2006}. Contact geometry has also been considered recently to construct $(3+1)$-integrable systems through the Lax pairs formalism \cite{Blaszak2017,Sergyeyev2018,Sergyeyev2025}.

Starting from G.~Reeb's celebrated work \cite{Reeb1952}, the global study of contact structures has led to significant advances in geometry and topology \cite{Blair2010,Geiges2008}. In differential geometry, much effort has been devoted to contact metric structures, with Sasakian structures as a special case. Originally, these structures described compatibility between a contact form and a Riemannian metric,  but more recent developments have extended this framework to the pseudo-Riemannian realm \cite{Calvaruso2010}, generalizing classical results \cite{Takahashi1969}. This construction, as noted in \cite{Duggal1990}, has natural applications in general relativity. In recent years, contact geometry has gained increasing attention in the study of dissipative systems \cite{Bravetti2017,Ciaglia2018,deLeon2019,Bravetti2020,Grabowska2022,LopezGordon2024,Rivas2021,deLeon2023}. Notably, contact geometry has been applied to analyze the so-called {\it contact Lie systems} \cite{deLucas2023}, a special class of systems of ordinary differential equations (ODEs in short) with broad applications in physics: the class of Lie systems. Indeed, contact Lie systems can also be regarded as a particular case of the so-called Jacobi--Lie systems introduced in \cite{Jacobi2015}.

A {\it Lie system} on a smooth manifold $M$ is a time-dependent vector field $\X: \R \times M \to \T M$, $(t,x) \mapsto  \X(t,x)$ such that $\R \ni t \mapsto  \mathbf{X}_{t}:= \X(t, \cdot)$ is a curve taking values in a finite-dimensional Lie algebra of vector fields $V \subset \cv(M)$, called a {\it Vessiot--Guldberg (VG) Lie algebra} of $\X$ (see \cite{deLucas2020} for more details). The main property of Lie systems is that their associated non-autonomous { system of ODEs admits a {\it superposition rule}, i.e., a time-independent map through which the general solution of the { system of ODEs} is expressed in terms of a generic finite family of particular solutions and certain constants related to the initial conditions. This holds if and only if the smallest Lie algebra containing $\set{\X_{t}}_{t \in \R}$, denoted by $V^{X}$, is finite-dimensional. This characterization is known as the Lie--Scheffers Theorem \cite[Section~3.8]{deLucas2020}.

Determining a superposition rule generally involves dealing with { systems of PDEs}, but the problem simplifies when a Lie system is compatible with specific geometric structures. In recent years, particular focus has been given to Lie systems compatible with symplectic structures, where techniques from superintegrable systems enable us to give an algorithmic derivation of superposition rules \cite{Ballesteros2013}. A {\it Lie--Hamilton (LH) system} on a symplectic manifold $(M, \omega)$ is a Lie system $\mathbf{X}$ on $M$ that admits a VG Lie algebra consisting of Hamiltonian vector fields relative to the symplectic form $\omega$. Thus, the integral curves of $\mathbf{X}$ correspond to the solutions of the Hamilton equations of a time-dependent Hamiltonian function $h: \R \times M \to \R$, given by $h(t,x) := h_{t}(x)$, where  $\iota_{\X_{t}} \omega = \dd h_{t}$ for all $t \in \R$.  The family $\set{h_{t}}_{t \in \R}$ spans a finite-dimensional Lie algebra $\mathcal{H}_{\omega}$ with respect to the Poisson bracket $\set{\cdot, \cdot}_{\omega}$ induced by $\omega$, called a {\it LH algebra associated with $\X$} (see \cite[Chapter 4]{deLucas2020}). We denote such LH system by $(M, \omega, \X)$.

A {\it contact Lie system} \cite{deLucas2023} is a triple $(M, \eta, \X)$, where $(M, \eta)$ is a (co-orientable) contact manifold and $\mathbf{X}$ is a Lie system possessing a VG Lie algebra formed by contact Hamiltonian vector fields relative to $\eta$. More specifically, $\eta \in \Omega^{1}(M)$, { i.e., the contact form of the contact manifold $(M, \eta)$}, satisfies $\eta \wedge (\dd \eta)^{n} \neq 0$, with $\dim M = 2n + 1$, and for every $t \in \R $, there exists $h_{t} \in C^{\infty}(M)$ such that $\eta(\X_{t}) = -h_{t}$  and $\iota_{\X_{t}} \dd \eta = \dd h_{t} - (\cR h_{t})\eta$, where $\cR$ is the so-called Reeb vector field associated with $\eta$ (see \cite{Kushner2006} for details).
In analogy to LH systems, the integral curves of $\X$ are the solutions of the contact Hamilton equations of the time-dependent Hamiltonian $h: \R \times M \to \R$ induced by the family $\{h_{t}\}_{t \in \R}$. The contact Lie system is said to be of {\it Liouville type} if the family $\{h_{t}\}_{t \in \R}$ consists of first integrals of $\cR$. { Note that the term `Liouville' arises from the fact that these systems satisfy an analogue of Liouville Theorem in symplectic geometry; namely, the vector fields of $V^{X}$ are symmetries of the volume form $\eta \wedge (\dd \eta)^{n}$ induced by the contact form $\eta$ (see \cite[Proposition 3.5]{deLucas2023})}.

The classification problem of Lie systems under local diffeomorphisms, as originally proposed by S.~Lie \cite{Lie1893}, has been strictly connected to that of homogeneous spaces and the determination of all Lie groups acting transitively on them \cite{Mostow1950,Gorbatsevich1977,Shnider1984,Shnider1984a}. Lie systems on $\R$ were originally classified by S.~Lie himself \cite[Section I]{Lie1893}, while their classification on $\R^{2}$ was completed in \cite{GonzalezLopez1992}. In this respect, it is worth stressing that LH systems on $\R^{2}$ have also been classified \cite{Ballesteros2015}. The situation on $\R^{n}$, for $n \geq 3$, is much more difficult. Transitive non-solvable Lie algebras of vector fields on $\R^{3}$ were classified in \cite{Doubrov2017}, while the nilpotent case was studied in \cite{Gorbatsevich2024}. In both works, it was observed that the problem for the nilpotent case is {\it wild} (i.e., it contains the subproblem of classifying pairs of $r \times r$-matrices of the same size up to the $\mathrm{GL}(r, \R)$-action by conjugacy). The same pathology has been observed for nilpotent Lie algebras of vector fields on $\R^{4}$ \cite{Bodarenko2019}. Taking into account these obstructions, it seems reasonable to consider the classification problem subjected to some constraint. In this context, the two following problems emerge within the frame of Lie systems:

\begin{problem}\label{problemA}
Construct {\it intrinsic} higher-dimensional Lie systems, {\it i.e.}, Lie systems on $\R^{n}$ with $n \geq 3$, which are not locally diffeomorphic to Lie systems on either $\R$ or $\R^{2}$, in the sense of \cite{Campoamor2024}.
\end{problem}

\begin{problem} \label{problemB}
Construct Lie systems on homogeneous manifolds which are not diffeomorphic to $\R^{n}$.
\end{problem}

Problem~\ref{problemA} was recently addressed to in \cite{Campoamor2024,Carballal2024sp6}, where it was shown that the representation theory of Lie algebras can be successfully applied to construct new higher-dimensional systems. On the other hand, Problem~\ref{problemB} has been faced in \cite{Herranz2017, Campoamor2024conformes} for the case of LH systems. It was shown there that the so-called {\it two-dimensional Cayley--Klein spaces} (2D CK in short) constitute a suitable framework to study LH systems on homogeneous spaces.

In this work, motivated by the $\sp(4,\R)$-LH systems studied in \cite{Campoamor2024}, we construct contact Lie systems based on $\sp(4,\R)$ on the 27 {\it 3D CK spaces}~\cite{Ballesteros1994,Herranz2006}. These are homogeneous spaces of constant curvature associated with the CK Lie groups or quasi-simple orthogonal groups~\cite{Gromov1990,Gromov1992,Herranz1997,Azcarraga1998,GH2021symmetry}. The CK framework involves  three real graded contraction parameters $\kappa_{1}$ (sectional curvature) and $\kappa_{2}, { \kappa_{3}}$ (determining the signature), thus the 3D CK spaces are collectively denoted by $\S^{3}_{[\kappa_{1}],\kappa_{2}, \kappa_{3}}$.     The relevance of this approach is that it is the first time that contact Lie systems \cite{deLucas2023} have been considered on homogeneous spaces. Indeed, contact geometry is particularly suited for studying Lie systems in this setting, as only $\S^{2}$ among even-dimensional spheres admits a symplectic structure, whereas all odd-dimensional spheres $\S^{2n+1}$ naturally carry a contact structure.

This paper is structured as follows. In Section~\ref{section:s2}, we extend {\it scaling symmetries} for autonomous Hamiltonian systems \cite{Sloan2018,Grabowska2022,Bravetti2023,Bravetti2024} to LH systems on symplectic manifolds, enabling their reduction to contact Lie systems. As examples, we study the reduction of a time-dependent harmonic oscillator (Section~\ref{ex:reduction:oscillator}) and apply this formalism to thermodynamics (Section~\ref{subsection:application2}), marking the first known use of Lie systems in this field. In Section~\ref{section:s3}, $\sp(4, \R)$-LH systems from \cite{Campoamor2024} are reduced to contact Lie systems on $\S^3$, which can be projected to LH systems on $\S^2$ via the Hopf fibration, as a particular application of Theorem~\ref{th:s3:reduction}. These systems form part of a larger family of contact Lie systems on the 3D CK spaces. Section~\ref{section:s4} introduces the 3D CK spaces, encompassing classical Riemannian spaces (sphere, Euclidean, hyperbolic) as well as  Lorentzian  (Minkowski and (anti)-de Sitter) and Newtonian or non-relativistic (Galilei and oscillating/expanding Newton--Hooke) spacetimes of constant curvature. Section~\ref{section:3DCK:contactstructure} establishes that all 3D CK spaces are contact manifolds, with their Reeb vector field leading to a geodesic foliation (Proposition~\ref{prop:3DCK:Reeb_foliation}). Theorem~\ref{th:3DCK:fibrebundles} shows that, except for the de Sitter spacetime, the contact structures of the nine aforementioned spaces define principal bundles associated with 1D CK Lie groups.  Section~\ref{section:contactmetric} reviews contact metric spaces, proving that when the Reeb vector field is Killing, first integrals can be obtained geometrically (Theorem~\ref{th:contactmetric:firstintegrals}), which applies to $\S^3$ and the $(2+1)$D anti-de Sitter spacetime (Section~\ref{subsection:contactmetric:sasakian}). In Section~\ref{section:contactck}, we construct contact Lie systems on all 3D CK spaces from the $\sp(4, \R)$-LH systems of \cite{Campoamor2024}, recovering those on $\S^3$ analyzed in Section~\ref{section:s3} as a special case. Sections~\ref{subsection:contactck:liouville} and \ref{subsection:contactck:liouville2} focus on Liouville-type subsystems, while Section~\ref{section:reduction} examines their reduction using the curvature-dependent principal bundles presented in Section~\ref{section:3DCK:contactstructure}  reproducing, in addition,  known LH systems on $\S^2$ and the real plane \cite{Herranz2017,Ballesteros2015,Blasco2015}.

At this stage,  the interdisciplinary nature of our construction is summarized in the following diagram:
\[
\adjustbox{scale=0.8,center}{
\begin{tikzcd}[every label/.append
  style={font=\normalsize}]
\fbox{$\begin{array}{c}
\textup{Representation theory:} \\
\textup{$\sp(4,\R)$-LH systems on $\R^{4}$}
\end{array}$} \arrow[rr,"\textup{Scaling symmetries for LH systems}"] &  &      \fbox{$\begin{array}{c}
\textup{Contact Lie systems} \\
\textup{on $\S^{3}$}
\end{array}$} \arrow[hookrightarrow]{dl}\\
& \fbox{$\begin{array}{c}
\textup{Contact Lie systems} \\
\textup{on 3D CK spaces $\S^{3}_{[\kappa_{1}],\kappa_{2},\kappa_{3}}$}
\end{array}$} \\
\fbox{$\begin{array}{c}
\textup{Contact Lie systems} \\
\textup{of Liouville type on $\S^{3}_{[\kappa_{1}],\kappa_{2},+}$, } \\
\textup{$(\kappa_{1}, \kappa_{2}) \neq (-1, -1)$}
\end{array}$} \arrow[hookrightarrow]{ur}  \arrow[rr,"\textup{$\S^{1}$- and $\R$-reduction}"]
&  &    \fbox{$\begin{array}{c}
\textup{LH systems} \\
\textup{  on $\S^{2}$, $\C\H^{1}$ and $\R^{2}$}
\end{array}$}
\end{tikzcd}} \]
Finally, in Section~\ref{conclusions} some conclusions are drawn, and
potential extensions of this work are briefly discussed.

%%%%%%%%%%%%%%%%%%%%%%%%%%%%%%%%%%%%%%%%%%%%%%

\section{Reduction of Lie--Hamilton systems by scaling symmetries} \label{section:scaling}
\label{section:s2}

Scaling symmetries of symplectic Hamiltonian systems are a particular type of `non-standard symmetries', i.e., symmetries which do not necessarily preserve the symplectic structure. The motivation for the study of these symmetries can be traced back to Poincar{\'e}'s idea that any minimal description of the universe should be scale-invariant \cite{Poincare1908, Gryb2021}.  The general framework for the study of scaling symmetries was established in \cite{Sloan2018}, where it was shown for the first time that the reduction of a symplectic Hamiltonian system through a scaling symmetry produces a contact Hamiltonian system. This reduction procedure was later on formalized in \cite{Grabowska2022,Bravetti2023,Bravetti2024}.

From now on $\R^{\times}$ stands for { the} multiplicative group $\R^{\times} = \R^{+}$ or $\R^{\times} = \R - \set{0}$. The following definition is an adaptation of the notion of  scaling symmetry for the case of LH systems:
\begin{defi}
\label{def:reduction:scaling_symmetry}
A \textit{scaling symmetry} for a LH system $(M, \omega, \X)$ is a principal $\R^{\times}$-action
\begin{equation}
\Phi: \R^{\times} \times M \to M
\nonumber
\end{equation}
on $M$ such that there exists a $t$-dependent Hamiltonian $h: \R \times M \to \R$ associated with $\X$ satisfying
\begin{equation}
\Phi_{s}^{*} \omega = s \,\omega, \qquad \Phi_{s}^{*} h_{t} = s \, h_{t}, \qquad s \in \R^{\times}, \quad\: t \in \R.
\label{eq:reduction:scaling}
\end{equation}
In this case, we say that the $t$-dependent Hamiltonian $h$ is \textit{compatible} with the scaling symmetry.
\end{defi}

The above concept is a particular case of $\ell$-homogeneity, which we  present below.
\begin{defi}
\label{def:reduction:homogeneity}
Given an integer $\ell \in \mathbb{Z}$, a function $f \in C^{\infty}(M)$ and a { differential $k$-form} $\alpha$ on $M$ is said to be \textit{$\ell$-homogeneous} with respect to the principal action $\Phi$ if $\Phi_{s}^{*} f = s^{\ell} \, f$ and $\Phi_{s}^{*} \alpha = s^{\ell} \, \alpha$ for all $s \in \R^{\times}$, respectively. Similarly, a vector field $\Y$ on $M$ is \textit{$\ell$-homogeneous} with respect to $\Phi$ if $(\Phi_{s})_{*} \Y = s^{\ell} \, \Y$ for all $s \in \R^{\times}$.
\end{defi}

\begin{remark}
Provided that $\R^{\times}$ is connected (i.e., $\R^{\times} = \R^{+}$), the $\ell$-homogeneity of $f$, $\alpha$ and $\Y$ is equivalent to
\begin{equation}
\cL_{\mathbf{\Delta}} f = \ell \, f, \qquad \cL_{\mathbf{\Delta}} \alpha = \ell \, \alpha, \qquad \cL_{\mathbf{\Delta}} \Y = [\mathbf{\Delta}, \Y] = \ell \, \Y,
\label{eq:reduction:scaling_infinitesimal}
\end{equation}
correspondingly, with $\mathbf{\Delta} \in \cv(M)$ being the infinitesimal generator of the $\R^{\times}$-action.
When $\R^{\times} = \R - \set{0}$, $f \in C^{\infty}(M)$, $\alpha \in \Omega^{k}(M)$ and $\Y \in \cv(M)$ are $\ell$-homogeneous with respect to $\Phi$ if and only if they satisfy  \eqref{eq:reduction:scaling_infinitesimal} and
\begin{equation}
(\Phi_{-1})^{*} f = (-1)^{\ell} f, \qquad (\Phi_{-1})^{*} \alpha = (-1)^{\ell} \alpha, \qquad (\Phi_{-1})_{*} \Y = (-1)^{\ell} \Y,
\nonumber
\end{equation}
respectively (see \cite[Chapter V]{Libermann1987} for details, where { their} $\ell$-homogeneous vector fields  are $(\ell -1)$-homogeneous vector fields in our terminology, which we have adopted from \cite{Grabowska2022}).
\end{remark}

From \eqref{eq:reduction:scaling} we see that the symplectic form $\omega$ and the Hamiltonian functions $h_{t}$ are $1$-homoge\-neous with respect to the principal action $\Phi$. Indeed, the homogeneity of $\omega$ implies that  $\omega = -\dd \lambda$ is an exact symplectic form, with the symplectic potential $\lambda := - \iota_{\mathbf{\Delta}} \omega$ being also $1$-homogeneous with respect to the $\R^{\times}$-action. It follows that $\mathbf{\Delta}$ is a \textit{Liouville vector field} of the exact symplectic manifold $(M, \omega = -\dd  \lambda)$. Taking into account the $t$-dependent vector field $\X$, the conditions \eqref{eq:reduction:scaling_infinitesimal} yield
\begin{equation}
\iota_{[\mathbf{\Delta}, \X_{t}]} \omega  = \cL_{\mathbf{\Delta}} \iota_{\X_{t}} \omega - \iota_{\X_{t}} \cL_{\mathbf{\Delta}} \omega =  0, \qquad t \in \R
\nonumber
\end{equation}
and, from the nondegeneracy of $\omega$, we conclude that  $[\mathbf{\Delta}, \X_{t}] = 0$ for all $t \in \R$. In particular, $\mathbf{\Delta}$ commutes with all the vector fields belonging to the VG Lie algebra $V^{X}$ of the LH system $\X$,  so it is a \textit{symmetry} of  $\X$.

In terms of the principal action $\Phi$, it is also worth noting that the homogeneity condition \eqref{eq:reduction:scaling} for a compatible $t$-dependent Hamiltonian $h$ can be equivalently formulated for the $t$-dependent vector field $\X$.

\begin{lema}[\cite{Libermann1987}]\label{prop:reduction:properties}
Let $\Phi: \R^{\times} \times M \to M$  be a principal $\R^{\times}$-action on a symplectic manifold $(M, \omega)$ such that $\omega$ is $1$-homogeneous. Then,
\begin{itemize}
\item[(1)] $f \in C^{\infty}(M)$ is $1$-homogeneous if and only if its associated Hamiltonian vector field $\X_{f}$ is $0$-homogeneous.
\item[(2)] The Poisson bracket $\set{f_{1}, f_{2}}_{\omega}$ of $1$-homogeneous functions $f_{1}, f_{2} \in C^{\infty}(M)$ is $1$-homogeneous. Thus the Lie bracket  $[\X_{f_{1}}, \X_{f_{2}}]$ of their associated Hamiltonian vector fields is $0$-homogeneous.
\end{itemize}
\end{lema}

\begin{prop} \label{lema:reduction:scaling_vf}
A principal $\R^{\times}$-action $\Phi: \R^{\times} \times M \to M$ is a scaling symmetry for a LH system $(M, \omega, \X)$ if and only if $\omega$ is $1$-homogeneous and $\X_{t}$ is $0$-homogeneous for all $t \in \R$.
\end{prop}

\begin{cor}
A principal $\R^{\times}$-action $\Phi: \R^{\times} \times M \to M$ is a scaling symmetry for a LH system $(M, \omega, \X)$ if and only if the VG Lie algebra $V^{X}$ of $\X$ is spanned by $0$-homogeneous vector fields.
\end{cor}

Suppose now that $\Phi: \R^{\times} \times M \to M$ is a scaling symmetry for a LH system $(M, \omega, \X)$ which is compatible with a $t$-dependent Hamiltonian $h: \R \times M \to \R$. Then, for every $t \in \R$, the function $h_{t}$ is nonconstant. Taking into account the LH algebra $\cH_{\omega} = \mathrm{Lie}\bigl(\set{h_{t}}_{t \in \R}, \set{\cdot, \cdot}_{\omega} \bigr)$ associated with $h$, Lemma~\ref{prop:reduction:properties} shows that $\cH_{\omega}$ cannot contain nonzero constant functions, i.e., $\cH_{\omega} \cap \R = \set{0}$.  From this, we obtain the following statement.

\begin{prop}\label{prop:reduction:gen_lh}
A principal $\R^{\times}$-action $\Phi: \R^{\times} \times M  \to M$ is a scaling symmetry for a LH system $(M, \omega, \X)$ if and only if $\omega$ is $1$-homogeneous and there exists a LH algebra $\cH_{\omega}$ associated with $\X$, with $\cH_{\omega} \cap \R = \set{0}$, spanned by $1$-homogeneous Hamiltonian functions. Moreover, if $M$ is connected, the LH algebra $\cH_{\omega}$ and the VG Lie algebra $V^{X}$ of $\X$ are isomorphic as real Lie algebras.
\end{prop}
\begin{proof}
Recall that, if $M$ is connected, the sequence of Lie algebras
\begin{equation}
\begin{tikzcd}
0 \hookrightarrow \cH_{\omega} \cap \R \hookrightarrow \cH_{\omega} \ar[r,"\varphi"] & V^{X} \ar[r] & 0
\end{tikzcd}
\nonumber
\end{equation}
is exact, where $\varphi: \cH_{\omega} \to V^{X}, \: f \mapsto - \X_{f}$ maps every Hamiltonian function of $\cH_{\omega}$ to minus its Hamiltonian vector field.
\end{proof}

Now we apply the reduction theory by scaling symmetries (see \cite{Bravetti2024,Grabowska2022}) for the case of LH systems. Consider a scaling symmetry $\Phi: \R^{\times} \times M \to M$ for a LH system $(M, \omega, \X)$, and denote by $\pi: M \to M / \R^{\times}$ the projection. From Proposition~\ref{lema:reduction:scaling_vf}, for all $t \in \R$, the vector field $\X_{t} \in \cv(M)$ is invariant with respect to the principal $\R^{\times}$-action $\Phi$ and, therefore, is $\pi$-projectable. This gives rise to a $t$-dependent vector field $\pi_{*}\X$ on the reduced space  $M/\R^{\times}$ given by
\begin{equation}
(\pi_{*}\X)_{t} = \pi_{*}\X_{t}, \qquad t \in \R.
\nonumber
\end{equation}
These projected vector fields span a finite-dimensional real Lie algebra $V^{\pi_{*}X}$, the projection of the VG Lie algebra $V^{X}$ of $\X$, which is isomorphic to $V^{X}$ as a real Lie algebra. Thus, the $t$-dependent vector field $\pi_{*}\X$ on $M/ \R^{\times}$ determines a Lie system.

The question now is if there exists any geometric structure on $M/\R^{\times}$ compatible with the reduced Lie system $\pi_{*}\X$, provided that $\dim M \geq 4$. The symplectic potential $\lambda =- \iota_{\mathbf{\Delta}} \omega$ determines a  codistribution $\langle \lambda \rangle \subset \T^{*}M$ such that its annihilator $\langle \lambda \rangle^{\circ} \subset \T M$ can be $\pi$-projected to a distribution
\begin{equation}
\cD := \pi_{*} (\langle \lambda \rangle^{\circ}) \subset \T(M/ \R^{\times})
\label{eq:reduction:contactdis}
\end{equation}
on $M/ \R^{\times}$. This distribution $\cD$ is a \textit{contact structure}, turning the reduced space into a \textit{contact manifold} $(M/\R^{\times}, \cD)$. In the most general case, $\cD$ is only locally described by the annihilator of the vector subbundle of $\T^{*}(M/ \R^{\times})$ spanned by a (locally defined) contact form. Nevertheless, provided that $\cD$ is \textit{co-orientable}, i.e., when there exists a (globally defined) contact form $\eta \in \Omega^{1}(M/\R^{\times})$ such that $\cD = \langle \eta \rangle^{\circ}$, every vector field of $V^{\pi_{*}X}$ is univocally associated with a smooth function. The existence of such a contact form $\eta$ was characterized in \cite{Bravetti2024} as follows.

\begin{prop}[\cite{Bravetti2024}] \label{prop:reduction:geometry_red}
Consider the contact structure $\cD$ \eqref{eq:reduction:contactdis} on $M/ \R^{\times}$. The following statements are equivalent:
\begin{itemize}
\item[(1)] There exists some globally defined contact form $\eta \in \Omega^{1}(M/ \R^{\times})$ such that $\cD = \langle \eta \rangle^{\circ}$.
\item[(2)] The principal $\R^{\times}$-bundle $\pi: M \to M / \R^{\times}$ is trivial, i.e., $M \simeq \R^{\times} \times (M / \R^{\times})$.
\item[(3)] There exists some nonvanishing $1$-homogeneous smooth function $F: M \to \R^{\times}$.
\end{itemize}
\end{prop}
In this situation, the trivialization of the bundle $\pi: M \to M/ \R^{\times}$ is provided by the diffeomorphism $(F, \pi): M \to \R^{\times} \times (M/ \R^{\times})$. The mapping $\sigma: M / \R^{\times} \to M$ defined by
\begin{equation}
\sigma(y) := (F, \pi)^{-1}(1, y), \qquad  y \in M/\R^{\times}
\nonumber
\end{equation}
is a global section of the bundle $\pi: M \to M / \R^{\times}$. Then,
\begin{equation}
\eta := \sigma^{*} \iota_{\mathbf{\Delta}} \omega
\label{eq:reduction:contact}
\end{equation}
is a contact form on $M / \R^{\times}$ such that $\cD = \langle \eta \rangle^{\circ}$.

Every vector field of the VG Lie algebra $V^{\pi_{*}X}$ is of the form $\pi_{*}\X_{f}$, being $\X_{f}$ the Hamiltonian vector field of $f \in \cH_{\omega}$ with respect to $\omega$. Moreover, the vector field $\pi_{*}\X_{f}$ is the Hamiltonian vector field of $\hat{f} \in C^{\infty}(M/\R^{\times})$ with respect to the contact form \eqref{eq:reduction:contact}, where $\pi^{*} \hat{f} = f/F$. This proves the following

\begin{thm} \label{th:scaling:reduction}
Let $\Phi: \R^{\times} \times M \to M$ be a scaling symmetry for a LH system $(M, \omega, \X)$ with VG Lie algebra $V^{X}$, and denote by  $\pi: M \to M / \R^{\times}$  the projection. Then, $\pi_{*}\X$ is a Lie system on $M/\R^{\times}$ whose VG Lie algebra $V^{\pi_{*}X}$ is isomorphic to $V^{X}$ as a real Lie algebra. If, in addition, any of the conditions of Proposition~\ref{prop:reduction:geometry_red} is satisfied, $\pi_{*}\X$ is a contact Lie system with respect to the contact form \eqref{eq:reduction:contact}.
\end{thm}

In order to illustrate this result, we present two applications of physical interest covering the possibilities of the existence or not of a (globally defined)  contact form $\eta$. The former concerns
the $t$-dependent 2D oscillator system, related to the Lie algebra $\sl(2,\R)$, which is the $t$-dependent analogue of the one developed in \cite{Bravetti2024}. In particular, we show that there does exist a contact form $\eta$, so that the corresponding oscillator LH system can be reduced to a contact Lie system. The latter encompasses a LH system based on a subalgebra of the symplectic Lie algebra $\mathfrak{sp}(6,\mathbb{R})$, for which there is no such contact form. However, we show explicitly that even in this case, it is possible to deduce contact Lie systems through restriction of  reduced Lie systems, deriving in this example a $t$-dependent thermodynamic system.

%%%%%%%%%%%%%%%%%%%%%%%%%%%%%%%%%%%%%%%%%%%%%%

\subsection{Application to the harmonic oscillator with time-dependent frequency} \label{ex:reduction:oscillator}

The 2D harmonic  oscillator on $\R^{2}_{0} := \R^{2} - \set{0}$ with a time-dependent frequency $\Omega(t)$ and unit mass is  given by the Hamiltonian
\begin{equation}
h := \frac{1}{2}\bigl(p_{1}^{2} + p_{2}^{2}\bigr) + \frac{\Omega^{2}(t)}{2}\bigl(q_{1}^{2} + q_{2}^{2}\bigr) = h_{1} + \Omega^{2}(t) h_{3}.
\nonumber
\end{equation}
The kinetic energy $h_{1} := \frac{1}{2} (p_{1}^{2} + p_{2}^{2})$ and the isotropic oscillator potential $h_{3} := \frac{1}{2} (q_{1}^{2} + q_{2}^{2})$ span, along with the Hamiltonian function $h_{2} := q_{1}p_{1} + q_{2}p_{2}$, an $\sl(2,\R)$-Lie algebra with respect to the Poisson bracket $\set{\cdot, \cdot}_{\omega}$ induced by the canonical symplectic form
\begin{equation}
\omega = \dd q_{1} \wedge \dd p_{1} + \dd q_{2} \wedge \dd p_{2}
\label{eq:scaling_symplectic}
\end{equation}
on $\T^{*}\R^{2}_{0}$. The corresponding Poisson brackets read
\begin{equation}
\set{h_{1}, h_{2}}_{\omega} = - 2h_{1}, \qquad \set{h_{1}, h_{3}}_{\omega} = -h_{2}, \qquad \set{h_{2}, h_{3}}_{\omega} = -2h_{3}.
\nonumber
\end{equation}
The Hamiltonian vector fields associated with these Hamiltonian functions, determined by the inner product condition
\be
\iota_{\X_{i}} \omega = \dd h_{i}
\label{inner}
\ee
with $1 \leq i \leq 3$, are expressed within the  global coordinates $(q_{1}, q_{2}, p_{1}, p_{2})$ of $\T^{*}\R^{2}_{0}$ as
\begin{equation}
\begin{split}
\X_{1} &= p_{1} \pdv{q_{1}} + p_{2} \pdv{q_{2}}, \qquad   \X_{3} = - q_{1}\pdv{p_{1}} - q_{2} \pdv{p_{2}},\\[2pt]
\X_{2} &= q_{1} \pdv{q_{1}} + q_{2} \pdv{q_{2}} - p_{1} \pdv{p_{1}} - p_{2} \pdv{p_{2}},
\end{split}
\label{eq:scaling:vf}
\end{equation}
which   obey the commutation relations
\begin{equation}
[\X_{1}, \X_{2}] = 2 \X_{1}, \qquad [\X_{1}, \X_{3}] = \X_{2}, \qquad [\X_{2}, \X_{3}] = 2 \X_{3}.
\label{eq:scaling:VG}
\end{equation}
Therefore, the $t$-dependent vector field on $\T^{*}\R^{2}_{0}$ given by
\begin{equation}
\X := \X_{1} + \Omega^{2}(t) \X_{3},
\nonumber
\end{equation}
which is associated with the first-order system of differential equations on $\T^{*}\R^{2}_{0}$
\begin{equation}
\begin{pmatrix}
\dot{q}_{1} \\[1pt]
\dot{q}_{2} \\[1pt]
\dot{p}_{1} \\[1pt]
\, \dot{p}_{2}
\end{pmatrix} = \begin{pmatrix}
0 & 0 & 1 & 0 \\[1pt]
0 & 0 & 0 & 1 \\[1pt]
- \Omega^{2}(t) & 0 & 0 & 0 \\[1pt]
0 & - \Omega^{2}(t) & 0 & 0
\end{pmatrix}
\begin{pmatrix}
q_{1} \\[1pt]
q_{2} \\[1pt]
p_{1} \\[1pt]
\, p_{2}
\end{pmatrix},
\label{eq:scaling:lh}
\end{equation}
 is a LH system, were both the VG Lie algebra and the LH algebra are isomorphic to $\sl(2, \R)$.

Let us  denote by $\mathbf{0}_{\R^{2}_{0}}$   the zero section of the bundle $\T^{*}\R^{2}_{0} \to \R^{2}_{0}$.  From the diffeomorphism
\begin{equation}
\R^{2}_{0} \to \R^{+} \times \S^{1}, \qquad \mathbf{x} \mapsto \left( \norm{\mathbf{x}}, \frac{\mathbf{x}}{\norm{\mathbf{x}}} \right),
\nonumber
\end{equation}
where $ \mathbf{x}$ stands for $(q_1,q_2)$ or $(p_1,p_2)$,  we find that the open subset $\T^{*}\R^{2}_{0} - \mathbf{0}_{\R^{2}_{0}} \subset \T^{*}\R^{2}_{0}$ is diffeomorphic to
\begin{equation}
\T^{*}\R^{2}_{0} - \mathbf{0}_{\R^{2}_{0}} \simeq \R^{2}_{0} \times \R^{2}_{0} \simeq \R^{+} \times \S^{1} \times \R^{+} \times \S^{1}.
\nonumber
\end{equation}
Let now $(r_{1}, \theta_{1}, r_{2}, \theta_{2})$ be polar coordinates on $\R^{+} \times \S^{1} \times \R^{+} \times \S^{1}$, such that
\be
q_1=r_1\cos\theta_{1},\qquad q_2=r_1\sin\theta_{1}, \qquad p_1=r_2\cos\theta_{2},\qquad p_2=r_2\sin\theta_{2},
\nonumber
\ee
and consider the diffeomorphism
\begin{equation}
\begin{split}
&\R^{+} \times \S^{1} \times \R^{+} \times \S^{1} \to \R^{+} \times \S^{1} \times \R^{+} \times \S^{1}, \\& (r_{1}, \theta_{1}, r_{2}, \theta_{2}) \mapsto (\rho_{1}:= r_{1}, \theta_{1}, \rho_{2}:= r_{2}/r_{1}, \theta_{2}).
\nonumber
\end{split}
\end{equation}
Within the coordinate system $(\rho_{1}, \theta_{1}, \rho_{2}, \theta_{2})$, the Hamiltonian vector fields (\ref{eq:scaling:vf})   adopt the form
\be
\label{eq:scaling:vf2}
\begin{split}
&\X_{1} =\rho_{1}  \rho_{2} \cos(\theta_{1} - \theta_{2})  \pdv{\rho_{1}} -
  \rho_{2}^2 \cos(\theta_{1} - \theta_{2})  \pdv{\rho_{2}} - \rho_{2} \sin(\theta_{1} - \theta_{2})  \pdv{\theta_{1}},\\[2pt]
  &  \X_{2} = \rho_{1}   \pdv{\rho_{1}  } -2 \rho_{2}  \pdv{\rho_{2}}  ,  \qquad
\X_{3} = - \cos(\theta_{1} - \theta_{2})  \pdv{\rho_{2}} - \frac{\sin(\theta_{1} - \theta_{2}) }{\rho_{2} }  \pdv{\theta_{2}}.
\end{split}
\ee
The  symplectic form \eqref{eq:scaling_symplectic} turns out to be
\begin{equation}
\begin{split}
\omega &= \rho_{1}   \sin (\theta_{1}- \theta_{2}) \bigl(  \rho_{2}  \dd \rho_{1} \wedge \dd \theta_{1} + \rho_{2} \dd \rho_{1} \wedge \dd \theta_{2}   + \rho_{1} \dd \rho_{2} \wedge \dd  \theta_{1} \bigr)   \\[2pt]
&\qquad  + \rho_{1} \cos(\theta_{1}- \theta_{2})\bigl(   \dd \rho_{1} \wedge \dd \rho_{2}  + \rho_{1}\rho_{2}   \dd \theta_{1} \wedge \dd \theta_{2}   \bigr),
\end{split}
\nonumber
\end{equation}
while the Hamiltonian functions $h_{1}, h_{2}$ and $h_{3}$ become
\begin{equation}
h_{1} = \frac{1}{2}\rho_{1}^{2} \rho_{2}^{2}, \qquad h_{2} = \rho_{1}^{2} \rho_{2} \cos(\theta_{1} - \theta_{2}), \qquad h_{3} = \frac{1}{2} \rho_{1}^{2}.
\nonumber
\end{equation}

Let us now consider the principal $\R^{+}$-action on $\R^{+} \times \S^{1} \times \R^{+} \times \S^{1}$ whose infinitesimal generator is given by
\begin{equation}
\mathbf{\Delta} := \frac{1}{2} \rho_{1} \pdv{\rho_{1}} \cdot
\nonumber
\end{equation}
As it satisfies the identities
\begin{equation}
\cL_{\mathbf{\Delta}} \omega = \omega, \qquad \cL_{\mathbf{\Delta}} h_{i} = h_{i}   ,\qquad  1 \leq i \leq 3,
\nonumber
\end{equation}
we see from Proposition~\ref{prop:reduction:gen_lh}  that it defines a scaling symmetry for the LH system~\eqref{eq:scaling:lh} on $\T^{*}\R^{2}_{0}- \mathbf{0}_{\R^{2}_{0}} \simeq \R^{+} \times \S^{1} \times \R^{+} \times \S^{1}$. The reduced space is diffeomorphic to $\R^{+} \times \S^{1} \times \S^{1}$,  which can be identified with the hypersurface of contact type
\begin{equation}
i: \R^{+} \times \S^{1} \times \S^{1} \equiv \set{\rho_{1} = 1} \hookrightarrow \R^{+} \times \S^{1} \times \R^{+} \times \S^{1}.
\nonumber
\end{equation}
It is straightforward to verify that
\begin{equation}
\eta := i^{*}(\iota_{\mathbf{\Delta}} \omega)
\nonumber
\end{equation}
is a contact form on $\R^{+} \times \S^{1} \times \S^{1}$ which, in the coordinates $(\rho:= \rho_{2}, \theta_{1}, \theta_{2})$ on $\R^{+} \times \S^{1} \times \S^{1}$, reads
\begin{equation}
\eta = \frac{1}{2} \bigl( \cos(\theta_{1}- \theta_{2}) \dd \rho + \rho \sin (\theta_{1}- \theta_{2}) ( \dd \theta_{1} + \dd \theta_{2}) \bigr).
\label{eq:scaling:contactform}
\end{equation}
Therefore, applying the reduction procedure of Theorem~\ref{th:scaling:reduction}, the LH system \eqref{eq:scaling:lh} on $\T^{*}\R^{2}_{0} - \mathbf{0}_{\R^{2}_{0}}$ can be reduced to a contact Lie system $(\R^{+} \times \S^{1} \times \S^{1}, \eta, \pi_{*}\X)$, where $\pi_{*}\X$ is the $t$-dependent vector field
\begin{equation}
\pi_{*}\X = \pi_{*}\X_{1} + \Omega^{2}(t) \pi_{*}\X_{3},
\label{eq:scaling:projtvf}
\end{equation}
being
\begin{equation}
\begin{split}
\pi_{*}\X_{1} &= - \rho^2  \cos(\theta_{1}- \theta_{2}) \pdv{\rho} -\rho  \sin(\theta_{1}- \theta_{2}) \pdv{\theta_{1}}  , \qquad \pi_{*}\X_{2} = - 2 \rho \pdv{\rho}, \\
\pi_{*}\X_{3} & = - \cos(\theta_{1} - \theta_{2}) \pdv{\rho} - \frac{\sin( \theta_{1} - \theta_{2})}{\rho} \pdv{\theta_{2}}
\end{split}
\label{eq:scaling:proj_vf}
\end{equation}
the projections of the generators (\ref{eq:scaling:vf2}) of the VG Lie algebra \eqref{eq:scaling:VG}.  The first-order system of ODEs on $\R^{+} \times \S^{1} \times \S^{1}$ associated with \eqref{eq:scaling:projtvf} is finally given by
\begin{equation}
\dv{\rho}{t} = - \cos(\theta_{1}- \theta_{2}) ( \rho^{2} + \Omega^{2}(t)), \qquad \dv{\theta_{1}}{t} = - \rho \sin(\theta_{1}- \theta_{2}), \qquad \dv{\theta_{2}}{t}= - \Omega^{2}(t) \frac{ \sin(\theta_{1}- \theta_{2})}{\rho}
\nonumber
\end{equation}
and its VG Lie algebra, spanned by the vector fields \eqref{eq:scaling:proj_vf}, is again isomorphic to $\sl(2,\R)$. Note that the third generator of this VG Lie algebra is related to the  Reeb vector  field $\cR$ associated with the contact form \eqref{eq:scaling:contactform} through $\pi_{*}\X_{3} = - \frac{1}{2} \cR$.

%%%%%%%%%%%%%%%%%%%%%%%%%%%%%%%%%%%%%%%%%%%%%%

\subsection{Application to a    time-dependent   thermodynamic system} \label{subsection:application2}
\label{section:s22}

In the previous case, as the contact distribution on the reduced space is co-orientable, every Hamiltonian function belonging to the LH algebra is associated with a (contact) Hamiltonian function on the reduced space. Nevertheless, in the most general case, such a function on the reduced space may not exist, as the following LH system shows.

 On the open subset   $\T^{*}\R^{3} - \mathbf{0}_{\R^{3}} \subset \T^{*}\R^{3}$, where $\mathbf{0}_{\R^{3}}$ is the zero section of the  bundle $\T^{*}\R^{3} \to \R^{3}$, consider the  following Hamiltonian functions $h_{i}$ $(1 \leq i \leq 9)$:
\begin{equation}
\begin{split}
&h_{1} := q_{1}p_{1}, \qquad h_{2} := q_{1}p_{2}, \qquad h_{3} := q_{1}p_{3}, \qquad h_{4}:= q_{2}p_{1}, \qquad h_{5}:= q_{2}p_{2}, \\
&h_{6}:= q_{2}p_{3}, \qquad h_{7} := q_{3}p_{1}, \qquad h_{8}:= q_{3}p_{2}, \qquad h_{9}:= q_{3}p_{3}.
\end{split}
\label{eq:reduction:hamfum_proj}
\end{equation}
With respect to the Poisson bracket $\set{\cdot, \cdot}_{\omega}$ induced by the canonical symplectic form
\begin{equation}
\omega = \dd q_{1} \wedge \dd p_{1} + \dd q_{2} \wedge \dd p_{2} + \dd q_{3} \wedge \dd p_{3},
\label{eq:reduction:symplectic3d}
\end{equation}
they have nonvanishing commutation relations given by
\begin{equation}
\begin{array}{lll}
 \set{h_{1}, h_{2}}_{\omega} = - h_{2},  &\quad\set{h_{1}, h_{3}}_{\omega} = - h_{3},   &\quad\set{h_{1}, h_{4}}_{\omega} = h_{4},  \\[2pt]
  \set{h_{1}, h_{7}}_{\omega} = h_{7},
&\quad \set{h_{2}, h_{4}}_{\omega} = - h_{1} + h_{5},   &\quad\set{h_{2}, h_{5}}_{\omega} = - h_{2},   \\[2pt]
 \set{h_{2}, h_{6}}_{\omega} = - h_{3},
  &\quad \set{h_{2}, h_{7}}_{\omega} = h_{8},  &\quad
 \set{h_{3}, h_{4}}_{\omega} = h_{6},     \\[2pt]
  \set{h_{3}, h_{7}}_{\omega} = - h_{1} + h_{9},
 & \quad \set{h_{3}, h_{8}}_{\omega} = - h_{2},   &\quad \set{h_{3}, h_{9}}_{\omega} = - h_{3},   \\[2pt]
 \set{h_{4}, h_{5}}_{\omega} = h_{4},   \quad
&\quad  \set{h_{4}, h_{8}}_{\omega} = h_{7},   &\quad\set{h_{5}, h_{6}}_{\omega} = - h_{6},   \\[2pt]
 \set{h_{5}, h_{8}}_{\omega} = h_{8},
&\quad \set{h_{6}, h_{7}}_{\omega} = - h_{4},   &\quad\set{h_{6}, h_{8}}_{\omega} = - h_{5} + h_{9},\\[2pt]
     \set{h_{6}, h_{9}}_{\omega} = - h_{6},
 &\quad\set{h_{7}, h_{9}}_{\omega} = h_{7},
&\quad \set{h_{8}, h_{9}}_{\omega} = h_{8},
\end{array}
\nonumber
\end{equation}
 spanning a 9D real LH algebra $\cH_{\omega}$, which is   isomorphic to the reductive Lie algebra $\mathfrak{sl}(3,\mathbb{R})\oplus\mathbb{R}$. This shows that the
 $t$-dependent vector field
\begin{equation}
\X := \sum_{i = 1}^{9} b_{i}(t) \X_{i},
\label{eq:reduction:tdep_proj}
\end{equation}
where $b_{i} \in C^{\infty}(\R)$ is an arbitrary $t$-dependent function and $\X_{i}$ is the Hamiltonian vector field associated with $h_{i}$  via (\ref{inner}) $(1 \leq i \leq 9)$, is a LH system on $\T^{*}\R^{3} - \mathbf{0}_{\R^{3}}$. More precisely, the corresponding Hamiltonian vector fields $\X_{i}$ turn out to be
\begin{equation}
\begin{split}
&\X_{1} = q_{1} \pdv{q_{1}} - p_{1} \pdv{p_{1}}, \qquad \X_{2} = q_{1} \pdv{q_{2}} - p_{2} \pdv{p_{1}}, \qquad \X_{3} = q_{1} \pdv{q_{3}} - p_{3} \pdv{p_{1}}, \\[2pt]
&\X_{4} = q_{2} \pdv{q_{1}} - p_{1} \pdv{p_{2}}, \qquad \X_{5} = q_{2} \pdv{q_{2}} - p_{2} \pdv{p_{2}}, \qquad \X_{6} = q_{2} \pdv{q_{3}} - p_{3} \pdv{p_{2}}, \\[2pt]
& \X_{7} = q_{3} \pdv{q_{1}} - p_{1} \pdv{p_{3}}, \qquad \X_{8} = q_{3} \pdv{q_{2}} - p_{2} \pdv{p_{3}}, \qquad \X_{9} = q_{3} \pdv{q_{3}} - p_{3} \pdv{p_{3}},
\end{split}
\label{eq:change1}
\end{equation}
which close as a subalgebra of the symplectic Lie algebra $\mathfrak{sp}(6,\mathbb{R})$. Hence,  the { system of ODEs} on $\T^{*}\R^{3} - \mathbf{0}_{\R^{3}}$ associated with the $t$-dependent vector field \eqref{eq:reduction:tdep_proj} reads
\begin{equation}
\begin{pmatrix}
\dot{q}_{1} \\[1pt]
\dot{q}_{2} \\[1pt]
\dot{q}_{3}\\[1pt]
\dot{p}_{1} \\[1pt]
\dot{p}_{2} \\[1pt]
\dot{p}_{3}
\end{pmatrix} = \begin{pmatrix}
b_{1}(t) & b_{4}(t) & b_{7}(t) & 0 & 0 & 0 \\[1pt]
b_{2}(t) & b_{5}(t) & b_{8}(t) & 0 & 0 & 0 \\[1pt]
b_{3}(t) & b_{6}(t) & b_{9}(t) & 0 & 0 & 0 \\[1pt]
0 & 0 & 0 & - b_{1}(t) & - b_{2}(t) & - b_{3}(t) \\[1pt]
0 & 0 & 0 & - b_{4}(t) & - b_{5}(t) & - b_{6}(t) \\[1pt]
0 & 0 & 0 & - b_{7}(t) & - b_{8}(t) &- b_{9}(t)
\end{pmatrix} \begin{pmatrix}
q_{1} \\[1pt]
q_{2} \\[1pt]
q_{3} \\[1pt]
p_{1} \\[1pt]
p_{2} \\[1pt]
p_{3}
\end{pmatrix}.
\nonumber
\end{equation}
We now consider the principal action of  the multiplicative group $\R - \set{0}$ on $\T^{*}\R^{3} - \mathbf{0}_{\R^{3}}$ defined by
\begin{equation}
\Phi: \bigl(\R - \set{0} \bigr) \times \bigl(\T^{*}\R^{3} - \mathbf{0}_{\R^{3}}\bigr) \to \T^{*}\R^{3} - \mathbf{0}_{\R^{3}}, \qquad (s, \alpha) \mapsto \Phi(s, \alpha ) := s \, \alpha.
\nonumber
\end{equation}
A routine computation shows that the Hamiltonian functions \eqref{eq:reduction:hamfum_proj} and the symplectic form \eqref{eq:reduction:symplectic3d} are $1$-homogeneous with respect to this action:
\begin{equation}
\Phi_{s}^{*} \omega = s \,\omega, \qquad \Phi_{s}^{*} h_{i} = s\, h_{i}, \qquad 1 \leq i \leq 9, \qquad s\in \R-\set{0} \!.
\nonumber
\end{equation}
Thus, from Proposition~\ref{prop:reduction:gen_lh}, this principal action is a scaling symmetry for the LH system \eqref{eq:reduction:tdep_proj}.

In this case, the reduced space $\bigl(\T^{*}\R^{3} - \mathbf{0}_{\R^{3}}\bigr) / \bigl(\R - \set{0}\bigr)$ is the so-called \textit{projective cotangent bundle} $\mathbb{P}\bigl(\T^{*}\R^{3}\bigr)$ of $\R^{3}$ \cite[Appendix 4]{Arnold1978}.  As done before, denote by $\pi: \T^{*}\R^{3} - \mathbf{0}_{\R^{3}} \to \mathbb{P}\bigl(\T^{*}\R^{3}\bigr)$ the projection.  From Theorem~\ref{th:scaling:reduction}, the LH system $\X$ \eqref{eq:reduction:tdep_proj} can be projected to a Lie system $\pi_{*}\X$ on $\mathbb{P}\bigl(\T^{*}\R^{3}\bigr)$, with $
\pi_{*}\X = \sum_{i = 1}^{9} b_i(t)\pi_{*}\X_{i}$.
Recall that the $t$-independent projected vector fields $\pi_{*}\X_{i}$ span { a} VG Lie algebra { $\pi_{*}V$} of the reduced Lie system $\pi_{*}\X$, which is isomorphic{ , as a real Lie algebra, to the VG Lie algebra $V$ spanned by the vector fields \eqref{eq:change1} associated with the LH system $\X$}.

In contrast to the oscillator system studied in Section~\ref{ex:reduction:oscillator}, where the reduced space was diffeomorphic to $\R^{+} \times \S^{1} \times \S^{1}$, the projective cotangent bundle $\mathbb{P}\bigl(\T^{*}\R^{3}\bigr)$ does not admit a globally defined contact form $\eta$ (\ref{eq:reduction:contact}), as it is nonorientable. Observe that $\mathbb{P}\bigl(\T^{*}\R^{3}\bigr)$ is diffeomorphic to $\R^{3} \times \R \mathbb{P}^{2}$, where $\R \mathbb{P}^{2}$ denotes the canonical 2D real projective space, which is nonorientable. Nevertheless,   restricting the reduced Lie system $\pi_{*}\X$ to suitable affine charts of $\mathbb{P}\bigl(\T^{*}\R^{3}\bigr)$ we can obtain interesting contact Lie systems.

Explicitly, on the affine chart
\begin{equation}
\mathscr{U}:= \mathbb{P}\bigl(\T^{*}\R^{3}\bigr) - \set{[(q_{1}, q_2, q_{3}, p_{1}, p_2, p_{3})] \in  \mathbb{P}\bigl(\T^{*}\R^{3}\bigr) : p_{1} \neq 0} \simeq \R^{5} ,
\nonumber
\end{equation}
let us consider the affine coordinates $(U, S, V, T, P) \in \R^{5}$ defined by
\begin{equation}
U:= q_{1}, \qquad S := q_{2}, \qquad V:= q_{3}, \qquad T:= - \frac{p_{2}}{p_{1}}, \qquad P:= \frac{p_{3}}{p_{1}} \cdot
\label{eq:reduction:coord_aff}
\end{equation}
The contact distribution $\cD$ of $\mathbb{P}(\T^{*}\R^{3})$ is locally described on $\mathscr{U}$  as $\cD \vert_{\mathscr{U}} = \langle \eta_{\mathscr{U}} \rangle^{\circ}$, where $\eta_{\mathscr{U}} \in \Omega^{1}(\mathscr{U})$ is the contact form on $\mathscr{U}$ given by
\begin{equation}
\eta_{\mathscr{U}} := \dd U - T \dd S + P \dd V.
\nonumber
\end{equation}
The projective cotangent bundle $\mathbb{P}(\T^{*}\R^{3})$ can be regarded as a \textit{thermodynamic phase space} \cite{Balian2001}, where the  extensive variables $(U, S, V)$ represent, in this order, the internal energy, the entropy and the volume, while the intensive variables $(T, P)$ are the temperature and the pressure, respectively. In this context, the    affine chart $\mathscr{U} \simeq    \R^{5}$ gives the energy representation of the \textit{Gibbs one-form} $\eta_{\mathscr{U}}$ (see \cite{Bazarov1964} for details).
The restriction  of the reduced Lie system $\pi_{*}\X$ to the affine chart $\mathscr{U} \simeq \R^{5}$ is a contact Lie system which, within the thermodynamic framework, may be interpreted as a mapping of one thermodynamical system into another~\cite{Mrugala1991,Mrugala2000,Eberard2007} (e.g.~when transforming an ideal gas into a real or a Van der Waals gas).   Specifically, the generators of the VG Lie algebra { $\pi_{*}V$} are expressed in terms of the affine coordinates \eqref{eq:reduction:coord_aff} as
\begin{equation}
\begin{array}{ll}
\displaystyle \pi_{*} \X_{1} = U \pdv{U} + T \pdv{T} + P \pdv{P},  &\qquad \displaystyle\pi_{*}\X_{2} = U \pdv{S} - T^{2} \pdv{T}- TP \pdv{P},   \\[10pt]
\displaystyle\pi_{*}\X_{3} = U \pdv{V} + TP \pdv{T} +  P^{2} \pdv{P},  &\qquad \displaystyle\pi_{*}\X_{4} =  S \pdv{U} + \pdv{T},    \\[10pt]
\displaystyle \pi_{*}\X_{5} = S \pdv{S} - T \pdv{T}, &\qquad \displaystyle  \pi_{*}\X_{6} = S \pdv{V} + P \pdv{T}, \\[10pt]
\displaystyle \pi_{*}\X_{7}= V \pdv{U} - \pdv{P}, & \qquad \displaystyle \pi_{*}\X_{8} = V \pdv{S} + T \pdv{P},   \\[10pt]
\displaystyle \pi_{*}\X_{9} = V \pdv{V} - P \pdv{P} \cdot
\end{array}
\nonumber
\end{equation}
Thus, the associated { system of ODEs} reads
\begin{equation}
\begin{split}
&\dv{U}{t} = b_{1}(t) U + b_{4}(t) S + b_{7}(t) V, \\[2pt]
&\dv{S}{t} = b_{2}(t) U + b_{5}(t) S + b_{8}(t) V, \\[2pt]
& \dv{V}{t} = b_{3}(t)U + b_{6}(t) S + b_{9}(t) V, \\[2pt]
&\dv{T}{t} = b_{4}(t)  + \bigl(b_{1}(t) - b_{5}(t) \bigr) T  - b_{2}(t) T^{2}+ b_{6}(t) P+  b_{3}(t) T   P, \\[2pt]
& \dv{P}{t} = - b_{7}(t) + \bigl(b_{1}(t) - b_{9}(t)\bigr)P +   b_{3}(t) P^{2} + b_{8}(t) T  -b_{2}(t) TP  .
\end{split}
\nonumber
\end{equation}
Therefore, according to the character of the thermodynamical variables, we obtain that the above system naturally splits into two subsets. On the one hand, the
derivatives of the extensive variables $(U, S, V)$ lead to three linear ODEs, involving the nine
arbitrary $t$-dependent coefficients $b_i(t)$, which can be formally solved
using standard methods{ ~\cite[Chapter 2]{Olver}. For example, if suitable conditions are imposed on the functions $b_{i}(t)$ so that the $t$-dependent vector field $\pi_{*}\mathbf{X}$ takes values in a VG Lie algebra that is solvable as a Lie algebra, then the system can be solved by quadratures \cite{Wei1963,Wei1964}.}  On the other, the differential equations for the intense variables $(T, P)$ correspond to two coupled equations of Riccati type.

 %%%%%%%%%%%%%%%%%%%%%%%%%%%%%%%%%%%%%%%%%%%%%%

\section{Contact Lie systems on the  sphere $\mathbf{S}^{3}$ from scaling symmetries} \label{section:s3}

It has recently been shown that a representation-theoretical approach can be applied to construct new intrinsic higher-dimensional LH systems. In particular, the following result was proved.

\begin{thm}[\cite{Campoamor2024}] \label{th:s3:sp4}
For every subalgebra $\g \subset \sp(4, \R)$ there exists a LH system on $\R^{4}$ with respect to the canonical symplectic form
\begin{equation}
\omega = \dd x^{0} \wedge \dd x^{1} + \dd x^{2} \wedge \dd x^{3} \in \Omega^{2}\bigl(\R^{4}\bigr),
\label{eq:s3:symplectic}
\end{equation}
expressed in the global coordinates $(x^{0}, x^{1}, x^{2}, x^{3})$ of $\R^{4}$, with a VG Lie algebra isomorphic to $\g$.
\end{thm}
This construction dealt with the fundamental representation of $\sp(4, \R)$, from which a realization of the 10D symplectic Lie algebra $\sp(4, \R)$ by means of Hamiltonian vector fields relative to \eqref{eq:s3:symplectic} was obtained, namely
\begin{align}
 & \X_{1} :=  x^{0} \pdv{x^{0}} - x^{1} \pdv{x^{1}}, & & \X_{2} :=x^{0} \pdv{x^{2}} - x^{3} \pdv{x^{1}}, & & \X_{3} :=  x^{2} \pdv{x^{0}} - x^{1} \pdv{x^{3}} , \nonumber  \\[2pt]
  &  \X_{4} :=  x^{2} \pdv{x^{2}} - x^{3} \pdv{x^{3}}, & &  \X_{5} := -x^{0} \pdv{x^{1}}, & &  \X_{6} :=  - x^{2} \pdv{x^{1}} - x^{0} \pdv{x^{3}}, \label{eq:s3:vf} \\[2pt]
 & \X_{7} :=  - x^{2} \pdv{x^{3}}, & & \X_{8} :=  x^{1} \pdv{x^{0}}, & & \X_{9} := x^{3} \pdv{x^{0}} + x^{1}\pdv{x^{2}}, \nonumber  \\[2pt]
  &  \X_{10} :=  x^{3} \pdv{x^{2}} \cdot \nonumber
\end{align}

%%%%%%%%%%%%%%%%%%%%%%%%%%%%%%%%%%%%%%%%%%%%%%

\begin{table}[t!]
\small
\caption{\small Lie brackets of the Hamiltonian vector fields $ \X_{i}$  $(1 \leq i \leq 10)$ (\ref{eq:s3:vf})
  that span a VG Lie algebra { $V \simeq \sp(4, \R)$}.}
\label{table:sp4}
\centering
\begin{tabular}{c|cccccccccc}
$[\,\cdot, \cdot\,]$ & $ \X_{1}$ & $ \X_{2}$ & $ \X_{3}$ & $ \X_{4}$ & $ \X_{5}$ & $ \X_{6}$ & $ \X_{7}$ & $ \X_{8}$ & $ \X_{9}$ & $ \X_{10}$ \\[4pt]
\hline
\\[-8pt]
	$ \X_{1}$ & 0 & $ \X_{2}$ & $- \X_{3}$ & $0$ & $2 \X_{5}$ & $ \X_{6}$ & $0$ & $-2 \X_{8}$ & $- \X_{9}$ & $0$ \\ [3pt]
$ \X_{2}$ &  & $0$ & $ \X_{1} -  \X_{4}$ & $ \X_{2}$ & $0$ & $2 \X_{5}$ & $ \X_{6}$ & $- \X_{9}$ & $-2 \X_{10}$ & $0$ \\ [3pt]
$ \X_{3}$ &  &  & $0$ & $- \X_{3}$ & $ \X_{6}$ & $2 \X_{7}$ & $0$ & $0$ & $-2 \X_{8}$ & $- \X_{9}$ \\[3pt]
$ \X_{4}$ &  &  &  & $0$ & $0$ & $ \X_{6}$ & $2 \X_{7}$ & $0$ & $- \X_{9}$ & $-2 \X_{10}$ \\ [3pt]
$ \X_{5}$ & & &  &  & $0$ & $0$ & $0$ & $- \X_{1}$ & $- \X_{2}$ & $0$ \\ [3pt]
$ \X_{6}$ &  &  &  &  &  & $0$  & $0$ & $- \X_{3}$ & $- \X_{1} -  \X_{4}$ & $- \X_{2}$ \\ [3pt]
$ \X_{7}$ &  &  &  &  &  &  & $0$ & $0$ & $- \X_{3}$ & $- \X_{4}$ \\  [3pt]
$ \X_{8}$ &  &  &  &  &  &  &  & $0$ & $0$ & $0$ \\  [3pt]
$ \X_{9}$ &  & &  &  &  &  &  &  & $0$ & $0$ \\ [3pt]
$ \X_{10}$ &  &  &  &  &  &  &  &  &  & $0$
 \end{tabular}
\end{table}

%%%%%%%%%%%%%%%%%%%%%%%%%%%%%%%%%%%%%%%%%%%%%%

\noindent
Their commutation relations are displayed in Table~\ref{table:sp4}.
This realization spans { a VG Lie algebra $V \simeq \sp(4, \R)$} of a LH system on $\R^{4}$ associated with the $t$-dependent vector field
\begin{equation}
\X := \sum_{i = 1}^{10} b_{i}(t) \X_{i}, \qquad b_{i} \in C^{\infty}(\R), \qquad 1 \leq i \leq 10.
\label{eq:s3:tvf}
\end{equation}
The corresponding first-order system of ODEs on $\R^{4}$ reads
\begin{equation}
\begin{split}
& \dv{x^{0}}{t} = b_{1}(t) x^{0}+ b_{8}(t) x^{1} + b_{3}(t) x^{2}  + b_{9}(t) x^{3} , \\[1pt]
& \dv{x^{1}}{t} =- b_{5}(t) x^{0} - b_{1}(t) x^{1} - b_{6}(t) x^{2}- b_{2}(t) x^{3}  , \\[1pt]
& \dv{x^{2}}{t} = b_{2}(t) x^{0} + b_{9}(t) x^{1} + b_{4}(t) x^{2} + b_{10}(t) x^{3}, \\[1pt]
& \dv{x^{3}}{t} =- b_{6}(t) x^{0} - b_{3}(t) x^{1}  -b_{7}(t) x^{2}- b_{4}(t) x^{3} .
\end{split}
\label{eq:s3:sp4_system}
\end{equation}
The Hamiltonian functions associated with the generators \eqref{eq:s3:vf} of the VG Lie algebra ${ V \simeq \sp(4, \R)}$, determined by the contraction condition (\ref{inner}), are given by
\begin{equation}
\begin{array}{llllll}
& \displaystyle h_{1} = x^{0} x^{1},  &\quad \displaystyle h_{2} = x^{0}x^{3},   &\quad\displaystyle h_{3} = x^{1} x^{2},   &\quad \displaystyle h_{4} = x^{2}x^{3},   &\quad \displaystyle h_{5} =  \frac{1}{2}  (x^{0} )^{2}, \\[10pt]
& \displaystyle h_{6} = x^{0} x^{2},    &\quad\displaystyle h_{7} = \frac{1}{2}  (x^{2} )^{2},   &\quad \displaystyle h_{8} = \frac{1}{2}  (x^{1} )^{2},\qquad &\quad \displaystyle h_{9} = x^{1}x^{3},   &\quad\displaystyle h_{10} = \frac{1}{2}(x^{3})^{2},
\end{array}
\label{eq:s3:ham_sym}
\end{equation}
and they span a LH algebra $\cH_{\omega} \simeq \sp(4, \R)$ with respect to the Poisson bracket $\set{\cdot, \cdot}_{\omega}$ induced by the symplectic form \eqref{eq:s3:symplectic} on $C^{\infty}\bigl(\R^{4}\bigr)$. Note that their Poisson brackets are formally the same as the Lie brackets given in Table~\ref{table:sp4} for the $ \X_{i}$'s, up to a global minus sign.

\begin{remark}
With respect to the notation appearing in \cite{Campoamor2024} we have set $(q_{1}, q_{2}, p_{1}, p_{2}) \equiv (x^{0}, x^{2}, x^{1}, x^{3})$. The reason for this change   will be explained in Section~\ref{section:contactmetric}.
\end{remark}

Now let us show how the contact reduction procedure established in the Section~\ref{section:scaling}  can be applied to these LH systems when restricted to the   open subset $\R^{4}_{0} := \R^{4} - \set{0} \subset \R^{4}$. Consider the principal $\R^{+}$-action on $\R^{4}_{0}$ whose infinitesimal generator is the radial vector field
\begin{equation}
\mathbf{\Delta} := \frac{1}{2} \left( x^{0} \pdv{x^{0}} + x^{1} \pdv{x^{1}} + x^{2} \pdv{x^{2}} + x^{3} \pdv{x^{3}} \right).
\label{eq:s3:radial}
\end{equation}
Note that this vector field can be regarded as the generator of dilations on $\R^{4}_{0}$  with respect to the usual Euclidean metric.
The symplectic form \eqref{eq:s3:symplectic} and the Hamiltonian functions \eqref{eq:s3:ham_sym} are  $1$-homogeneous with respect to the $\R^{+}$-action, as they satisfy
\begin{equation}
\cL_{\mathbf{\Delta}} \omega = \omega, \qquad  \cL_{\mathbf{\Delta}} h_{i} = h_{i} , \qquad 1 \leq i \leq 10.
\nonumber
\end{equation}
Thus, applying Proposition~\ref{prop:reduction:gen_lh}, $\mathbf{\Delta}$ defines a scaling symmetry for the LH system \eqref{eq:s3:sp4_system} on $\R^{4}_{0}$. The reduced space $\R^{4}_{0} / \R^{+}$ is the 3D sphere $\S^{3}$, which we identify with  the hypersurface of $\R^{4}_{0}$ given by $\S^{3} = F^{-1}(1)$, where $F:=  (x^{0})^{2} + (x^{1})^{2} + (x^{2})^{2} + (x^{3})^{2} \in C^{\infty}\bigl(\R^{4}_{0}\bigr)$ satisfies the condition (3) in Proposition~\ref{prop:reduction:geometry_red}.

Indeed, if we consider the inclusion $i: \S^{3} \hookrightarrow \R^{4}_{0}$, we find that
\begin{equation}
\eta := i^{*}(\iota_{\mathbf{\Delta}} \omega) =  \frac{1}{2}  \left( - x^{1} \dd x^{0} +x^{0} \dd x^{1} - x^{3} \dd x^{2} + x^{2} \dd x^{3} \right) \in \Omega^{1}\bigl(\S^{3} \bigr)
\label{eq:s3:contactform}
\end{equation}
is a contact form on $\S^{3}$, turning $(\S^{3}, \eta)$ into a (co-orientable) contact manifold.
Therefore, applying the reduction procedure of Theorem~\ref{th:scaling:reduction}, the LH system \eqref{eq:s3:sp4_system} on $\R^{4}_{0}$ can be reduced to a contact Lie system $(\S^{3}, \eta, \pi_{*}\X)$, where $\pi_{*}\X$ is the projection to $\S^{3}$ of the $t$-dependent vector field \eqref{eq:s3:tvf}:
\begin{equation}
\pi_{*}\X  = \sum_{i = 1}^{10} b_{i}(t) \pi_{*}\X_{i}.
\nonumber
\end{equation}
The projected vector fields for the generators of the VG Lie algebra \eqref{eq:s3:vf} are presented in  Table~\ref{table:s3:proj_vf}, which obviously { satisfy} the same Lie brackets shown in Table~\ref{table:sp4}.

%%%%%%%%%%%%%%%%%%%%%%%%%%%%%%%%%%%%%%%%%%%%%%

\begin{table}[t!]
\centering
 { \small
 \caption{\small The projected vector fields  $\pi_{*}\X_{i} \in \cv(\S^{3})$ $(1 \leq i \leq 10)$ for the generators of the VG Lie algebra \eqref{eq:s3:vf} expressed in the global coordinates $(x^{0}, x^{1}, x^{2}, x^{3})$ of $\R^{4}$, where $\S^{3}$ is identified with the hypersurface of $\R^{4}$ given by $(x^{0})^{2} + (x^{1})^{2} + (x^{2})^{2} + (x^{3})^{2} = 1$ and $\partial_{a} := \pdv{x^{a}}$ $(0 \leq a \leq 3)$. They span the VG Lie algebra of the reduced contact Lie system on $\S^{3}$ with respect to the contact form $\eta$ \eqref{eq:s3:contactform}, which is isomorphic to $\sp(4,\R)$.}
 \begin{adjustbox}{max width=\linewidth}
 \begin{tabular}{ll}
 \hline

 \hline
 \\[-4pt]
& $\displaystyle \begin{array}{ll}
\displaystyle &\pi_{*}\X_{1} = x^{0} \bigl(2 (x^{1})^{2} + (x^{2})^{2} + (x^{3})^{2}\bigr) \partial_{0} - x^{1} \bigl(2 (x^{0})^{2} + (x^{2})^{2} + (x^{3})^{2}\bigr) \partial_{1} \\[4pt]
&\qquad\qquad\quad -x^{2}\bigl(   (x^{0})^{2} - (x^{1})^{2} \bigr) \partial_{2}  - x^{3} \bigl(   (x^{0})^{2} - (x^{1})^{2}\bigr) \partial_{3}
\end{array}$ \\[16pt]
& $\displaystyle \begin{array}{ll}
&\pi_{*}\X_{2}  = -x^{0}\bigl(   x^{0} x^{2} - x^{1}x^{3}\bigr) \partial_{0} - \bigl\{  x^{3} \bigl( (x^{0})^{2} + (x^{2})^{2} + (x^{3})^{2} \bigr) + x^{0} x^{1} x^{2} \bigr\} \partial_{1} \\[4pt]
&\qquad\qquad\quad + \bigl\{   x^{0} \bigl( (x^{0})^{2}+(x^{1})^{2} + (x^{3})^{2}\bigr) + x^{1} x^{2} x^{3} \bigr\} \partial_{2} - x^{3}( x^{0} x^{2} - x^{1} x^{3}) \partial_{3}
\end{array}$ \\[16pt]
& $\displaystyle \begin{array}{ll}
&\pi_{*}\X_{3}  = \bigl\{   x^{2} \bigl(  (x^{1})^2  + (x^{2})^{3}+(x^{3})^{2}  \bigr) +x^{0} x^{1} x^{3}\bigr\}  \partial_{0} - x^{1} ( x^{0} x^{2}-x^{1} x^{3})  \partial_{1}  \\[4pt]
&\qquad \qquad\quad  - x^{2} ( x^{0} x^{2}-x^{1} x^{3}) \partial_{2} - \bigl\{ x^{1} \bigl( (x^{0})^2+(x^{1})^2+(x^{2})^2\bigr)+x^{0} x^{2} x^{3} \bigr\}   \partial_{3}
\end{array}$  \\[16pt]
& $\displaystyle \begin{array}{ll}
&\pi_{*}\X_{4}  = -x^{0} \bigl(   (x^{2})^{2} - (x^{3})^{2} \bigr) \partial_{0}
-  x^{1} \bigl(  (x^{2})^{2} - (x^{3})^{2} \bigr) \partial_{1}\\[4pt]
&\qquad\qquad \quad
+x^{2} \bigl( (x^{0})^2+ (x^{1})^2+2 (x^{3})^2 \bigr) \partial_{2}- x^{3} \bigl( (x^{0})^{2} + (x^{1})^{2} + 2 (x^{2})^{2} \bigr) \partial_{3}
\end{array}$\\[16pt]
& $\displaystyle \begin{array}{ll}
&\pi_{*}\X_{5}  =  x^{1} (x^{0})^2\partial_{0}  - x^{0} \bigl((x^{0})^2 + (x^{2})^2+ (x^{3})^2 \bigr) \partial_{1}  + x^{0} x^{1} x^{2}  \partial_{2} + x^{0} x^{1} x^{3} \partial_{3}
\end{array}$\\[8pt]
& $\displaystyle \begin{array}{ll}
&\pi_{*} \X_{6} = x^{0} (x^{0} x^{3}+x^{1} x^{2}) \partial_{0}
-\bigl\{ x^{2} \bigl( (x^{0})^2  + (x^{2})^2 + (x^{3})^2\bigr)- x^{0} x^{1} x^{3} \bigr\} \partial_{1}  \\[4pt]
&\qquad \qquad\quad+x^{2} (x^{0} x^{3}+x^{1} x^{2})  \partial_{2} - \bigl\{   x^{0} \bigl( (x^{0})^2+(x^{1})^2+ (x^{2})^2 \bigr)-x^{1} x^{2} x^{3}) \bigr\}\partial_{3}
\end{array}$ \\[16pt]
& $\displaystyle \begin{array}{ll}
&\pi_{*}\X_{7}  =x^{0} x^{2} x^{3} \partial_{0} +x^{1} x^{2} x^{3} \partial_{1}  + x^{3}(x^{2})^2 \partial_{2}  -  x^{2} \bigl((x^{0})^2 + (x^{1})^2 + (x^{2})^2\bigl) \partial_{3}
\end{array}$ \\[8pt]
&$\displaystyle \begin{array}{ll}
&\pi_{*}\X_{8}   = x^{1} \bigl( (x^{1})^2+(x^{2})^2+(x^{3})^2 \bigr) \partial_{0} -x^{0} (x^{1})^2 \partial_{1}-x^{0} x^{1} x^{2} \partial_{2}-x^{0} x^{1} x^{3} \partial_{3}
\end{array}$ \\[8pt]
&$\displaystyle \begin{array}{ll}
&\pi_{*}\X_{9}  = \bigl\{ x^{3}\bigl( (x^{1})^2+(x^{2})^2  +(x^{3})^2 \bigr) -x^{0} x^{1} x^{2} \bigr\} \partial_{0} -x^{1} (x^{0} x^{3}+x^{1} x^{2}) \partial_{1} \\[4pt]
&\qquad\qquad\quad +  \bigl\{x^{1}   \bigl( (x^{0})^2  +(x^{1})^2+ (x^{3})^2 \bigr) -x^{0} x^{2} x^{3} \bigr\}  \partial_{2} -x^{3} (x^{0} x^{3} +x^{1} x^{2}) \partial_{3}
\end{array}$ \\[16pt]
&  $\displaystyle \begin{array}{ll}
&\!\!\! \pi_{*}\X_{10}    =
 - x^{0} x^{2} x^{3} \partial_{0} - x^{1} x^{2} x^{3} \partial_{1} + x^{3} \bigl( ( x^{0})^{2} + (x^{1})^{2} + (x^{3})^{2} \bigr) \partial_{2} - x^{2} (x^{3})^{2} \partial_{3}
\end{array}$   \\[8pt]
 \hline

\hline
\end{tabular}
\end{adjustbox}
 \label{table:s3:proj_vf}
}
\end{table}

%%%%%%%%%%%%%%%%%%%%%%%%%%%%%%%%%%%%%%%%%%%%%%

Summarizing up, the following statement has been proved:

\begin{thm}\label{th:s3:contactLS}
For any Lie subalgebra $\g \subset \sp(4, \R)$, there exists a contact Lie system on $\bigl(\S^{3}, \eta\bigr)$ { with a} VG Lie algebra  isomorphic to $\g$.
\end{thm}
The Reeb vector field $\cR \in \cv(\S^{3})$ associated with the contact structure \eqref{eq:s3:contactform}, determined by the conditions
\begin{equation}
\eta (\cR) = 1, \qquad \iota_{\cR} \dd \eta = 0,
\nonumber
\end{equation}
turns out to be
\begin{equation}
\cR = 2 \left( - x^{1} \pdv{x^{0}} + x^{0} \pdv{x^{1}} - x^{3} \pdv{x^{2}} + x^{2} \pdv{x^{3}} \right).
\nonumber
\end{equation}
Note that,   in terms of generators of  the $\sp(4, \R)$-realization \eqref{eq:s3:vf}, it can be expressed as
\begin{equation}
\cR = -2 (\X_{5} + \X_{7} + \X_{8} + \X_{10}).
\label{eq:s3:desc_reeb}
\end{equation}
In addition, $\cR$ is a linear combination of two Killing vector fields relative to the canonical Euclidean metric on $\R^{4}$; namely, the vector field $-\X_{5} - \X_{8} = x^{0} \partial_{x^{1}} - x^{1} \partial_{x^{0}} $, representing a rotation in the $(0,1)$-plane, and $-\X_{7}- \X_{10} = x^{2} \partial_{x^{3}} - x^{3} \partial_{x^{2}}$, which is a rotation in the $(2,3)$-plane.

By means of the mapping $\R^{4} \ni (x^{0}, x^{1}, x^{2}, x^{3}) \mapsto (z_{0} := x^{0} + \mathrm{i} x^{1}, z_{1} := x^{2} + \mathrm{i} x^{3}) \in \C^{2}$, with $\mathrm{i} := \sqrt{-1}$ being the imaginary unit, we can identify $\S^{3}$ with the unit sphere on $\C^{2}$:
\begin{equation}
\S^{3} \equiv \bigl\{(z_{0}, z_{1}) \in \C^{2}: \vert z_0 \vert^{2} + \vert z_{1} \vert^{2} = 1 \bigr\}.
\nonumber
\end{equation}
Then, the integral curves of $\cR$ through a point $(z_{0}, z_{1}) \in \S^{3}$ are of the form
\begin{equation}
\R \ni t \mapsto \bigl(\e^{2\mathrm{i}t} z_{0}, \e^{2\mathrm{i}t} z_{1}\bigr) \in \S^{3}
\nonumber
\end{equation}
and all of them are diffeomorphic to $\S^{1}$. Moreover, they define the fibres of the well-known \textit{Hopf fibration}
\begin{equation}
\begin{tikzcd}
\S^{1} \arrow[r,hook] & \S^{3} \arrow[r, "\pi_{\cR}"] & \S^{2} \simeq \S^{3}/\cR ,
\end{tikzcd}
\label{eq:s3:Hopffibration}
\end{equation}
where $\S^{2}$ is the unit sphere in $\R \times \C$ and $\pi_{\cR}: \S^{3} \to \S^{2}$ is the \textit{Hopf map} defined in complex coordinates $(z_{0}, z_{1}) \in \S^{3}$ by
\begin{equation}
\pi_{\cR}(z_{0}, z_{1}) := \bigl(\abs{z_{0}}^{2} - \abs{z_{1}}^{2}, 2 z_{0} \overline{z}_{1}\bigr),
\nonumber
\end{equation}
or, equivalently, via the previous identification $\R^{4} \equiv \C^{2}$, in ambient coordinates $(x^{0}, x^{1}, x^{2}, x^{3}) \in \S^{3}$ as
\begin{equation}
\pi_{\cR}(x^{0}, x^{1}, x^{2}, x^{3}) = \bigl( (x^{0})^{2} + (x^{1})^{2} - (x^{2})^{2} - (x^{3})^{2},2(x^{0}x^{2} + x^{1}x^{3}), 2 (- x^{0}x^{3} + x^{1} x^{2})\bigr).
\nonumber
\end{equation}
With respect to the contact form \eqref{eq:s3:contactform}, as $\cL_{\cR} \dd \eta = 0$ holds, there exists a unique symplectic form $\omega \in \Omega^{2}(\S^{2})$ such that $\pi_{\cR}^{*}(\omega) = \dd \eta$. Thus, applying \cite[Proposition~3.9]{deLucas2023}, every contact Lie system of Liouville type on $(\S^{3}, \eta)$ is projected onto a LH system on $(\S^{2}, \omega)$. After a    routine calculation  one finds that such symplectic form $\omega$ on $\S^{2} \equiv \set{(x^{0}, x^{1}, x^{2}) \in \R^{3}: (x^{0})^{2} + (x^{1})^{2} + (x^{2})^{2} = 1}$ reads as
\begin{equation}
\omega =  - \frac{1}{4} (x^{2} \dd x^{0} \wedge \dd x^{1} - x^{1} \dd x^{0} \wedge \dd x^{2} + x^{0} \dd x^{1} \wedge \dd x^{2}),
\label{eq:s3:vols2}
\end{equation}
so it is a (scalar) multiple of the usual volume form on $\S^{2}$, representing the magnetic field of a magnetic monopole centered at the origin of $\R^{3}$ \cite{Ryder1980}.

In the most general setting, a contact manifold $(M, \eta)$ is called \textit{regular} if the space $M / \cR$ of integral curves of the Reeb vector field carries a smooth manifold structure such that the canonical projection $\pi_{\cR}: M \to M / \cR$ is a smooth fibration. Under some topological assumptions, it was proved in \cite{Kegel2021,Grabowska2024}  that $\pi_{\cR}: M \to M / \cR$ is either an $\S^{1}$-principal bundle or an $\R$-principal bundle. The application of these results to the case of contact Lie systems of Liouville type follows immediately from  \cite[Proposition~3.9]{deLucas2023}.

\begin{thm} \label{th:s3:reduction}
Let $(M, \eta, \X)$ be a contact Lie system of Liouville type on a regular and connected contact manifold $(M, \eta)$ whose Reeb vector field $\cR$ is complete, and let $\pi_{\cR}: M \to M / \cR$ be the corresponding smooth fibration. Then, $\pi_{\cR}: M \to M/ \cR$ is either an $\S^{1}$-principal bundle or an $\R$-principal bundle. In both cases, $(M/\cR, \omega, \pi_{\cR_{*}}\X)$, where $\omega$ is the unique differential form on $M/ \cR$ such that $\pi_{\cR}^{*}(\omega) = \dd \eta$, is a LH system. Moreover, $\eta$	is a principal connection form of the principal bundle $\pi_{\cR}: M \to M / \cR$ and $\omega$ is its curvature form.
\end{thm}

We advance that in Section~\ref{section:contactmetric} we will show how to construct contact Lie systems of Liouville type on $(\S^{3}, \eta)$ using Sasakian geometry and,  furthermore, obtaining them explicitly in Section~\ref{section:contactck}. Then, by means of the Hopf fibration \eqref{eq:s3:Hopffibration}, we will finally study the corresponding reduced LH systems on $(\S^{2}, \omega)$ in Section~\ref{section:reduction}.

 %%%%%%%%%%%%%%%%%%%%%%%%%%%%%%%%%%%%%%%%%%%%%%%%%

  \section{Three-dimensional Cayley--Klein spaces}
\label{section:s4}

The contact Lie systems on $(\S^{3}, \eta)$ obtained in Theorem~\ref{th:s3:contactLS} are   particular cases of a more general family of contact Lie systems on 3D spaces of constant curvature that will be developed further in Section~\ref{section:contactck}. For this purpose, we   highlight  in this section  some of  the algebraic and geometric aspects of the so-called \textit{3D Cayley--Klein (CK) spaces}~\cite{Ballesteros1994,Herranz2006}, which are constructed as homogeneous spaces from the
CK Lie  groups or quasi-simple orthogonal groups (for arbitrary dimension see~\cite{Gromov1990,Gromov1992,Herranz1997,Azcarraga1998,GH2021symmetry} and references therein). We remark that they comprise, among others and in a uniform setting, Riemannian,  Lorentzian and Newtonian spaces of constant curvature.

 %%%%%%%%%%%%%%%%%%%%%%%%%%%%%%%%%%%%%%%%%%%%%%%%%

  \subsection{Cayley--Klein Lie algebras and groups}
\label{section:s41}

To start with, let us consider the 6D real Lie algebra $\so(4)$ over a basis $\set{J_{ab}: 0 \leq a < b \leq 3}$ { satisfying} the following non-vanishing commutation relations
\begin{equation}
[J_{ab}, J_{ac}] =   J_{bc} ,\qquad [J_{ab}, J_{bc}] =  - J_{ac} ,\qquad  [J_{ac}, J_{bc}] =  J_{ab},\qquad a<b<c ,
\label{eq:3DCK:brackets}
\end{equation}
and with the two independent second-order Casimir operators given by
\begin{equation}
\begin{split}
C_{1} &:= J_{01}^{2} + J_{02}^{2} + J_{03}^{2} + J_{12}^{2} + J_{13}^{2} + J_{23}^{2}, \\[2pt]
C_{2} &:= J_{01}J_{23} - J_{02} J_{13}+J_{03} J_{12}  .
\label{cas1}
\end{split}
\end{equation}
The automorphisms $\Theta_{0}, \Theta_{01}, \Theta_{012}: \so(4) \to \so(4)$ defined as
\begin{equation}
\begin{split}
\Theta_{0}&: (J_{01}, J_{02}, J_{03}, J_{12}, J_{13}, J_{23}) \mapsto (-J_{01}, -J_{02}, -J_{03}, J_{12}, J_{13}, J_{23}), \\
\Theta_{01}&: (J_{01}, J_{02}, J_{03}, J_{12}, J_{13}, J_{23}) \mapsto (J_{01}, -J_{02}, -J_{03}, -J_{12}, -J_{13}, J_{23}) ,\\
\Theta_{012}&:(J_{01}, J_{02}, J_{03}, J_{12}, J_{13}, J_{23}) \mapsto (J_{01}, J_{02}, -J_{03}, J_{12}, -J_{13}, -J_{23})
\end{split}
\label{eq:3DCK:automorphism}
\end{equation}
generate a $\Z_{2} \times \Z_{2} \times \Z_{2}$ group of commuting involutive automorphisms of $\so(4)$, inducing thus the following $(\Z_{2} \times \Z_{2} \times \Z_{2})$-grading of $\so(4)$:
\begin{equation}
\so(4) = \bigoplus_{(\alpha, \beta, \gamma) } E_{(\alpha, \beta, \gamma)},\qquad  (\alpha, \beta, \gamma) \in \Z_{2} \times \Z_{2} \times \Z_{2},
\nonumber
\end{equation}
where the subspaces
\begin{equation}
\begin{split}
&E_{(1,0,0)} := \langle J_{01} \rangle, \qquad E_{(1,1,0)} := \langle J_{02} \rangle, \qquad E_{(1,1,1)} := \langle J_{03} \rangle, \\
&E_{(0,1,0)} := \langle J_{12} \rangle,\qquad   E_{(0,1,1)} := \langle J_{13} \rangle, \qquad E_{(0,0,1)} := \langle J_{23} \rangle, \\
&E_{(0,0,0)} := \set{0}
\end{split}
\nonumber
\end{equation}
satisfy $[E_{(\alpha_{1}, \beta_{1}, \gamma_{1})}, E_{(\alpha_{2}, \beta_{2}, \gamma_{2})}] = E_{(\alpha_{1} + \alpha_{2},\beta_{1} + \beta_{2}, \gamma_{1} + \gamma_{2})}$ for every $\alpha_{l}, \beta_l, \gamma_{l} \in \Z_{2}$ ($1 \leq l \leq 2$).  A particular solution of the set of $(\Z_{2} \times \Z_{2} \times \Z_{2})$-graded contractions~\cite{Montigny1991, Moody1991} from $\so(4)$ (\ref{eq:3DCK:brackets}) leads to a family of 6D real Lie algebras depending on three {\em real  graded contraction parameters} $\kappa_{1}, \kappa_{2}$ and $\kappa_{3}$. They are called {\em CK Lie algebras} (or  quasi-simple orthogonal algebras) which are collectively denoted by $\so_{
\k}(4)$ with $\k := (\kappa_{1}, \kappa_{2}, \kappa_{3})$ and satisfy the commutation rules given by~\cite{Herranz1994}:
\begin{equation}
\begin{array}{llll}
&[J_{01}, J_{02}] = \kappa_{1} J_{12} ,& \qquad [J_{01}, J_{12}] =- J_{02} ,  & \qquad [J_{02}, J_{12}] =  \kappa_{2} J_{01}    , \\[2pt]
&  [J_{01}, J_{03}] = \kappa_{1} J_{13},    & \qquad   [J_{01}, J_{13}] = -J_{03}, & \qquad [J_{03}, J_{13}] =  \kappa_{2} \kappa_{3} J_{01},  \\[2pt]
&[J_{02}, J_{03}] = \kappa_{1}  \kappa_{2}  J_{23} ,   & \qquad [J_{02}, J_{23}]  =- J_{03}, &\qquad  [J_{03}, J_{23}] =   \kappa_{3} J_{02}, \\[2pt]
& [J_{12}, J_{13}] = \kappa_{2} J_{23},     &  \qquad  [J_{12}, J_{23}] = - J_{13}, & \qquad [J_{13}, J_{23}] = \kappa_{3} J_{12},\\[2pt]
& [J_{01}, J_{23}] =0,     &  \qquad  [J_{02}, J_{13}] = 0, & \qquad [J_{03}, J_{12}] = 0.
\end{array}
\label{eq:3DCK:commrel_Cartan}
\end{equation}

 The graded contractions of the Casimir operators (\ref{cas1}) yield those of the CK Lie algebra
  $\so_{\k}(4)$~\cite{Herranz1997}, namely
\begin{equation}
\begin{split}
C_{1, \k} &:= \kappa_{2} \kappa_{3} J_{01}^{2} + \kappa_{3} J_{02}^{2} + J_{03}^{2} + \kappa_{1} \kappa_{3} J_{12}^{2} + \kappa_{1} J_{13}^{2} + \kappa_{1} \kappa_{2} J_{23}^{2}, \\
C_{2, \k} &:= \kappa_{2} J_{01} J_{23} - J_{02}J_{13}+J_{03} J_{12} .
\end{split}
\qquad
\label{eq:3DCK:casimirs}
\end{equation}
Each contraction parameter  $\kappa_m$ can be reduced to the `normalized' values  $+1, 0$ or $-1$ by rescaling  the Lie algebra generators. When all are nonzero we obtain the semisimple  real Lie algebras  $\mathfrak{so}(4)$, $\mathfrak{so}(3,1)$ and $\mathfrak{so}(2,2)$ ({\em i.e.}, all possible real forms at this dimension), while the vanishing of one (or more) $\kappa_{m}$ gives rise to really contracted Lie algebras; observe that to set $\kappa_{m}=0$ is equivalent to applying an \textit{In\"on\"u--Wigner contraction} \cite{Inonu1953}. Then, the CK family $\so_{\k}(4)$ contains $3^{3} = 27$ specific Lie algebras (some of them isomorphic).  According to the values of $\k$, the Lie algebras contained within $\so_{\k}(4)$ are displayed in Figure~\ref{fig:3DCK:cube} (see~\cite{Herranz1997,Azcarraga1998} for their algebra structure).

%%%%%%%%%%%%%%%% figure1 %%%%%%%%%%%%%%%%%
\begin{figure}[t!]
\centering
\includegraphics[width=16.5cm,height=8.2cm]{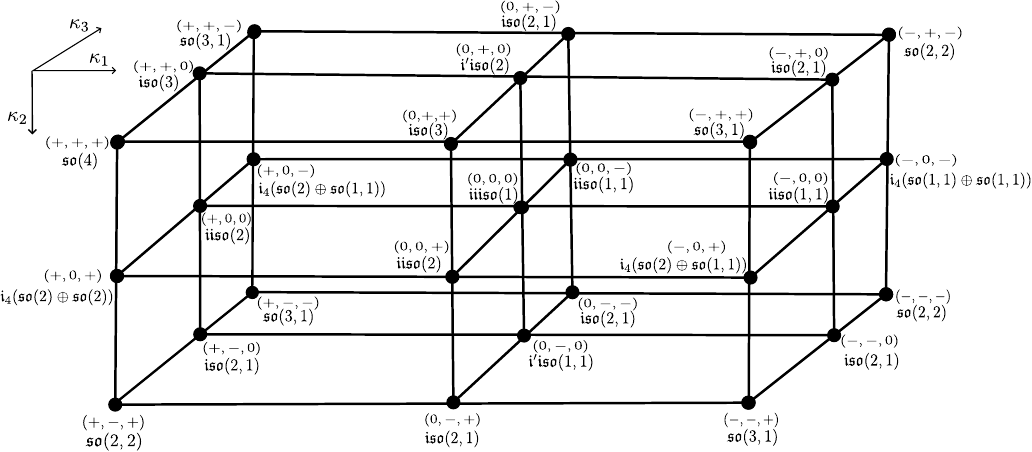}
\caption{\small The CK Lie algebras $\so_{\k}(4)$ with commutation relations \eqref{eq:3DCK:commrel_Cartan}. Each point corresponds to a set of values of the graded contraction parameters $\k=(\kappa_{1}, \kappa_{2}, \kappa_{3})$ with $\kappa_{m} \in \set{+, 0, -}$ and the semisimple algebras placed at the vertices. An In\"on\"u--Wigner contraction of the parameter $\kappa_{m} \to 0$ moves from the sides to the centre slice of the `cube' following the direction $\kappa_{m}$.}
\label{fig:3DCK:cube}
\end{figure}

  %%%%%%%%%%%%%%%%%%%%%%%%%%%%%%%%%%%%%%%%%%%%

Let us now introduce the   two-indices contraction parameters $\kappa_{ab}$ $(0 \leq a < b \leq 3)$ defined as
\begin{equation}
\begin{split}
&\kappa_{01} := \kappa_{1}, \qquad \kappa_{02} := \kappa_{1} \kappa_{2}, \qquad \kappa_{03}:= \kappa_{1} \kappa_{2} \kappa_{3}, \\
& \kappa_{12} := \kappa_{2}, \qquad \kappa_{13} := \kappa_{2} \kappa_{3}, \qquad \kappa_{23}:= \kappa_{3},
\end{split}
\nonumber
\end{equation}
and  consider   the faithful representation $\Gamma: \so_{\k}(4) \to \gl(4, \R)$ of $\so_{\k}(4)$ given by \cite{Herranz1997,Ballesteros1994, Azcarraga1998}
\begin{equation}
\Gamma(J_{ab}) := - \kappa_{ab} E_{ab} + E_{ba}, \qquad 0 \leq a < b \leq 3,
\label{eq:3DCK:rep}
\end{equation}
where $E_{ab}$ is the elementary matrix $(E_{ab})_{c d} = \delta_{ac} \delta_{bd}$. This representation establishes an isomorphism between $\so_{\k}(4)$ and the subalgebra of $\gl(4, \R)$ consisting of the matrices $X \in \gl(4, \R)$ satisfying
\begin{equation}
X^{T} \Iq_{\k} + \Iq_{\k} X = 0,
\nonumber
\end{equation}
where $\Iq_{\k}$ is the diagonal matrix given by
\begin{equation}
\Iq_{\k} := \diag(1, \kappa_{01}, \kappa_{02}, \kappa_{03}) = \diag(1, \kappa_{1}, \kappa_{1} \kappa_{2}, \kappa_{1} \kappa_{2} \kappa_{3}).
\label{eq:3DCK:quadratic_form}
\end{equation}
 The elements of the representation $\Gamma$ \eqref{eq:3DCK:rep} generate by matrix exponentiation the referred to as \textit{CK Lie group} $\SO_{\k}(4)$, whose one-parameter subgroups can be easily computed as
\begin{equation}
\exp(x \Gamma(J_{ab})) = \sum_{s = 0, s \neq a,b}^{3} E_{ss} + \Cos_{\kappa_{ab}}(x) (E_{aa} + E_{bb}) + \Sin_{\kappa_{ab}}(x) (- \kappa_{ab} E_{ab} + E_{ba}),
\label{eq:3DCK:rep_group}
\end{equation}
where we have introduced the so-called \textit{$\kappa$-dependent cosine} and \textit{sine} functions, which respectively read \cite{Herranz1997,Herranz2000}
\begin{equation}
\Cos_{\kappa}(x) :=  \begin{cases}
\cos{\sqrt{\kappa} \, x} & \  \kappa > 0, \\
\qquad 1 & \  \kappa = 0, \\
\cosh{\sqrt{-\kappa} \, x} & \  \kappa < 0.
\end{cases}  \qquad
\Sin_{\kappa} (x) :=   \begin{cases}
\frac{1}{\sqrt{\kappa}}\sin{\sqrt{\kappa} \, x} & \  \kappa > 0, \\
\qquad x & \  \kappa = 0, \\
\frac{1}{\sqrt{-\kappa}}\sinh{\sqrt{-\kappa} \, x} & \  \kappa < 0.
\end{cases}
\label{eq:kdep:cos_sin}
\end{equation}
Likewise,  the \textit{$\kappa$-tangent} is defined as
\begin{equation}
\Tan_{\kappa}(x) := \frac{\Sin_{\kappa}(x)}{\Cos_{\kappa}(x)}.
\label{eq:kdep:tan_ver}
\end{equation}
These $\kappa$-functions cover the usual circular $(\kappa >0)$ as well as the hyperbolic $(\kappa < 0)$ trigonometric functions, while the parabolic or Galilean ones $(\kappa = 0)$ are obtained from the contraction $\kappa \to 0$ as $\Cos_{0}(x)= 1$, $\Sin_{0}(x) = \Tan_{0}(x) = x$.  The derivatives of the functions \eqref{eq:kdep:cos_sin} and \eqref{eq:kdep:tan_ver} read
\begin{equation}
\dv{x}  \Cos_{\kappa} (x) = - \kappa \Sin_{\kappa} (x), \qquad \dv{x} \Sin_{\kappa}(x) = \Cos_{\kappa} (x), \qquad \dv{x}  \Tan_{\kappa}(x) = \frac{1}{\Cos_{\kappa}^{2}(x)}
\nonumber
\end{equation}
and some straightforward relations are, for instance  (cf.~\cite{Herranz2000} for many others)
\begin{equation}
\Cos_{\kappa}^{2}(x) + \kappa \Sin_{\kappa}^{2}(x) = 1, \qquad \Cos_{\kappa}(2x) = \Cos_{\kappa}^{2}(x) - \kappa \Sin_{\kappa}^{2}(x), \qquad \Sin_{\kappa}(2x) = 2 \Sin_{\kappa}(x) \Cos_{\kappa}(x).
\nonumber
\end{equation}

Note that the representation $\Gamma$ \eqref{eq:3DCK:rep} associates  each generator
$J_{ab}$ with the contraction parameter $\kappa_{ab}$ and the one-parameter subgroup (\ref{eq:3DCK:rep_group}) is isomorphic to ${\rm SO}(2)$, ${\rm SO}(1,1)$ or ${\rm ISO}(1)\equiv \R$  for positive, negative and zero values of $\kappa_{ab}$, respectively.

%%%%%%%%%%%%%%%%%%%%%%%%%%%%%%%%%%%%%%%%%%%%%%%%%

  \subsection{Cayley--Klein spaces and coordinate systems}
\label{section:s42}

The family $\so_{\k}(4)$ contains all the Lie algebras of the motion groups of the   3D  CK homogeneous spaces~\cite{Herranz2006,Herranz1997,GH2021symmetry}.   Let us now construct them explicitly.

The automorphism $\Theta_{0}$ \eqref{eq:3DCK:automorphism} gives rise to the following Cartan decomposition of the CK algebra $\so_{\k}(4)$ understood as a vector space:
\begin{equation}
\so_{\k}(4) = \h_{0} \oplus \p_{0}, \qquad \h_{0} := \langle J_{12}, J_{13}, J_{23} \rangle, \qquad \p_{0} := \langle J_{01}, J_{02}, J_{03} \rangle ,
\label{eq:3DCK:Cartan}
\end{equation}
where $\h_{0}=\so_{\kappa_2,\kappa_3}(3)$ is always a 3D CK Lie subalgebra depending on $\kappa_2$ and $\kappa_3$, while  the  subspace $\p_{0}$, whose commutators involve the remaining parameter $\kappa_1$, is only a subalgebra for the contracted cases with $\kappa_1=0$ and so abelian.  Consider now the Lie subgroup $H_{0}:= \SO_{\kappa_{2}, \kappa_{3}}(3)$ associated with the Lie subalgebra $\h_{0} $. The \textit{CK family of 3D symmetrical homogeneous spaces} is defined  from the  CK Lie group  $\SO_{\k}(4)$ by the quotient
\begin{equation}
\S^{3}_{[\kappa_{1}], \kappa_{2}, \kappa_{3}} := \SO_{\kappa_{1}, \kappa_{2}, \kappa_{3}}(4) / \SO_{\kappa_{2}, \kappa_{3}}(3),
\label{eq:3DCK:family}
\end{equation}
thus $ \SO_{\kappa_{2}, \kappa_{3}}(3)$ is the isotropy subgroup of a point.
From the Cartan decomposition \eqref{eq:3DCK:Cartan} it follows that the tangent space to $\S^{3}_{[\kappa_{1}],\kappa_{2}, \kappa_{3}}$ at the element corresponding to the identity  $ e \in \SO_{\k}(4)$ is canonically identified with the subspace $\p_{0} $. The Killing--Cartan form $g^{(0)}$ of the CK algebra $\so_{\k}(4)$ when restricted to  $\p_{0}$ is given by
\begin{equation}
g^{(0)}(J_{0a}, J_{0b}) =-4 \delta_{ab} \kappa_{0a}, \qquad 1 \leq a,b \leq 3.
\nonumber
\end{equation}
As $g^{(0)}$ identically vanishes when $\kappa_{1} = 0$, we introduce the (possibly degenerate) metric $g^{(1)}$ given by $g^{(0)} = -4 \kappa_{1} g^{(1)}$,   whose relation with the quadratic form $\Iq_{\k}$ \eqref{eq:3DCK:quadratic_form} is clear. We call  \textit{main metric}  $g_{\k}$ of $ \S^{3}_{[\kappa_{1}],\kappa_{2}, \kappa_{3}}$ the projection to $ \S^{3}_{[\kappa_{1}],\kappa_{2}, \kappa_{3}}$ of the metric obtained on $\SO_{\k}(4)$ by   right translation  of $g^{(1)}$. Then, the CK spaces \eqref{eq:3DCK:family} become a family of symmetric homogeneous spaces relative to the obtained metric. The contraction parameter $\kappa_{1}$ becomes the (sectional) \textit{curvature}, while $\kappa_{2}$ and $\kappa_{3}$ determine the \textit{signature} of the metric through $\diag(1, \kappa_{2}, \kappa_{2} \kappa_{3})$.
The notation in (\ref{eq:3DCK:family}) { indicates} these facts with the curvature $\kappa_1$ written in brackets.

 The linear representation \eqref{eq:3DCK:rep_group} of $\SO_{\k}(4)$ shows that every element $v  \in \SO_{\k}(4)$ satisfies  $v^{T} \Iq_{\k} v = \Iq_{\k}$,
so we can consider the Lie group action of $\SO_{\k}(4)$ on $\R^{4}$ as isometries of $\Iq_{\k}$. This action is not transitive as it preserves the quadratic form induced by $\Iq_{\k}$. Also, the subgroup $\SO_{\kappa_{2}, \kappa_{3}}(3) \subset \SO_{\k}(4)$ is the isotropy group of the point $O:=(1,0,0,0) \in \R^{4}$, which is thus taken as the \textit{origin} in the space $\S^{3}_{[\kappa_{1}],\kappa_{2}, \kappa_{3}}$. Nevertheless, the action becomes transitive when we restrict ourselves to the orbit of the point $O$; namely, the connected component of the submanifold
\begin{equation}
\Sigma_{\k} := \set{\x: = (x^{0}, x^{1}, x^{2}, x^{3}) \in \R^{4}: \Iq_{\k}(\x, \x) = (x^{0})^{2} + \kappa_{01} (x^{1})^{2} + \kappa_{02} (x^{2})^{2} +\kappa_{03} (x^{3})^{2} = 1}
\label{eq:3DCK:model}
\end{equation}
containing the origin $O$, allowing us to identify the space $ \S^{3}_{[\kappa_{1}],\kappa_{2}, \kappa_{3}}$ with the latter orbit.

The coordinates $(x^{0}, x^{1}, x^{2}, x^{3})$ on $\R^{4}$ satisfying the constraint \eqref{eq:3DCK:model} on $\Sigma_{\k}$ are called \textit{ambient} or \textit{Weierstrass} coordinates. In terms of these coordinates, the (main) metric $g_{\k}$ of $ \S^{3}_{[\kappa_{1}],\kappa_{2}, \kappa_{3}}$ is obtained from   the flat ambient metric in $\R^{4}$, denoted  $\widetilde{g}_{\k}$, and divided by the curvature $\kappa_1$:\begin{equation}
 {g}_{\k} := \frac{1}{\kappa_{1}} \, \widetilde{g}_{\k}= \frac{1}{\kappa_{1}} \left( \dd x^{0} \otimes \dd x^{0} + \kappa_{01} \dd x^{1} \otimes \dd x^{1} + \kappa_{02} \dd x^{2} \otimes \dd x^{2} + \kappa_{03} \dd x^{3} \otimes \dd x^{3} \right)
\label{eq:3DCK:metric}
\end{equation}
as $g_{\k} = i^{*} \widetilde{g}_{\k}$, where $i: \S^{3}_{[\kappa_{1}],\kappa_{2},\kappa_{3}} \hookrightarrow \R^{4}$ is the inclusion.
Recall that when $\kappa_{1} = 0$  the submanifold $\Sigma_{\k}$ is flat,  has two connected components and the metric $g_{\k}$ is well-defined.

We introduce two relevant types of intrinsic geodesic coordinate systems on $\S^{3}_{[\kappa_{1}], \kappa_{2}, \kappa_{3}}$. The CK Lie group $\SO_{\k}(4)$ is an isometry group of the space $\S^{3}_{[\kappa_{1}], \kappa_{2}, \kappa_{3}}$ in such a manner that   $J_{01}, J_{02}$ and $J_{03}$ generate translations moving the origin $O$ along three geodesics $l_{1}, l_{2}$ and $l_{3}$ orthogonal at  $O$, while $J_{12}, J_{13}$ and $J_{23}$ behave as  rotations  around $l_3$, $l_2$ and $l_1$, respectively. The so-called \textit{geodesic parallel coordinates} $(x, y, z)$ and \textit{geodesic polar coordinates} $(r, \theta, \phi)$ of a point $Q:= (x^{0}, x^{1}, x^{2}, x^{3})$ are defined through the action of the one-parameter subgroups \eqref{eq:3DCK:rep_group} on $O=(1,0,0,0)$ as follows~\cite{Herranz2002,Ballesteros2003}:
\begin{equation}
\begin{split}
Q &= \exp\bigl(x\, \Gamma(J_{01}) \bigr) \exp \bigl(y \,\Gamma(J_{02}) \bigr) \exp \bigl(z \,\Gamma(J_{03})\bigr) O\\[2pt]
&= \exp \bigl(\phi\, \Gamma(J_{23})\bigr) \exp\bigl(\theta\, \Gamma(J_{12}))  \exp \bigl(r \,\Gamma(J_{01})\bigr)  O,
\end{split}
\nonumber
\end{equation}
obtaining that
\begin{equation}
\begin{split}
x^{0} &= \Cos_{\kappa_{01}}(x) \Cos_{\kappa_{02}}(y) \Cos_{\kappa_{03}}(z) = \Cos_{\kappa_{1}}(r), \\[2pt]
x^{1} & = \Sin_{\kappa_{01}} (x) \Cos_{\kappa_{02}}(y) \Cos_{\kappa_{03}}(z) = \Sin_{\kappa_{1}}(r) \Cos_{\kappa_{2}}(\theta), \\[2pt]
x^{2} & = \Sin_{\kappa_{02}}(y) \Cos_{\kappa_{03}}(z) = \Sin_{\kappa_{1}}(r) \Sin_{\kappa_{2}}(\theta) \Cos_{\kappa_{3}}(\phi), \\[2pt]
x^{3} & = \Sin_{\kappa_{03}}(z) = \Sin_{\kappa_{1}}(r) \Sin_{\kappa_{2}}(\theta) \Sin_{\kappa_{3}}(\phi).
\end{split}
\label{coord}
\end{equation}
The metric \eqref{eq:3DCK:metric} in these coordinates turns out to be
\begin{equation}
\begin{split}
g_{\k} &= \Cos_{\kappa_{02}}^{2}(y) \Cos_{\kappa_{03}}^{2}(z) \,\dd x \otimes \dd x + \kappa_{2}\Cos_{\kappa_{03}}^{2}(z)\, \dd y \otimes \dd y + \kappa_{2} \kappa_{3}\, \dd z \otimes \dd z \\[2pt]
&=\dd r \otimes \dd r + \kappa_{2} \Sin_{\kappa_{1}}^{2}(r)\, \dd \theta \otimes \dd \theta + \kappa_{2} \kappa_{3} \Sin_{\kappa_{1}}^{2}(r) \Sin_{\kappa_{2}}^{2}(\theta) \,\dd \phi \otimes \dd \phi.
\end{split}
\label{eq:3DCK:coord}
\end{equation}
The referred to as the {\em connection} of  $\S^{3}_{[\kappa_{1}],\kappa_{2},\kappa_{3}}$  is the $\SO_{\k}(4)$-invariant affine connection $\nabla$ whose nonzero coefficients read as follows in geodesic polar coordinates:
\begin{equation}
\begin{split}
&\Gamma_{\theta r}^{\theta} =      \Gamma_{\phi r}^{\phi} := 1 / \Tan_{\kappa_{1}}(r), \qquad \Gamma_{\phi \theta}^{\phi} := 1 / \Tan_{\kappa_{2}}(\theta), \qquad \Gamma_{\theta \theta}^{r} := - \kappa_{2} \Cos_{\kappa_{1}}(r) \Sin_{\kappa_{1}}(r) ,\\[2pt]
&\Gamma_{\phi \phi}^{r} := - \kappa_{2} \kappa_{3} \Cos_{\kappa_{1}}(r) \Sin_{\kappa_{1}}(r) \Sin_{\kappa_{2}}^{2}(\theta), \qquad \Gamma_{\phi \phi}^{\theta} := - \kappa_{3} \Cos_{\kappa_{2}}(\theta) \Sin_{\kappa_{2}}(\theta).
\end{split}
\label{eq:3DCK:connection}
\end{equation}
In particular, when the   metric $g_{\k}$ \eqref{eq:3DCK:coord} is nondegenerate (i.e., when both $\kappa_{2} ,\kappa_3\neq 0$), we recover the Levi-Civita  connection of $g_{\k}$. If the metric is degenerate the connection is obtained by direct contraction; e.g., if $\kappa_3=0$, then $\Gamma_{\phi \phi}^{r}=\Gamma_{\phi \phi}^{\theta}=0$.

  In the nine flat CK spaces with $\kappa_1=0$, placed at the middle in the cube of Figure~\ref{fig:3DCK:cube}, the coordinates (\ref{coord}) reduce to
\begin{equation}
x^{1}  =x =  r \Cos_{\kappa_{2}}(\theta), \qquad
x^{2}  = y =  r  \Sin_{\kappa_{2}}(\theta) \Cos_{\kappa_{3}}(\phi), \qquad
x^{3}  = z= r  \Sin_{\kappa_{2}}(\theta) \Sin_{\kappa_{3}}(\phi)
\nonumber
\end{equation}
on the hyperplane $x^0=+1$. Thus, the geodesic parallel and polar coordinates can be regarded as a curved generalization of the usual Cartesian and pseudo-spherical ones, respectively~\cite{Ballesteros2003}. Under this flat contraction, the metric (\ref{eq:3DCK:coord}) becomes
\begin{equation}
\begin{split}
 g_{\k}&= \dd x \otimes \dd x + \kappa_{2} \, \dd y \otimes \dd y + \kappa_{2} \kappa_{3}\, \dd z \otimes \dd z  \\[2pt]
 &=\dd r \otimes \dd r + \kappa_{2}\, r^2\, \dd \theta \otimes \dd \theta + \kappa_{2} \kappa_{3} \, r^2 \Sin_{\kappa_{2}}^{2}(\theta) \,\dd \phi \otimes \dd \phi .
\end{split}
\nonumber
\end{equation}

A realization of the CK algebra $\so_{\k}(4)$  (\ref{eq:3DCK:commrel_Cartan})  of Killing vector fields on $\S^{3}_{[\kappa_{1}],\kappa_{2},\kappa_{3}}$ is  induced by the representation  $\Gamma$~\eqref{eq:3DCK:rep}, which is spanned by the following six vector fields expressed in Weierstrass coordinates:
\begin{equation}
\mathbf{J}_{ab} = \kappa_{ab} \,x^{b} \pdv{x^a} - x^a\pdv{x^{b}}, \qquad 0 \leq a< b \leq 3.
\label{eq:3DCK:Killing}
\end{equation}
The parametrization (\ref{coord}) allows to obtain their expressions in either of the two geodesic coordinate systems~\cite{Herranz2006,Herranz2002}.

 It is worth observing that, in addition to the three classical Riemannian spaces of constant curvature,
the family $\S^{3}_{[\kappa_{1}],\kappa_{2},\kappa_{3}}$ (\ref{eq:3DCK:family}) encompasses all possible models of $(2+1)$D  spacetimes, whose kinematical Lie algebras were introduced in~\cite{Bacry1968}, with the exception of the static one (see~\cite{GH2021symmetry} and references therein).
For our purposes let us focus on the nine most relevant spaces which correspond to set $\kappa_3=+1$,   so their Lie algebras appear on the front face of the cube in~Figure~\ref{fig:3DCK:cube} (for their 2D   counterpart in relation to LH systems we refer to~\cite{Herranz2017,Campoamor2024conformes} and references therein). These are~\cite{Ballesteros1994}:

 \begin{itemize}
\item \textit{Riemannian spaces} when $\kappa_{2} > 0$: The sphere $\S^{3}$, Euclidean $\E^{3}$ and hyperbolic $\H^{3}$ spaces for a positive/zero/negative value of the curvature $\kappa_1$, correspondingly, where the isotropy subgroup in (\ref{eq:3DCK:family}) is $ \SO_{+, +}(3)={\rm SO}(3)$.

 \item \textit{Semi-Riemannian spaces} or \textit{Newtonian spacetimes} for $\kappa_{2} = 0$: The  oscillating Newton--Hooke (NH) $\NH_{+}^{2+1}$, Galilean  $\G^{2+1}$ and expanding NH $\NH_{-}^{2+1}$ spacetimes again for a positive/zero/negative value of   $\kappa_1$, where now $ \SO_{0, +}(3)={\rm ISO}(2)$ (the 3D Euclidean group).

\item  \textit{Pseudo-Riemannian spaces} or \textit{Lorentzian spacetimes} for $\kappa_{2} < 0$:
The  anti-de Sitter $\AdS^{2+1}$, Minkowskian $\M^{2+1}$ and de Sitter $\dS^{2+1}$ spacetimes for a positive/zero/negative value of   $\kappa_1$ and  $ \SO_{-, +}(3)={\rm SO}(2,1)$  (the 3D Lorentz group).
\end{itemize}

Therefore, the six CK spaces with $\kappa_2\le 0$   provide $(2+1)$D  spacetimes of constant curvature. The precise physical interpretation is achieved by considering the usual kinematical basis
   $H$, $P_{\alpha}$, $K_{\alpha}$ and $J$ $(1 \leq \alpha \leq 2)$ corresponding, in this order, to the
 the  generators of time translations, space translations, inertial transformations (boosts) and spatial rotations,
 together with the following identifications \cite{Ballesteros1994}:
\begin{equation}
H = J_{01}, \qquad J_{02} = P_{1}, \qquad J_{03} = P_{2}, \qquad K_{1} = J_{12}, \qquad K_{2} = J_{13}, \qquad J = J_{23}.
\nonumber
\end{equation}
The  contraction parameters   are related to the cosmological constant $\Lambda$ and the speed of light $c$ through  $\kappa_{1} = - \Lambda$ and $ \kappa_{2} = - 1 / c^{2}$. Thus, the commutation relations \eqref{eq:3DCK:commrel_Cartan} now read
\begin{equation}
\begin{array}{lllll}
&\!\!\displaystyle [J, K_{\alpha}] = \epsilon_{\alpha\beta} K_{\beta},  &\quad  \displaystyle [K_{1}, K_{2}] = - \frac{1}{c^{2}} J,  & \quad \displaystyle [H, K_{\alpha}] = - P_{\alpha},  &\quad \displaystyle [P_\alpha, K_\beta] =- \frac{1}{c^{2}} \delta_{\alpha\beta}H, \\[6pt]
&\!\! [J, P_{\alpha}] = \epsilon_{\alpha\beta} P_{\beta},  &\quad \displaystyle [P_{1}, P_{2}] = \frac{\Lambda}{c^{2}}J,  &\quad \displaystyle [H, P_{\alpha}] = - \Lambda K_{\alpha},  &\quad \displaystyle [H, J] = 0,
\end{array}
\nonumber
\end{equation}
where $\epsilon_{\alpha\beta}$ is the skew-symmetric tensor given by $\epsilon_{12} = 1$ and $1 \leq \alpha, \beta \leq 2$.

 Within this framework, the Casimir operators $C_{1,\k}$ and $C_{2,\k}$ (\ref{eq:3DCK:casimirs}) correspond to the energy and the angular momentum of a free particle on the spacetime, respectively.
 The geodesic parallel coordinates $(x,y,z)$ are just the time $x \equiv t$ and the spatial $(y,z)$ ones. With respect to the geodesic polar coordinates $(r, \theta, \phi)$, the variable $r$ has dimensions of a time-like length, whilst $\theta$ corresponds to a rapidity and $\phi$ is an ordinary angle \cite{Herranz2006}.

Finally, observe that the three Newtonian spacetimes with $\kappa_2=0$ are obtained via the non-relativistic limit $c\to \infty$ from the three relativistic or Lorentzian ones.  In this contracted case, we obtain homogeneous spaces foliated by the foliation $\cF_{1}$ coming from the kernel of the degenerate (main) metric $g_{\k}$ (\ref{eq:3DCK:coord}). The leaves of this foliation (in geodesic parallel coordinates) are of the form $x = x_{0}$ (constant) and they are invariant under the action of the CK group $\SO_{\kappa_{1},0,+}(4)$. On each leaf of $\cF_{1}$ we define a nondegenerate subsidiary metric given by
\begin{equation}
g_{\k}':= \dd y \otimes \dd y +   \dd z \otimes \dd z \qquad \textup{on} \qquad x = x_{0}.
\label{subsidiary}
\end{equation}
  Thus, the main metric $g_{\k}= \dd t\otimes \dd t$ provides `absolute-time' since $t\equiv x$ and the leaves of the invariant foliation $\cF_{1}$ are the 2D `absolute-space' at $t = t_{0}$.

%%%%%%%%%%%%%%%%%%%%%%%%%%%%%%%%%%%%%%%%%%%%%%%%%

\section{Contact structure on Cayley--Klein spaces and principal bundles}
\label{section:3DCK:contactstructure}

We begin showing that \textit{all} the 3D CK spaces $\S^{3}_{[\kappa_{1}],\kappa_{2}, \kappa_{3}}$ are, in a natural way, contact manifolds.

The radial vector field $\mathbf{\Delta}$ \eqref{eq:s3:radial} is a Liouville vector field of the exact symplectic manifold $\bigl(\R^{4}_{0}, \omega\bigr)$, with $\omega$ being the exact symplectic form \eqref{eq:s3:symplectic}. Moreover, $\mathbf{\Delta}$ is transversal to the hypersurface $i: \S_{[\kappa_{1}],\kappa_{2},\kappa_{3}}^{3} \hookrightarrow \R^{4}_{0}$ as, taking into account the ambient description \eqref{eq:3DCK:model}, we have that
\begin{equation}
\cL_{\mathbf{\Delta}}\Iq_{\k}(\x,\x) = \Iq_{\k}(\x,\x) = 1 \neq 0, \qquad \x = (x^{0}, x^{1}, x^{2}, x^{3}) \in \S^{3}_{[\kappa_{1}], \kappa_{2}, \kappa_{3}}.
\nonumber
\end{equation}
Thus, $i: \S_{[\kappa_{1}],\kappa_{2},\kappa_{3}}^{3} \hookrightarrow \R^{4}_{0}$ is a hypersurface of contact type of $\bigl(\R^{4}_{0}, \omega \bigr)$ with associated contact form
\begin{equation}
\eta_{\k} := i^{*}(\iota_{\mathbf{\Delta}} \omega),
\label{eq:3DCK:contactform}
\end{equation}
turning $\bigl(\S^{3}_{[\kappa_{1}],\kappa_{2}, \kappa_{3}}, \eta_{\k} \bigr)$ into a (co-orientable) contact manifold. The expressions of the contact form $\eta_{\k}$ and its associated Reeb vector field $\cR_{\k}$ in the different coordinates of $\S^{3}_{[\kappa_{1}],\kappa_{2} , \kappa_{3}}$ introduced in Section~\ref{section:s42} are presented in Table~\ref{table:3DCK:contactgen} and explicitly described for the nine relevant 3D CK spaces $\S^{3}_{[\kappa_{1}],\kappa_{2},+1}$ in Table~\ref{table:3DCK:contactnine},  together with their metric structure. In particular, when $\k = (+1,+1,+1)$ we recover the contact structure \eqref{eq:s3:contactform} of the sphere $\S^{3}$ studied in Section~\ref{section:s3}.

In analogy to the decomposition \eqref{eq:s3:desc_reeb}, the Reeb vector field $\cR_{\k}$ is also a linear combination of the vector fields \eqref{eq:s3:vf} spanning the $\sp(4,\R)$-realization on $\R^{4}$:
\begin{equation}
\cR_{\k} = -2 ( \X_{5} + \kappa_{02} \X_{7} + \kappa_{01} \X_{8} + \kappa_{03} \X_{10}).
\nonumber
\end{equation}
It can also be expressed in terms of the Killing vector fields \eqref{eq:3DCK:Killing} as
\begin{equation}
\cR_{\k} = -2 (\mathbf{J}_{01} + \kappa_{02} \mathbf{J}_{23}),
\label{eq:3DCK:Reeb_desc}
\end{equation}
which shows that $\cR_{\k}$ is a Killing vector field of $\S^{3}_{[\kappa_{1}],\kappa_{2},\kappa_{3}}$.

Moreover, with respect to the connection $\nabla$  \eqref{eq:3DCK:connection} of $\S^{3}_{[\kappa_{1}],\kappa_{2},\kappa_{3}}$, we have that
\begin{equation}
\nabla_{\cR_{\k}} \cR_{\k} = 0,
\nonumber
\end{equation}
showing that the integral curves of the Reeb vector field $\cR_{\k}$ are all geodesics. In addition, by means of \eqref{eq:3DCK:Reeb_desc}, the flow $\mathrm{Fl}_{t}^{\cR_{\k}}$ of $\cR_{\k}$ can be easily computed in Weierstrass coordinates $\x = (x^{0}, x^{1}, x^{2}, x^{3})$ through the one-parameter subgroups \eqref{eq:3DCK:rep_group} as
\begin{equation}
\mathrm{Fl}_{t}^{\cR_{\k}} (\x) = \exp\bigl(2t\Gamma(J_{01})\bigr)  \exp\bigl(2\kappa_{02}t\Gamma(J_{23}) \bigr) (\x) = \begin{pmatrix}
x^{0} \Cos_{\kappa_{1}}(2t) - \kappa_{1} x^{1} \Sin_{\kappa_{1}}(2t) \\[2pt]
x^{0} \Sin_{\kappa_{1}}(2t) + x^{1} \Cos_{\kappa_{1}}(2t) \\[2pt]
x^{2} \Cos_{\kappa_{3}}(2 \kappa_{02} t) - \kappa_{3}x^{3} \Sin_{\kappa_{3}} (2 \kappa_{02}t)\\[2pt]
x^{2} \Sin_{\kappa_{3}}( 2 \kappa_{02}t) + x^{3} \Cos_{\kappa_{3}} ( 2 \kappa_{02}t)
\end{pmatrix} .
\label{eq:3DCK:flowReeb}
\end{equation}

The following result follows immediately from all the  previous observations.
\begin{prop} \label{prop:3DCK:Reeb_foliation}
Let $\cR_{\k}$ be the Reeb vector field of $\S^{3}_{[\kappa_{1}],\kappa_{2},\kappa_{3}}$. Then:
\begin{itemize}
\item[(1)] $\cR_{\k}$ is a Killing vector field.
\item[(2)] The flow of $\cR_{\k}$ induces an $\R$-action on $\S^{3}_{[\kappa_{1}],\kappa_{2},\kappa_{3}}$, whose orbits define a one-dimensional foliation of $\S^{3}_{[\kappa_{1}],\kappa_{2},\kappa_{3}}$ by geodesics.
\end{itemize}
\end{prop}

%%%%%%%%%%%%%%%%%%%%%%%%%%%%%%%%%%%

\begin{table}[t!]
\small
\caption{\small The contact form $\eta_{\k}$~\eqref{eq:3DCK:contactform} of the 3D CK spaces $\S^{3}_{[\kappa_{1}],\kappa_{2}, \kappa_{3}}$ (\ref{eq:3DCK:family}) and their associated Reeb vector field $\cR_{\k}$ (\ref{eq:3DCK:Reeb_desc}) expressed in Weierstrass  $(x^{0}, x^{1}, x^{2}, x^{3})$ (\ref{eq:3DCK:model}), geodesic parallel  $(x,y,z)$ and   polar  $(r, \theta, \phi)$ coordinates (\ref{coord}).}
\label{table:3DCK:contactgen}
\centering
\begin{tabular}{ll}
\hline

\hline \\[-6pt]
\quad Coordinates &\qquad Contact form $\eta_{\k}$ and Reeb vector field $\cR_{\k}$  \\[4pt]
\hline
 \\[-6pt]
$\bullet$ Weierstrass &\quad $\displaystyle \eta_{\k} = \frac{1}{2}  \left( - x^{1} \dd x^{0} + x^{0} \dd x^{1} - x^{3} \dd x^{2} + x^{2} \dd x^{3} \right)$    \\[10pt]
\quad\!$(x^{0},x^{1},x^{2}, x^{3})$&\quad $\displaystyle \cR_{\k} = 2 \left( - \kappa_{01} x^{1} \partial_{0} + x^{0} \partial_{1} - \kappa_{03} x^{3} \partial_{2}
+ \kappa_{02} x^{2} \partial_{3} \right)$
 \\ [8pt]
 $\bullet$ Geodesic parallel  &\quad $\displaystyle \eta_{\k} =  \frac{1}{2} \left( \!\Cos_{\kappa_{02}}^{2}(y) \Cos_{\kappa_{03}}^{2}(z)\, \dd x - \frac{1}{2}\Cos_{\kappa_{02}}(y)  \Sin_{\kappa_{03}}(2z)\,  \dd y + \Sin_{\kappa_{02}}(y) \, \dd z \right)$\\[12pt]
\quad\!$(x,y,z)$ &\quad  $ \displaystyle \cR_{\k} =  2 \bigl( \partial_{x} - \kappa_{03} \Cos_{\kappa_{02}}(y) \Tan_{\kappa_{03}}(z) \, \partial_{y} + \kappa_{02} \Sin_{\kappa_{02}}(y)\,  \partial_{z}  \bigr) $
 \\ [8pt]
$ \bullet $ Geodesic polar &\quad $\displaystyle \eta_{\k} = \frac{1}{2} \left( \!\Cos_{\kappa_{2}}(\theta)\,  \dd r - \frac{1}{2} \,\kappa_{2}  \Sin_{\kappa_{1}}(2r) \Sin_{\kappa_{2}}(\theta)\,  \dd \theta + \Sin_{\kappa_{1}}^{2}(r) \Sin_{\kappa_{2}}^{2}(\theta) \, \dd \phi \right) $ \\[10pt]
\quad\!$(r, \theta, \phi)$ &\quad$ \displaystyle \cR_{\k} =  2 \left(\! \Cos_{\kappa_{2}}(\theta) \, \partial_{r} - \frac{\Sin_{\kappa_{2}}(\theta)}{\Tan_{\kappa_{1}}(r)}\,  \partial_{\theta} + \kappa_{02}\, \partial_{\phi}  \right)$  \vspace*{0.2cm} \\ \hline

  \hline
\end{tabular}
\end{table}

%%%%%%%%%%%%%%%%%%%%%%%%%%%%%%%%%%%

So far we have proved that $\bigl(\S^{3}_{[\kappa_{1}],\kappa_{2},\kappa_{3}}, \eta_{\k}\bigr)$ is a connected  contact manifold whose Reeb vector field $\mathcal{R}_{\k}$ is complete. For the particular case of the sphere $\S^{3}$, as pointed out in Section~\ref{section:s3}, the contact structure is regular, inducing the Hopf fibration \eqref{eq:s3:Hopffibration}. This suggests studying the regularity of the contact structures of the rest of the 3D CK spaces. In the following, we restrict ourselves to the nine spaces $\S^{3}_{[\kappa_{1}] ,\kappa_{2},+1}$ in Table~\ref{table:3DCK:contactnine}, as they are the most representative ones within the 3D CK family and the situation in the most general case turns out to be very cumbersome.
For the sake of clarity, hereafter we drop the index $\k = (\kappa_{1}, \kappa_{2},+1)$ with $\kappa_{1}, \kappa_{2} \in \set{+1, 0, -1}$ when dealing with a particular 3D CK space.

%%%%%%%%%%%%%%%%%%%%%%%%%%%%%%%%%%%%

\begin{landscape}

\begin{table}[t]
\vskip-0.35cm
\footnotesize
 \caption{\small The contact structure on each of the nine CK spaces  $\S^{3}_{[\kappa_{1}],\kappa_{2},+1}$ (\ref{eq:3DCK:family})  according to the `normalized values' $\kappa_{1}, \kappa_{2} \in \set{+1, 0, -1}$. For each of them it is shown, in geodesic parallel  $(x, y,z)$ and polar  $(r, \theta, \phi)$  coordinates (\ref{coord}), the main metric $g_{\k}$ \eqref{eq:3DCK:coord} and the subsidiary one $g'_{\k}$ (\ref{subsidiary}), the contact form $\eta_{\k}$ \eqref{eq:3DCK:contactform} and its associated Reeb vector field $\cR_{\k}$ (\ref{eq:3DCK:Reeb_desc}) given in Table~\ref{table:3DCK:contactgen}. For the sake of clarity, we drop the index $\k = (\kappa_{1}, \kappa_{2}, +1)$.}
\label{table:3DCK:contactnine}
 \!\!\!\!\begin{tabular}{lll}
\hline

\hline \\[-6pt]
 & Geodesic parallel coordinates $(x,y,z)$ & \\[4pt]
 \hline \\[-6pt]
$\bullet$ Sphere $\S^{3}\equiv \S^{3}_{[+],+,+}$ & $\bullet$ Euclidean space $\E^{3}\equiv \S^{3}_{[0],+,+}$ & $\bullet$ Hyperbolic space $\H^{3}\equiv \S^{3}_{[-],+,+}$ \\[4pt]
$\displaystyle g=  \cos^{2}\!y  \cos^{2}\!  z\,  \dd x \otimes \dd x + \cos^{2}\!  z \, \dd y \otimes \dd y + \dd z \otimes \dd z $  & $\displaystyle g = \dd x \otimes \dd x + \dd y \otimes \dd y + \dd z \otimes \dd z$ & $\displaystyle g = \cosh^{2}\! y  \cosh^{2} \!z \, \dd x \otimes \dd x +\cosh^{2}\! z  \,\dd y \otimes \dd y + \dd z \otimes \dd z$\\[4pt]
$\displaystyle \eta = \tfrac{1}{2} \bigl(\cos^{2} \!y  \cos^{2} \!z \, \dd x - \tfrac{1}{2}\! \cos y  \sin 2z \, \dd y + \sin y \, \dd z \bigr)$ & $\displaystyle \eta = \tfrac{1}{2} ( \dd x - z\, \dd y + y \,\dd z)$& $\displaystyle \eta = \tfrac{1}{2}\bigl( \cosh^{2}\! y  \cosh^{2} \!z  \,\dd x -  \tfrac{1}{2} \!\cosh y  \sinh 2z \, \dd y + \sinh y  \,\dd z \bigr)$ \\[4pt]
$\displaystyle \cR = 2 (\partial_{x} - \cos y  \tan z \, \partial_{y} + \sin y \,  \partial_{z})$ & $\cR = 2 \partial_{x}$ & $\displaystyle \cR = 2 ( \partial_{x} + \cosh y  \tanh z\,  \partial_{y} - \sinh y  \, \partial_{z})$ \\[6pt]

$\bullet$ Oscillating NH space $\NH^{2+1}_{+}\equiv\S^{3}_{[+],0,+}$ & $\bullet$ Galilean space $\G^{2+1}\equiv \S^{3}_{[0],0,+}$ & $\bullet$ Expanding NH space $\NH^{2+1}_{-}\equiv \S^{3}_{[-],0,+}$ \\[4pt]
$\displaystyle g = \dd x \otimes \dd x$\quad  $  \displaystyle g' = \dd y \otimes \dd y + \dd z \otimes \dd z$\ \ on $x = x_0$ & $\displaystyle g = \dd x \otimes \dd x$ & $\displaystyle g = \dd x \otimes \dd x$ \quad $\displaystyle g' = \dd y \otimes \dd y + \dd z \otimes \dd z$\ \  on $x = x_0$ \\ [4pt]
$\displaystyle \eta = \tfrac{1}{2} ( \dd x - z \,\dd y + y\, \dd z) $ &  $\displaystyle g' = \dd y \otimes \dd y + \dd z \otimes \dd z$\ \  on $x = x_0$  & $\displaystyle \eta = \tfrac{1}{2} ( \dd x - z\, \dd y + y\, \dd z) $ \\[4pt]
$\displaystyle \cR = 2 \partial_{x}$ &$\displaystyle \eta = \tfrac{1}{2} ( \dd x - z \,\dd y + y \,\dd z) $\quad $\displaystyle \cR = 2 \partial_{x}$ & $\displaystyle \cR = 2 \partial_{x}$ \\[6pt]

$\bullet$ Anti-de Sitter space $\AdS^{2+1}\equiv\S^{3}_{[+],-,+}$ & $\bullet$ Minkowskian space $\M^{2+1}\equiv \S^{3}_{[0],-,+}$ & $\bullet$ de Sitter space $\dS^{2+1}\equiv \S^{3}_{[-],-,+}$ \\[4pt]
$\displaystyle g = \cosh^{2} \!y   \cosh^{2}\! z \, \dd x \otimes \dd x - \cosh^{2}\! z \, \dd y \otimes \dd y - \dd z \otimes \dd z$ & $\displaystyle g = \dd x \otimes \dd x - \dd y \otimes \dd y - \dd z \otimes \dd z$ & $g = \cos^{2} \!y  \cos^{2} \!z  \,\dd x \otimes \dd x - \cos^{2}\! z  \,\dd y \otimes \dd y - \dd z \otimes \dd z $ \\[4pt]
$\displaystyle \eta = \tfrac{1}{2}\bigl(\cosh^{2}\! y  \cosh^{2}\! z\,  \dd x - \tfrac{1}{2}\! \cosh y   \sinh 2 z\,  \dd y + \sinh y \, \dd z \bigr)$ & $\displaystyle \eta = \tfrac{1}{2}(\dd x - z\, \dd y + y\, \dd z)$ & $ \displaystyle \eta = \tfrac{1}{2}\bigl(\cos^{2}\! y  \cos^{2} \!z\,  \dd x -  \tfrac{1}{2}\! \cos y    \sin 2z\,  \dd y + \sin y \, \dd z\bigr)$\\[4pt]
$\displaystyle \cR = 2(\partial_{x} + \cosh y  \tanh z  \, \partial_{y} - \sinh y  \, \partial_{z})$ & $\displaystyle \cR = 2 \partial_{x}$ & $\cR = 2 (\partial_{x} - \cos y  \tan z \,  \partial_{y} + \sin y   \,\partial_{z})$ \\[4pt]

\hline \\[-6pt]
& Geodesic polar coordinates $(r, \theta, \phi)$ & \\[4pt]
\hline \\[-6pt]
$\bullet$ Sphere $\S^{3}\equiv \S^{3}_{[+],+,+}$ & $\bullet$ Euclidean space $\E^{3}\equiv \S^{3}_{[0],+,+}$ & $\bullet$ Hyperbolic space $\H^{3}\equiv \S^{3}_{[-],+,+}$ \\[4pt]
$\displaystyle g= \dd r \otimes \dd r + \sin^{2} \!r\,  \dd \theta \otimes \dd \theta + \sin^{2}\! r  \sin^{2} \!\theta  \,\dd \phi \otimes \dd \phi $  & $\displaystyle g = \dd r \otimes \dd r + r^{2} \dd \theta \otimes \dd \theta + r^{2} \!\sin^{2}\! \theta \, \dd \phi \otimes \dd \phi$ & $\displaystyle g = \dd r \otimes \dd r + \sinh^{2}\! r  \,\dd \theta \otimes \dd \theta + \sinh^{2}\! r  \sin^{2} \!\theta \, \dd \theta \otimes \dd \theta$\\[4pt]
$\displaystyle \eta = \tfrac{1}{2} \bigl(\cos \theta \, \dd r - \tfrac{1}{2}\!  \sin 2r \sin \theta\,  \dd \theta + \sin^{2} \!r  \sin^{2}\! \theta \, \dd \phi \bigr)$ & $\displaystyle \eta = \tfrac{1}{2} ( \cos \theta \, \dd r - r \sin \theta  \,\dd \theta + r^{2}\! \sin^{2}\! \theta \, \dd \phi )$& $\displaystyle \eta = \tfrac{1}{2}\bigl(\cos \theta\,  \dd r - \tfrac{1}{2}\!   \sinh 2r  \sin \theta \, \dd \theta + \sinh^{2}\! r  \sin^{2} \!\theta \, \dd \phi \bigr)$ \\[4pt]
$\displaystyle \cR = 2 (\cos \theta \,  \partial_{r} - \cot r  \sin \theta \,  \partial_{\theta} + \partial_{\phi} $ & $\displaystyle \cR = 2 \bigl(\cos  \theta  \, \partial_{r} - r^{-1}{\sin \theta }\,\partial_{\theta} \bigr)$ & $\displaystyle \cR = 2 ( \cos \theta  \, \partial_{r} - \coth r  \sin \theta  \, \partial_{\theta} - \partial_{\phi})$ \\[6pt]

$\bullet$ Oscillating NH space $\NH^{2+1}_{+}\equiv\S^{3}_{[+],0,+}$ & $\bullet$ Galilean space $\G^{2+1}\equiv\S^{3}_{[0],0,+}$ & $\bullet$ Expanding NH space $\NH^{2+1}_{-}\equiv \S^{3}_{[-],0,+}$ \\[4pt]
$\displaystyle g = \dd r \otimes \dd r$& $\displaystyle g = \dd r \otimes \dd r$ & $\displaystyle g = \dd r \otimes \dd r$ \\ [4pt]
  $\displaystyle g' =
\sin^{2} \!r\,  \dd \theta \otimes \dd \theta+ \theta^{2} \!\sin^{2} \!r\,   \dd \phi \otimes \dd \phi$\ \ on  $r = r_0$   & $\displaystyle g' = r^2 \dd \theta \otimes \dd \theta + r^2 \theta^{2} \dd \phi \otimes \dd \phi$\ \ on $r =  r_0$  & $\displaystyle g' =
\sinh^{2} \!r\,  \dd \theta \otimes \dd \theta+ \theta^{2} \sinh^{2} \!r\,   \dd \phi \otimes \dd \phi$\ \ on $r =  r_0$  \\[4pt]
$\displaystyle \eta = \tfrac{1}{2} (\dd r + \theta^{2} \!\sin^{2}\! r \, \dd \phi) $ \quad $\displaystyle \cR = 2 (\partial_{r} - \theta \cot r \,  \partial_{\theta})$ &$\displaystyle \eta = \tfrac{1}{2} ( \dd r + r^{2} \theta^{2} \dd \phi) $\quad $\displaystyle \cR = 2 \bigl( \partial_{r} - r^{-1}{\theta} \,\partial_{\theta} \bigr)$&$\displaystyle \eta = \tfrac{1}{2} ( \dd r + \theta^{2} \!\sinh^{2} \!r  \,\dd \phi)$\quad $\displaystyle \cR = 2 (\partial_{r} - \theta \coth r   \,\partial_{\theta})$ \\[6pt]

$\bullet$ Anti-de Sitter space $\AdS^{2+1}\equiv \S^{3}_{[+],-,+}$ & $\bullet$ Minkowskian space $\M^{2+1}\equiv\S^{3}_{[0],-,+}$ & $\bullet$ de Sitter space $\dS^{2+1}\equiv \S^{3}_{[-],-,+}$ \\[4pt]
$\displaystyle g = \dd r \otimes \dd r - \sin^{2}\! r \, \dd \theta \otimes \dd \theta - \sin^{2}\! r  \sinh^{2} \!\theta\,  \dd \phi \otimes \dd \phi$ & $\displaystyle g = \dd r \otimes \dd r - r^{2} \dd \theta \otimes \dd \theta - r^{2} \!\sinh^{2}\! \theta \, \dd \phi \otimes \dd \phi$ & $\displaystyle g = \dd r \otimes \dd r - \sinh^{2}\! r  \,\dd \theta \otimes \dd \theta - \sinh^{2} \!r  \sinh^{2}\! \theta \, \dd \phi \otimes \dd \phi$\\[4pt]
$\displaystyle \eta = \tfrac{1}{2}\bigl(\cosh \theta \, \dd r +  \tfrac{1}{2} \!\sin 2r  \sinh \theta\,  \dd \theta + \sin^{2} \!r  \sinh^{2}\! \theta  \,\dd \phi \bigr) $ & $\displaystyle \eta = \tfrac{1}{2}(\cosh \theta \, \dd r + r \sinh \theta \, \dd \theta + r^{2} \!\sinh^{2} \!\theta \, \dd \phi)$ & $ \displaystyle \eta = \tfrac{1}{2}\bigl(\cosh \theta\,  \dd r + \tfrac{1}{2} \!  \sinh 2r  \sinh \theta  \,\dd \theta + \sinh^{2}\! r  \sinh^{2}\! \theta  \, \dd \theta\bigr)$\\[4pt]
$\displaystyle \cR = 2 (\cosh \theta  \, \partial_{r} - \cot r  \sinh \theta  \, \partial_{\theta} - \partial_{\phi})$ &  $\displaystyle \cR = 2 \bigl( \cosh \theta  \, \partial_{r} - r^{-1} {\sinh \theta } \,\partial_{\theta} \bigr)$ & $\cR = 2 (\cosh \theta \,  \partial_{r} - \coth r  \sinh \theta   \,\partial_{\theta} + \partial_{\phi})$\\[6pt]
\hline

\hline
\end{tabular}
\end{table}

\end{landscape}

%%%%%%%%%%%%%%%%%%%%%%%%%%%%%%%%%%%%

%%%%%%%%%%%%%%%%%%%%%%%%%%%%%%%%%%%%

\subsection{Positive curvature: $\mathbf{S}^{1}$-principal bundles}
\label{subsection:fibre_pos}

Similarly to the sphere $\S^{3}$, the contact structure for the two remaining spaces with positive curvature,  the oscillating NH  $\NH_{+}^{2+1}$ and the anti-de Sitter  $\AdS^{2+1}$ spacetimes shown in the first column in Table~\ref{table:3DCK:contactnine} is also regular,   giving rise to the $\S^{1}$-principal bundles described below.

%%%%%%%%%%%%%%%%%%%%%%%%%%%%%%%%%%%%

\subsubsection{Principal bundle $\S^{1} \hookrightarrow \AdS^{2+1} \to \C \H^{1}$}
\label{subsection:AdS}

Taking into account the ambient description \eqref{eq:3DCK:model}, the anti-de Sitter space  $\AdS^{2+1}\equiv \S^{3}_{[+],-,+}$ can be identified with
\begin{equation}
\AdS^{2+1} = \set{ \bigl(z_{0} = x^{0} + \mathrm{i} x^{1}, z_{1} = x^{2} + \mathrm{i} x^{3} \bigr) \in \C^{2}: \abs{z_{0}}^{2} - \abs{z_{1}}^{2} = 1}.
\nonumber
\end{equation}
The Reeb vector field $\cR$,  in the Weierstrass coordinates given in Table~\ref{table:3DCK:contactgen}, reads as
\begin{equation}
\cR = 2 \left( - x^{1} \pdv{x^{0}} + x^{0} \pdv{x^{1}} + x^{3} \pdv{x^{2}} - x^{2} \pdv{x^{3}} \right).
\nonumber
\end{equation}
Its integral curves are of the form
\begin{equation}
\R \ni t \mapsto \bigl(\mathrm{e}^{2 \mathrm{it}} z_{0}, \mathrm{e}^{-2 \mathrm{i}t} z_{1} \bigr), \qquad (z_{0}, z_{1}) \in \AdS^{2+1},
\nonumber
\end{equation}
and all of them are diffeomorphic to $\S^{1}$. The space $\AdS^{2+1}/\cR$ of integral curves of the Reeb vector field is diffeomorphic to the \textit{complex hyperbolic line} $\C \H^{1}$. From now on, we only consider the so-called \textit{ball model} of $\C \H^{1}$, given by (see \cite[Chapter~3]{Goldman1999}):
\begin{equation}
\C \H^{1} \equiv \bigl\{ z \in \C: \abs{z}^{2} <1 \bigr\},
\nonumber
\end{equation}
also known as the \textit{Poincar\'e disk model} for the real hyperbolic plane $\H^2$.
 Then, we have an $\S^{1}$-principal bundle
\begin{equation}
\begin{tikzcd}
\S^{1} \arrow[r, hook]& \AdS^{2+1} \arrow[r, "\pi_{\cR}"] & \C \H^{1} \simeq \AdS^{2+1}/\cR
\end{tikzcd}
\label{fibAds}
\end{equation}
where the fibration $\pi_{\cR}: \AdS^{2+1} \to \C \H^{1}$ is given in complex coordinates $(z_{0}, z_{1}) \in \AdS^{2+1}$ by
\begin{equation}
\pi_{\cR}(z_{0}, z_{1}) := \frac{z_{1}}{\overline{z}_{0}}
\nonumber
\end{equation}
or, equivalently, in Weierstrass coordinates $\bigl(x^{0}, x^{1}, x^{2}, x^{3}\bigr) \in \AdS^{2+1}$, as
\begin{equation}
\pi_{\cR}\bigl(x^{0}, x^{1}, x^{2}, x^{3} \bigr) = \left( \frac{x^{0} x^{2} - x^{1} x^{3}}{(x^{0})^{2} + (x^{1})^{2}}, \frac{x^{0} x^{3} + x^{1} x^{2}}{(x^{0})^{2} + (x^{1})^{2}} \right).
\nonumber
\end{equation}
Within the global coordinates $(u, v)$ on $\C \H^{1}$, with $z = u + \mathrm{i}v$, let us consider the symplectic form
\begin{equation}
\omega := \frac{1}{(1 - u^{2} - v^{2})^{2}} \,\dd u \wedge \dd v \in \Omega^{2}\bigl(\C \H ^{1} \bigr),
\label{eq:3DCK:symplectic_complexhyper}
\end{equation}
which is a scalar multiple of the area form associated with the usual Poincar\'e metric in  $\H^2$. As it satisfies $\pi_{\cR}^{*}(\omega) = \dd \eta$, we see that any contact Lie system of Liouville type on $\AdS^{2+1}$ can be projected to a LH system on $\C \H^{1}$ with respect to the symplectic form \eqref{eq:3DCK:symplectic_complexhyper}.

%%%%%%%%%%%%%%%%%%%%%%%%%%%%%%%%%%%%

\subsubsection{Principal bundle $\S^{1} \hookrightarrow \NH_{+}^{2+1} \to \R^{2}$}

Topologically, the oscillating NH space $\NH^{2+1}_{+}\equiv \S^{3}_{[+],0,+}$  is just the `cylinder'
\begin{equation}
\NH^{2+1}_{+} \equiv \set{ \bigl(z = x^{0} + \mathrm{i} x^{1}, x^{2}, x^{3} \bigr) \in \C \times \R^{2}: \abs{z}^{2} = 1} \simeq \S^{1} \times \R^{2}.
\nonumber
\end{equation}
The Reeb vector field $\cR$ is expressed in Weierstrass coordinates as
\begin{equation}
\cR = 2 \left(- x^{1} \pdv{x^{0}} + x^{0} \pdv{x^{1}} \right).
\nonumber
\end{equation}
It follows that the integral curves take the form
\begin{equation}
\R \ni t \mapsto \bigl(\mathrm{e}^{2 \mathrm{i} t} z, x^{2}, x^{3} \bigr) \in \NH^{2+1}_{+}, \qquad \bigl(z, x^{2}, x^{3} \bigr) \in \NH^{2+1}_{+},
\nonumber
\end{equation}
all being diffeomorphic to $\S^{1}$. Thus, the Reeb vector field $\cR$  defines an $\S^{1}$-action on $\NH^{2+1}_{+}$ such that its space of integral curves $\NH^{2+1}_{+} / \cR$ is diffeomorphic to the real plane $\R^{2}$. Hence, we have a trivial $\S^{1}$-principal bundle
\begin{equation}
\begin{tikzcd}
\S^{1} \arrow[r, hook]& \NH^{2+1}_{+} \arrow[r, "\pi_{\cR}"] & \R^{2} \simeq \NH^{2+1}_{+} / \cR,
\end{tikzcd}
\nonumber
\end{equation}
where $\pi_{\cR}: \NH^{2+1}_{+} \to \R^{2}$ is just the projection onto the last factor
\begin{equation}
\pi_{\cR} \bigl(z, x^{2}, x^{3} \bigr) := \bigl(x^{2}, x^{3} \bigr).
\label{eq:3DCK:fibrationNHp}
\end{equation}
Within the global coordinates $(q, p)$ on $\R^{2}$, the canonical symplectic form
\begin{equation}
\omega := \dd q \wedge \dd p \in \Omega^{2} \bigl(\R^{2} \bigr)
\label{eq:3DCK:symplecticplane}
\end{equation}
satisfies $\pi_{\cR}^{*}(\omega) = \dd \eta$. Therefore, any contact Lie system of Liouville type on $\NH^{2+1}_{+}$ can be projected to a LH system on $\R^{2}$ with respect to the canonical symplectic form \eqref{eq:3DCK:symplecticplane}.

%%%%%%%%%%%%%%%%%%%%%%%%%%%%%%%%%%%%

\subsection{Nonpositive curvature: $\mathbb{R}$-principal bundles}
\label{subsection:fibre_nonpos}

In contrast to the positive curvature scenario, not all of the six spaces $\S^{3}_{[\kappa_{1}],\kappa_{2},+}$ with $\kappa_{1} \leq 0$, shown in second and third columns in Table~\ref{table:3DCK:contactnine}, are regular contact manifolds. Indeed, the contact structure of the de Sitter spacetime $\dS^{2+1}\equiv \S^{3}_{[-],-,+}$ cannot be regular, as the $\R$-action induced by the flow of its Reeb vector field has two non-diffeomorphic orbits (from \cite{Grabowska2024}, we recall that all the orbits of the $\R$-action must be diffeomorphic to each other). Actually, the  orbit $\mathcal{O}_{O}$ of the origin $O = (1,0,0,0)$ and the orbit $\mathcal{O}_{Q}$ of the point $Q:=(0,0,1,0)$ turn out to be
\begin{equation}
\mathcal{O}_{O} = \set{(\cosh(2t), \sinh(2t),0,0): t \in \R} \simeq \R, \qquad \mathcal{O}_{Q} = \set{(0,0,\cos(2t), \sin(2t)): t \in \R} \simeq \S^{1}.
\nonumber
\end{equation}
Nevertheless, the contact structure of the remaining nonpositive curvature spaces, i.e., the three flat spaces $\S^{3}_{[0],\kappa_{2},+}$, the hyperbolic space $\H^{3}$ and the NH spacetime $\NH_{-}^{2+1}$  is regular, leading in this case to $\R$-principal bundles.

%%%%%%%%%%%%%%%%%%%%%%%%%%%%%%%%%%%%

\subsubsection{Principal bundle $\R \hookrightarrow \{ \E^{3}, \G^{2+1}, \M^{2+1} \} \to \R^{2}$}

Within the ambient description \eqref{eq:3DCK:model}, the flat space $\S^{3}_{[0],\kappa_{2},+}$ is the affine hyperplane $  \set{x^{0} = 1} \subset \R^{4}$, which is canonically identified with $\R^{3}$, regardless of the value of $\kappa_{2}$. Recall that the spaces $\E^{3} \equiv \S^{3}_{[0],+,+}$, $\G^{2+1} \equiv \S^{3}_{[0],0,+}$ and $\M^{2+1} \equiv \S^{3}_{[0],-,+}$, despite being topologically equivalent, are distinguished by their metric structure \eqref{eq:3DCK:coord} through the contraction parameter $\kappa_{2}$, as shown in the middle column in Table~\ref{table:3DCK:contactnine}. The Reeb vector field $\cR$, in Weierstrass coordinates, is simply the translation (see Table~\ref{table:3DCK:contactgen})
\begin{equation}
\cR =2 \pdv{x^{1}} ,
\nonumber
\end{equation}
and its integral curves are the lines parallel to $ \R \ni t \mapsto (1,t, 0,0) \in \S^{2}_{[0],\kappa_{2},+}$. Note that, when $\kappa_{2} \leq 0$, these are time-like   geodesics. In particular, on $\G^{2+1} $ they give rise to the invariant foliation $\mathcal{F}_{1}$ which provides `absolute-time'.

Then, we have a trivial $\R$-principal bundle
\begin{equation}
\begin{tikzcd}
\R \arrow[r, hook]& \S^{3}_{[0],\kappa_{2},+} \arrow[r, "\pi_{\cR}"] & \R^{2} \simeq \S^{3}_{[0],\kappa_{2},+}/\cR,
\end{tikzcd}
\nonumber
\end{equation}
where $\pi_{\cR}: \S^{3}_{[0],\kappa_{2},+} \to \R^{2}$ is just the projection onto the last two coordinates
\begin{equation}
\pi_{\cR}\bigl(1, x^{1}, x^{2}, x^{3}\bigr) := \bigl(x^{2}, x^{3} \bigr).
\label{eq:3DCK:fibrationflat}
\end{equation}
Clearly, the canonical symplectic form $\omega$ \eqref{eq:3DCK:symplecticplane}  of $\R^{2}$ satisfies $\pi_{\cR}^{*}(\omega) = \dd \eta$. Hence, as in the case of $\NH^{2+1}_{+}$, every contact Lie system of Liouville type on $\S^{3}_{[0],\kappa_{2},+}$ can be projected to a LH system on $\R^{2}$ with respect to the canonical symplectic form \eqref{eq:3DCK:symplecticplane}.

%%%%%%%%%%%%%%%%%%%%%%%%%%%%%%%%%%%%

\subsubsection{Principal bundle $\R \hookrightarrow \H^{3} \to \R^{2}$}
\label{subsection:3DCK:H3}

The so-called \textit{hyperboloid model} of the hyperbolic space $\H^{3} \equiv \S^{3}_{[-],+,+}$ is given by the ambient description \eqref{eq:3DCK:model}:
\begin{equation}
\H^{3} = \set{ \bigl(x^{0}, x^{1}, x^{2},  x^{3} \bigr) \in \R^{4}: (x^{0})^{2} - (x^{1})^{2} - (x^{2})^{2} - (x^{3})^{2} = 1,\    x^{0} \ge 1 }.
\nonumber
\end{equation}
The Reeb vector field $\cR$ in Weierstrass coordinates is
\begin{equation}
\cR = 2 \left( x^{1} \pdv{x^{0}} + x^{0} \pdv{x^{1}} + x^{3} \pdv{x^{2}} - x^{2} \pdv{x^{3}} \right).
\label{eq:3DCK:ReebH3}
\end{equation}
Unlike the $\S^{3}$ and the $\AdS^{2+1}$ cases, the space $\H^{3}$ cannot be described in terms of complex numbers alone, but as a suitable combination of complex numbers and the so-called \textit{split-complex numbers} \cite{Yaglom1979,Gromov1990,Kisil2012}. The set of split-complex numbers is the Clifford algebra $\mathrm{Cl}_{1,0}(\R)$, which is isomorphic to the quotient ring $\R[T]/(T^{2}-1)$ as an $\R$-algebra, while isomorphic to the real plane $\R^{2}$ as a topological space. In terms of the so-called \textit{split-complex unit} $\mathrm{j}$ (which can be identified with the class of $T$ in $\R[T]/(T^{2}-1)$), that commutes with real numbers and satisfies $\mathrm{j}^{2} = +1$, the Clifford algebra $\mathrm{Cl}_{1,0}(\R)$ can be described as
\begin{equation}
\mathrm{Cl}_{1,0}(\R) = \set{z = \xi_{1} + \mathrm{j}\xi_{2}: \xi_{1}, \xi_{2} \in \R} .
\label{eq:3DCK:Clifford}
\end{equation}
By means of the Clifford norm
\[\abs{z}^{2}_{\mathrm{Cl}_{1,0}(\R)}  = \xi_{1}^{2} - \xi_{2}^{2}, \qquad z = \xi_{1} + \mathrm{j} \xi_{2} \in \mathrm{Cl}_{1,0}(\R) ,
\]
  the hyperbolic space $\H^{3}$ is just
\begin{equation}
\H^{3} = \set{ \bigl(z_{0} = x^{0} + \mathrm{j} x^{1}, z_{1} = x^{2} + \mathrm{i} x^{3} \bigr) \in \mathrm{Cl}_{1,0}(\R) \times \C: \abs{z_{0}}^{2}_{\mathrm{Cl}_{1,0}(\R)} - \abs{z_{1}}^{2} = 1,\  x^{0} \ge 1 }.
\nonumber
\end{equation}
Moreover, through the Euler's formula analogue
\begin{equation}
\mathrm{e}^{\mathrm{j}x} = \cosh x + \mathrm{j} \sinh x , \qquad x \in \R,
\label{eq:3DCK:Euler_hyperbolic}
\end{equation}
the  integral curves of the Reeb vector field \eqref{eq:3DCK:ReebH3} can be easily expressed as
\begin{equation}
\R \ni t \mapsto \bigl(\mathrm{e}^{2 \mathrm{j} t}z_{0}, \mathrm{e}^{- 2 \mathrm{i} t} z_{1} \bigr), \qquad (z_{0}, z_{1}) \in \H^{3}.
\nonumber
\end{equation}
The space of integral curves of the Reeb vector field $\H^{3} / \cR$ is diffeomorphic to $\C \simeq \R^{2}$. The diffeomorphism is induced by the smooth fibration
\begin{equation}
\pi_{\cR}: \H^{3} \to \C, \qquad (z_{0}, z_{1}) \mapsto  z_{1} \exp{ \frac{\mathrm{i}}{2} \log \left( \frac{x^{0} + x^{1}}{x^{0} - x^{1}}  \right)},
\nonumber
\end{equation}
yielding an $\R$-principal bundle
\begin{equation}
\begin{tikzcd}
\R \arrow[r, hook]& \H^{3} \arrow[r, "\pi_{\cR}"] & \R^{2} \simeq \H^{3}/\cR.
\end{tikzcd}
\nonumber
\end{equation}
A routine computation shows that the canonical symplectic form $\omega$ \eqref{eq:3DCK:symplecticplane} on $\R^{2}$ satisfies $\pi_{\cR}^{*}(\omega) = \dd \eta$. Hence, any contact Lie system of Liouville type on $\H^{3}$ can be projected to a LH system on $\R^{2}$ with respect to the canonical symplectic form \eqref{eq:3DCK:symplecticplane}.

%%%%%%%%%%%%%%%%%%%%%%%%%%%%%%%%%%%%

\subsubsection{Principal bundle $\R \hookrightarrow \NH_{-}^{2+1} \to \R^{2}$}

The ambient description of the expanding NH space $\NH_{-}^{2+1}\equiv \S^{3}_{[-],0,+}$ coming from \eqref{eq:3DCK:model} is
\begin{equation}
\NH_{-}^{2+1} = \set{ \bigl(x^{0}, x^{1}, x^{2}, x^{3}\bigr) \in \R^{4}: (x^0)^{2} - (x^{1})^{2} = 1, \   x^{0} \ge 1 }.
\nonumber
\end{equation}
Topologically, this is just the Euclidean space $\NH_{-}^{2+1} \simeq \R^{3}$. As in the case of the hyperbolic space $\H^{3}$ studied above, we can describe $\NH_{-}^{2+1}$ by means of split-complex numbers \eqref{eq:3DCK:Clifford} as
\begin{equation}
\NH_{-}^{2+1} = \set{ \bigl(z_{0} = x^{0} + \mathrm{j} x^{1}, x^{2}, x^{3} \bigl) \in \mathrm{Cl}_{1,0}(\R) \times \R^{2}: \abs{z_{0}}_{\mathrm{Cl}_{1,0}(\R)} = 1, \   x^{0} \ge 1 }.
\nonumber
\end{equation}
In this case, the Reeb vector field $\cR$ is just the hyperbolic rotation (i.e.~a boost in the spacetime)
\begin{equation}
\cR = 2 \left( x^{1} \pdv{x^{0}} + x^{0} \pdv{x^{1}} \right),
\nonumber
\end{equation}
and its integral curves, all of them diffeomorphic to $\R$, are directly expressed using \eqref{eq:3DCK:Euler_hyperbolic}:
\begin{equation}
\R \ni t \mapsto \bigl(\mathrm{e}^{2\mathrm{j}t}z_{0}, x^{2}, x^{3}\bigr), \qquad \bigl(z_{0}, x^{2}, x^{3}\bigr) \in \NH_{-}^{2+1}.
\nonumber
\end{equation}
This leads to the trivial $\R$-principal bundle
\begin{equation}
\begin{tikzcd}
\R \arrow[r, hook]& \NH_{-}^{2+1} \arrow[r, "\pi_{\cR}"] & \R^{2} \simeq \NH_{-}^{2+1}/\cR,
\end{tikzcd}
\nonumber
\end{equation}
where the smooth fibration $\pi_{\cR}: \NH_{-}^{2+1} \to \R^{2}$ is just the projection onto the last two coordinates
\begin{equation}
\pi_{\cR} \bigl(z_{0}, x^{2}, x^{3} \bigr) := \bigl(x^{2}, x^{3} \bigr).
\label{eq:3DCK:fibrationNHm}
\end{equation}
Clearly, the canonical symplectic form $\omega$ \eqref{eq:3DCK:symplecticplane} on $\R^{2}$ satisfies $\pi_{\cR}^{*} \omega = \dd \eta$. Thus, any contact Lie system of Liouville type on $\NH_{-}^{2+1}$ can be projected to a LH system on $\R^{2}$ with respect to the canonical symplectic form \eqref{eq:3DCK:symplecticplane}.

%%%%%%%%%%%%%%%%%%%%%%%%%%%%%%%%%%%%

\subsection{Curvature-dependent principal bundles from one-dimensional spaces}

Finally, let us show that all the principal bundles obtained from the regular contact structures can be interpreted within the CK geometries framework as follows. First of all, let us consider the projection
\begin{equation}
\R^{4} \ni \bigl(x^{0}, x^{1}, x^{2}, x^{3} \bigr) \mapsto \bigl(x^{0}, x^{1}\bigr) \in \R^{2}
\label{eq:3DCK:projection1DCK}
\end{equation}
onto the first two factors, so the image of the submanifold $\Sigma_{\k}$ \eqref{eq:3DCK:model} under this mapping is just
\begin{equation}
\Sigma_{\kappa_1} := \set{ \bigl(x^{0}, x^{1} \bigr) \in \R^{2}: (x^{0})^{2} + \kappa_1 (x^{1})^{2} = 1}.
\nonumber
\end{equation}
We call \textit{1D CK space $\S^{1}_{[\kappa_1]}$}~\cite{Campoamor2024conformes} the connected component of $\Sigma_{\kappa_1}$ containing the point $(1, 0)$, which is the image of the origin of $\S^{3}_{[\kappa_{1}],\kappa_{2},\kappa_{3}}$ under the projection \eqref{eq:3DCK:projection1DCK}. For $\kappa_1 > 0$ we obtain the circle $\S^{1}$, while the case $\kappa_1 < 0$ gives rise to the (real) hyperbolic line $\H^{1}$, which is the branch of the hyperbola $(x^{0})^{2} - (x^{1})^{2}$ with $x^{0} \geq 1$. The Euclidean line $\E^{1}$, identified with the affine line $x^{0} = +1$, appears after the contraction $\kappa_1 = 0$.

%%%%%%%%%%%%%%%%%%%%%%%%%%%%%%%%%%%%

\begin{figure}[t!]
\centering
\includegraphics[width = 0.85\textwidth]{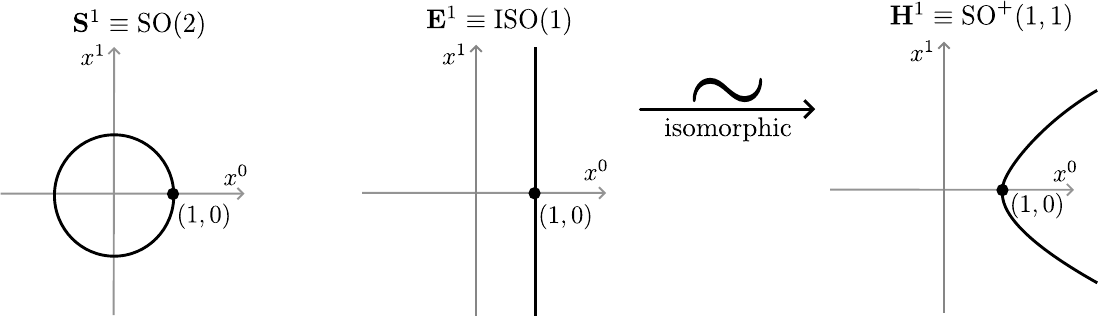}
\caption{\small The three 1D CK spaces $\S^{1}_{[\kappa_1]}$ according to the normalized values of $\kappa_1 \in \set{+1,0,-1}$ and their identifications as real 2D Lie groups. $\mathrm{SO}^{+}(1,1)$ denotes the identity component of $\mathrm{SO}(1,1)$.}
\label{fig:3DCK:1DCKgroups}
\end{figure}

%%%%%%%%%%%%%%%%%%%%%%%%%%%%%%%%%%%%

 In contrast to higher-dimensional CK spaces, the three 1D CK spaces are Lie groups within the identifications displayed in Figure~\ref{fig:3DCK:1DCKgroups}. In particular, we recall that the  Euler's formula analogue \eqref{eq:3DCK:Euler_hyperbolic} for the Clifford algebra $\mathrm{Cl}_{1,0}(\R)$ establishes such an isomorphism between $\mathrm{ISO}(1)$ and $\mathrm{SO}^{+}(1,1)$.

Now, reading backwards, we see that the fibres of the principal bundles studied in Sections~\ref{subsection:fibre_pos} and \ref{subsection:fibre_nonpos}, which are time-like geodesics provided that $\kappa_{2}\leq 0$, are completely determined by the curvature $\kappa_{1}$ of those regular CK spaces. Indeed, note that the orbit $\mathcal{O}_{O}$ of the origin $O = (1,0,0,0)$ under the $\R$-action induced by the flow \eqref{eq:3DCK:flowReeb} of the Reeb vector field $\cR_{\k}$ on $\S^{3}_{[\kappa_{1}],\kappa_{2},\kappa_{3}}$ is diffeomorphic to the 1D CK space $\S^{1}_{[\kappa_1]}$:
\begin{equation}
\mathcal{O}_{O} = \set{(\!\Cos_{\kappa_1}(2t), \Sin_{\kappa_1}(2t), 0, 0): t \in \R} \simeq \S^{1}_{[\kappa_1]}.
\nonumber
\end{equation}
   This allows us to state the following result regarding the principal bundles coming from the contact structure of $\S^{3}_{[\kappa_{1}],\kappa_{2},+}$:

\begin{thm}\label{th:3DCK:fibrebundles}
   The contact structure  $\eta_{\k}$ of $\S^{3}_{[\kappa_{1}],\kappa_{2},+}$ is regular if  $(\kappa_{1}, \kappa_{2}) \neq (-1, -1)$, thus excluding  $\dS^{2+1}$, and induces an $\S^{1}_{[\kappa_1]}$-principal bundle
 \begin{equation}
 \begin{tikzcd}
\S^{1}_{[\kappa_1]}\arrow[r, hook]& \S^{3}_{[\kappa_{1}],\kappa_{2},+} \arrow[r, "\pi_{\cR_{\k}}"] & \S^{3}_{[\kappa_{1}],\kappa_{2},+} / \cR_{\k}.
\end{tikzcd}
\nonumber
\end{equation}

\end{thm}

%%%%%%%%%%%%%%%%%%%%%%

\section{Contact metric geometry on Cayley--Klein spaces}
\label{section:contactmetric}

An \textit{almost contact structure} on a $(2n+1)$-dimensional manifold $M$ is a triplet $(\varphi, \cR, \eta)$, where $\varphi$ is a tensor field of type $(1,1)$, $\cR$ is a vector field and $\eta$ is a $1$-form satisfying
\begin{equation}
\varphi^{2} = - \mathrm{Id} + \eta \otimes \cR, \qquad \eta(\cR) = 1,
\label{eq:contactmetric:almost}
\end{equation}
with $\mathrm{Id}$ being the identity endomorphism on $\T M$.
Moreover, from \eqref{eq:contactmetric:almost} it can be proved that $\varphi$ has rank $2n$, as well as the following identities:
\begin{equation}
\varphi \cR = 0, \qquad \eta \circ \varphi = 0.
\label{eq:contactmetric:almostvis}
\end{equation}
A Riemannian or pseudo-Riemannian metric $g$ on $M$ is said to be \textit{compatible} with the almost contact structure $(\varphi, \cR, \eta)$ if
\begin{equation}
g(\varphi \X, \varphi \Y) = g(\X, \Y) - \varepsilon \eta (\X) \eta(\Y), \qquad \X, \Y \in \cv(M),
\label{eq:contactmetric:compatible}
\end{equation}
where $\varepsilon = 1$ in the Riemannian case and $\varepsilon = \pm 1$ if $g$ is pseudo-Riemmanian. In this situation, $M$ is called an \textit{almost contact (pseudo-)Riemannian metric manifold}. From \eqref{eq:contactmetric:almostvis}, setting $\mathbf{Y} = \cR$ in \eqref{eq:contactmetric:compatible} we see that
\begin{equation}
\eta(\X) = \varepsilon g(\X, \cR), \qquad \X \in \cv(M).
\label{eq:contactmetric:cform}
\end{equation}
In particular, $g(\cR, \cR) = \varepsilon$, showing that $\cR$ has a unit length in the Riemannian case and that $\cR$ can be either space-like $(\varepsilon = 1)$ or time-like $(\varepsilon = -1)$ in the pseudo-Riemannian scenario,   but it cannot be light-like.  If, in addition, the compatible metric $g$ satisfies
\begin{equation}
g(\X, \varphi \Y) =  \dd \eta (\X, \Y), \qquad \X, \Y \in \cv(M),
\nonumber
\end{equation}
then $\eta$ is a contact form whose Reeb vector field is $\cR$ and $(M, \varphi, \cR, \eta, g)$ (or simply $(M, \eta, g)$) is called a \textit{contact (pseudo-)Riemannian metric manifold}~\cite{Takahashi1969,Duggal1990,Bejancu1993,Calvaruso2010}.

Let now $(M, \eta, g)$ be a contact (pseudo-)Riemannian metric manifold and consider the Levi-Civita connection $\nabla$ associated with $g$. Then, as $\cL_{\cR} \eta = 0$, we have that
\begin{equation}
0 = \varepsilon (\cL_{\cR} \eta) \X = \cR g(\cR, \X) - g(\cR, [\cR, \X]) = g(\nabla_{\cR} \cR, \X), \qquad \X \in \cv(M).
\nonumber
\end{equation}
Thus, $\nabla_{\cR} \cR = 0$, showing that the integral curves of the Reeb vector field $\cR$ are all geodesics \cite{Calvaruso2010}.

Among all contact (pseudo-)Riemannian metric manifolds there are two types of special interest:

\begin{defi}
A contact (pseudo-)Riemannian metric manifold $(M, \eta, g)$ is said to be

\item[(i)] \textit{Sasakian} if it is normal, i.e., $N_{\varphi} + 2 \dd \eta \otimes \cR = 0$, where $N_{\varphi}$ is the Nijenhuis tensor of $\varphi$;
\item[(ii)] \textit{$K$-contact} if the Reeb vector field is a Killing vector field, i.e., $\cL_{\cR} g = 0$.
\end{defi}

The following result establishes when an almost contact (pseudo-)Riemannian metric manifold defines a Sasakian structure.

\begin{prop}[\cite{Bejancu1993,Calvaruso2010}] \label{prop:contactmetric:carac}
An almost contact (pseudo-)Riemannian metric manifold $(M, \varphi, \cR, \eta, g)$ is Sasakian if and only if
\begin{equation}
(\nabla_{\X} \varphi) \Y = g(\X, \Y) \cR - \varepsilon \eta(\Y) \X, \qquad \X, \Y \in \cv(M).
\nonumber
\end{equation}
In particular,  every Sasakian (pseudo-)Riemannian manifold is $K$-contact.
\end{prop}

Let us show that Killing vector fields on a $K$-contact (pseudo-)Riemannian manifold can be used to provide first integrals of the Reeb vector field.

\begin{thm} \label{th:contactmetric:firstintegrals}
Let $(M, \eta, g)$ be a $K$-contact  (pseudo-)Riemannian manifold and $\mathbf{K}$ a Killing vector field. Then, $\eta(\mathbf{K})$ is a first integral of the Reeb vector field $\cR$.
\end{thm}
\begin{proof}
First of all, as $\cL_{\cR} \eta = 0$, it can be easily concluded that
\begin{equation}
\cR \eta(\mathbf{K}) = (\cL_{\cR} \eta)\mathbf{K} + \eta ([\cR, \mathbf{K}]) = \eta ([\cR, \mathbf{K}]).
\label{eq:contactmetric:integral}
\end{equation}
Using \eqref{eq:contactmetric:cform}, it follows that
\begin{equation}
\eta([\cR, \mathbf{K}]) = \varepsilon([\cR, \mathbf{K}], \cR) = \varepsilon(g(\nabla_{\cR} \mathbf{K}, \cR) - g(\nabla_{\mathbf{K}} \cR, \cR)).
\label{eq:contactmetric:integral1}
\end{equation}
Any Killing vector field $\X \in \cv(M)$ satisfies
\begin{equation}
g(\nabla_{\Y} \X, \mathbf{Z}) + g(\Y, \nabla_{\mathbf{Z}} \X) = 0, \qquad \Y, \mathbf{Z} \in \cv(M).
\label{eq:contactmetric:killing}
\end{equation}
Taking $\X = \Y = \cR$ and $\mathbf{Z} = \mathbf{K}$ in \eqref{eq:contactmetric:killing}, as $\nabla_{\cR} \cR = 0$, we have that $g(\nabla_{\mathbf{K}} \cR, \cR) = 0$. Similarly, we get that $g(\nabla_{\cR} \mathbf{K}, \cR) = 0$ after the choices $\X = \mathbf{K}$ and $\Y = \mathbf{Z} = \cR$ in \eqref{eq:contactmetric:killing}. The result follows from expressions \eqref{eq:contactmetric:integral} and \eqref{eq:contactmetric:integral1}.
\end{proof}

For the 3D CK spaces $\S^{3}_{[\kappa_{1}],\kappa_{2},\kappa_{3}}$ with all $\kappa_{m} \neq 0$ (i.e., those for which the main metric $g_{\k}$ \eqref{eq:3DCK:coord} is nondegenerate), the compatibility condition
\begin{equation}
\eta_{\k}(\X) = \varepsilon g_{\k}(\X, \cR_{\k}), \qquad \X \in \cv(M), \qquad \varepsilon = \pm1 ,
\nonumber
\end{equation}
  is only satisfied when $\kappa_{1}\kappa_{3} = 1$.  In particular, among the nine spaces with $\kappa_{3} = +1$, the only candidates for which $\eta_{\k}$ and $g_{\k}$ may form part of a $K$-contact structure are the sphere $\S^{3}$ and the anti-de Sitter spacetime $\AdS^{2+1}$.

%%%%%%%%%%%%%%%%%%%%%%

\subsection{$\mathbf{S}^{3}$ and $\mathbf{AdS}^{2+1}$ as Sasakian manifolds}
\label{subsection:contactmetric:sasakian}

The first construction of a Sasakian structure on odd-dimensional spheres is due to Sasaki and Hatakeyama~\cite{Sasaki1962} (see also~\cite{Tashiro1963}). Later, using similar techniques, Takahashi obtained pseudo-Riemannian Sasakian structures on odd-dimensional pseudospheres and pseudohyperbolic spaces \cite{Takahashi1969}. Let us now sketch the main details of Takahashi's construction which, among the 3D CK spaces, shows that the sphere $\S^{3}$ and the anti-de Sitter $\AdS^{2+1} $ spaces are Sasakian manifolds in a natural way.
We restrict ourselves to the normalized values $\kappa_{2} = \pm 1$ for $\S^{3}_{[+],\kappa_{2},+}$ such that   $\k = (+1,\pm 1, +1)$.

First of all, let us consider the tensor field $\mathcal{J} $ of type $(1, 1)$ on $\R^{4}$ given by
\begin{equation}
\mathcal{J}  \pdv{x^{0}} :=  - \pdv{x^{1}}, \qquad \mathcal{J}  \pdv{x^{1}} :=  \pdv{x^{0}}, \qquad \mathcal{J}  \pdv{x^{2}} := - \kappa_{2} \pdv{x^{3}}, \qquad \mathcal{J}  \pdv{x^{3}} :=  \kappa_{2} \pdv{x^{2}}.
\nonumber
\end{equation}
Note that it defines an almost complex structure on $\R^{4}$ as it satisfies $\mathcal{J}^{2} = - \mathrm{Id}$.
With respect to the ambient metric  $\widetilde{g}_{\k}$ \eqref{eq:3DCK:metric},  inducing the metric $g_{\k}$ of $\S^{3}_{[+],\kappa_{2}, +}$, we see that
\begin{equation}
\widetilde{g}_{\k} (\mathcal{J} \X, \mathcal{J}  \Y) = \widetilde{g}_{\k} (\X, \Y), \qquad \X, \Y \in \cv\bigl(\R^{4}\bigr).
\nonumber
\end{equation}
This shows that $\widetilde{g}_{\k}$ is an almost Hermitian metric if $\kappa_{2} = +1$ and an almost pseudo-Hermitian metric if $\kappa_{2} = -1$. Now, the normal vector field to $\S^{3}_{[+], \kappa_{2},+}$ with respect to $\widetilde{g}_{\k}$ is $\nu := 2 \mathbf{\Delta}$, where $\mathbf{\Delta}$ is the scaling symmetry \eqref{eq:s3:radial} studied in Section~\ref{section:s3}. Then,  the vector field
\begin{equation}
\overline{\cR}_{\k} := -\mathcal{J}  \nu = -x^{1} \pdv{x^{0}} +  x^{0} \pdv{x^{1}} - \kappa_{2} x^{3} \pdv{x^{2}}+ \kappa_{2} x^{2} \pdv{x^{3}}
\nonumber
\end{equation}
is tangent to $\S^{3}_{[+],\kappa_{2},+}$ with $g_{\k}(\overline{\cR}_{\k}, \overline{\cR}_{\k}) = 1$. In particular, it is a space-like vector field for $\AdS^{2+1}$. Consider now the 1-form $\overline{\eta}_{\k}$ on $\S^{3}_{[+],\kappa_{2},+}$ defined  by
\begin{equation}
\overline{\eta}_{\k} (\X) :=  g_{\k}(\X, \overline{\cR}_{\k}), \qquad \X \in \cv\bigl(\S^{3}_{[+],\kappa_{2},+}\bigr)
\nonumber
\end{equation}
explicitly given in Weierstrass coordinates by
\begin{equation}
\overline{\eta}_{\k}= -x^{1} \dd x^{0} +  x^{0} \dd x^{1} -   x^{3} \dd x^{2} +  x^{2} \dd x^{3}.
\nonumber
\end{equation}
Denote by $\langle \overline{\cR}_{\k} \rangle \subset \T \S^{3}_{[+],\kappa_{2},+}$ the distribution spanned by $\overline{\cR}_{\k}$ and by $\langle \overline{\cR}_{\k} \rangle^{\perp} \subset \T \S^{3}_{[+],\kappa_{2},+} $ its orthogonal complement  with respect to $g_{\k}$, and let $\pr: \T\S^{3}_{[+],\kappa_{2},+} \to \langle \overline{\cR}_{\k} \rangle^{\perp}$ be the orthogonal projection morphism. Then we see that
\begin{equation}
\pr(\X) = \X - g_{\k}(\X, \overline{\cR}_{\k})\overline{\cR}_{\k} = \X - \overline{\eta}_{\k}(\X) \overline{\cR}_{\k}, \qquad \X \in \cv\bigl( \S^{3}_{[+],\kappa_{2},+} \bigr).
\nonumber
\end{equation}
Finally, if we consider the $(1,1)$-tensor field $\varphi_{\k}$ on $\S^{3}_{[+], \kappa_{2},+}$ given by $\varphi_{\k} := \pr \circ \mathcal{J} $, we have that  $\bigl(\varphi_{\k}, \overline{\cR}_{\k}, \overline{\eta}_{\k}, g_{\k} \bigr)$ is an almost contact (pseudo-)Riemannian metric structure on $\S^{3}_{[+],\kappa_{2},+}$.

From the Gauss--Weingarten equations of $\S^{3}_{[+],\kappa_{2},+}$ (see~\cite{Takahashi1969,Bejancu1993} for details), it can be proved that
\begin{equation}
(\nabla_{\X} \varphi_{\k})\Y = g_{\k}(\X, \Y) \overline{\cR}_{\k} -  \overline{\eta}_{\k}(\Y) \X, \qquad \X, \Y \in \cv \bigl(\S^{3}_{[+], \kappa_{2},+}\bigr),
\nonumber
\end{equation}
with $\nabla$ being the Levi-Civita connection on $\S^{3}_{[+], \kappa_{2},+}$ associated with the metric $g_{\k}$. From Proposition~\ref{prop:contactmetric:carac} it follows that $\bigl(\varphi_{\k}, \overline{\cR}_{\k}, \overline{\eta}_{\k}, g_{\k} \bigr)$ defines a Sasakian structure on  $\S^{3}_{[+],\kappa_{2},+}$. Recall that, with respect to the contact structure of $\S^{3}_{[+],\kappa_{2},+}$ obtained in Section~\ref{section:3DCK:contactstructure} (see Table~\ref{table:3DCK:contactgen}), we have that $\overline{\eta}_{\k} = 2 \eta_{\k}$ and $\overline{\cR}_{\k} = \frac{1}{2} \cR_{\k}$. Then,  $(\varphi_{\k},\cR_{\k}, \eta_{\k}, 2g_{\k})$ also defines a Sasakian structure on $\S^{3}_{[+], \kappa_{2},+}$.

 In Section~\ref{subsection:contactck:liouville}  we shall apply Theorem~\ref{th:contactmetric:firstintegrals} to construct contact Lie systems of Liouville type on $\S^{3}$ and  $\AdS^{2+1} $.

%%%%%%%%%%%%%%%%%%%%%%

\section{Contact Lie systems  on Cayley--Klein spaces}
\label{section:contactck}

The contact form $\eta_{\k}$ (see Table~\ref{table:3DCK:contactgen}) of $\S^{3}_{[\kappa_{1}],\kappa_{2}, \kappa_{3}}$ induces, in a natural way, a Jacobi bracket
\begin{equation}
\set{\cdot, \cdot}_{\eta_{\k}}: C^{\infty} \bigl( \S^{3}_{[\kappa_{1}],\kappa_{2}, \kappa_{3}} \bigr) \times C^{\infty}\bigl(\S^{3}_{[\kappa_{1}],\kappa_{2}, \kappa_{3}}\bigr) \to C^{\infty}\bigl(\S^{3}_{[\kappa_{1}],\kappa_{2}, \kappa_{3}}\bigr),
\nonumber
\end{equation}
  given by
\begin{equation}
\set{f, g}_{\eta_{\k}} = \X_{f} g + g \cR f = - \dd \eta (\X_{f}, \X_{g}) - f \cR g + g \cR f, \qquad f, g \in C^{\infty}\bigl(\S^{3}_{[\kappa_{1}],\kappa_{2}, \kappa_{3}}\bigr),
\label{eq:contactck:Jacobi}
\end{equation}
where $\X_{f}$ and $\X_{g}$ denote the contact Hamiltonian vector fields associated with $f$ and $g$, respectively. We write $\mathrm{Ham}(\S^{3}_{[\kappa_{1}],\kappa_{2}, \kappa_{3}}, \eta_{\k})$ for the space of contact Hamiltonian vector fields relative to $\eta_{\k}$, which is a Lie subalgebra of $\cv(\S^{3}_{[\kappa_{1}],\kappa_{2},\kappa_{3}})$.
Consider now the Hamiltonian functions $h_{i}$ \eqref{eq:s3:ham_sym} spanning the LH algebra of the $\sp(4,\R)$-LH system \eqref{eq:s3:sp4_system} on $\R^{4}_{0}$, and denote by $h_{\k,i} := i^{*}h_{i}$ $(1 \leq i \leq 10)$ their restrictions to the 3D CK space $i: \S^{3}_{[\kappa_{1}],\kappa_{2}, \kappa_{3}} \hookrightarrow \R^{4}_{0}$. From their commutation relations with respect to the Jacobi bracket \eqref{eq:contactck:Jacobi} shown in Table~\ref{table:contactck:commutation}, it follows that they span a Lie algebra isomorphic to $\sp(4, \R)$ (see Table~\ref{table:sp4}).

%%%%%%%%%%%%%%%%%%%%%%

\begin{table}[t!]
\small
\caption{\small Commutation relations of the Hamiltonian functions $h_{\k, i} = i^{*}h_{i}$ with respect to the Jacobi bracket \eqref{eq:contactck:Jacobi} induced by the contact form \eqref{eq:3DCK:contactform} on $ \S^{3}_{[\kappa_{1}],\kappa_{2}, \kappa_{3}}$, where $h_{i}$ are the generators of the LH algebra \eqref{eq:s3:ham_sym} on $\R^{4}_{0}$ $(1 \leq i \leq 10)$.}
\label{table:contactck:commutation}
\centering
\begin{tabular}{c|cccccccccc}
$\set{\cdot, \cdot}_{\eta_{\k}}$ & $h_{\k,1}$ & $h_{\k,2}$ & $h_{\k,3}$ & $h_{\k,4}$ & $h_{\k,5}$ & $h_{\k,6}$ & $h_{\k,7}$ & $h_{\k,8}$ & $h_{\k,9}$ & $h_{\k,10}$ \\[4pt]
\hline
\\[-8pt]
	$h_{\k,1}$ & 0 & $h_{\k,2}$ & $-h_{\k,3}$ & $0$ & $2h_{\k,5}$ & $h_{\k,6}$ & $0$ & $-2h_{\k,8}$ & $-h_{\k,9}$ & $0$ \\ [3pt]
$h_{\k,2}$ &  & $0$ & $h_{\k,1} - h_{\k,4}$ & $h_{\k,2}$ & $0$ & $2h_{\k,5}$ & $h_{\k,6}$ & $-h_{\k,9}$ & $-2h_{\k,10}$ & $0$ \\  [3pt]
$h_{\k,3}$ &  &  & $0$ & $-h_{\k,3}$ & $h_{\k,6}$ & $2h_{\k,7}$ & $0$ & $0$ & $-2h_{\k,8}$ & $-h_{\k,9}$ \\  [3pt]
$h_{\k,4}$ &  &  &  & $0$ & $0$ & $h_{\k,6}$ & $2h_{\k,7}$ & $0$ & $-h_{\k,9}$ & $-2h_{\k,10}$ \\  [3pt]
$h_{\k,5}$ & & &  &  & $0$ & $0$ & $0$ & $-h_{\k,1}$ & $-h_{\k,2}$ & $0$ \\  [3pt]
$h_{\k,6}$ &  &  &  &  &  & $0$  & $0$ & $-h_{\k,3}$ & $-h_{\k,1} - h_{\k,4}$ & $-h_{\k,2}$ \\  [3pt]
$h_{\k,7}$ &  &  &  &  &  &  & $0$ & $0$ & $-h_{\k,3}$ & $-h_{\k,4}$ \\   [3pt]
$h_{\k,8}$ &  &  &  &  &  &  &  & $0$ & $0$ & $0$ \\   [3pt]
$h_{\k,9}$ &  & &  &  &  &  &  &  & $0$ & $0$ \\  [3pt]
$h_{\k,10}$ &  &  &  &  &  &  &  &  &  & $0$
\end{tabular}
\end{table}

%%%%%%%%%%%%%%%%%%%%%%

The mapping
\begin{equation}
C^{\infty}\bigl(\S^{3}_{[\kappa_{1}],\kappa_{2}, \kappa_{3}}\bigr) \ni f \mapsto \X_{f} \in \mathrm{Ham}\bigl(\S^{3}_{[\kappa_{1}],\kappa_{2}, \kappa_{3}}, \eta_{\k}\bigr)
\nonumber
\end{equation}
is a Lie algebra isomorphism. Then, if $\X_{\k, i} \in \mathrm{Ham}\bigl(\S^{3}_{[\kappa_{1}],\kappa_{2}, \kappa_{3}}, \eta_{\k} \bigr)$ is the  Hamiltonian vector field associated with the  Hamiltonian function $h_{\k,i}$ $(1 \leq i \leq 10)$, the $t$-dependent vector field
\begin{equation}
\X_{\k} := \sum_{i = 1}^{10} b_{i}(t) \X_{\k, i}, \qquad b_{i} \in C^{\infty}(\R),
\label{eq:contactck:system}
\end{equation}
is a contact Lie system on $(\S^{3}_{[\kappa_{1}],\kappa_{2}, \kappa_{3}}, \eta_{\k})$ { with a VG Lie algebra, spanned by the Hamiltonian vector fields $\mathbf{X}_{\k, i}$ $(1 \leq i \leq 10)$, isomorphic to $\sp(4,\R)$}. Thus it yields the curved analogue of Theorem~\ref{th:s3:sp4}.
\begin{thm}
For every  subalgebra $\g \subset \sp(4, \R)$ there exists a contact Lie system on $\bigl( \S^{3}_{[\kappa_{1}],\kappa_{2}, \kappa_{3}}, \eta_{\k} \bigr)$ { with a} VG Lie algebra is isomorphic to $\g$.
\end{thm}

   The explicit expressions for the  Hamiltonian functions $h_{\k, i}$ and their associated Hamiltonian vector fields $\X_{\k, i}$ are presented in Table~\ref{table:contactck:sp4} in geodesic parallel coordinates $(x,y,z)$ (see  (\ref{coord})). Note that $\X_{\k, i}$ $(1 \leq i \leq 10)$ { satisfy} commutation relations that are formally the same as those for $h_{\k, i}$ displayed in Table~\ref{table:contactck:commutation}.

%%%%%%%%%%%%%%%%%%%%%%
 \newpage

\begin{table}[h!]
\small
   \caption{\small The Hamiltonian functions $h_{\k, i}$ and their associated Hamiltonian vector fields $\X_{\k, i}$ $(1 \leq i \leq 10)$ relative to the contact  form $\eta_{\k}$ \eqref{eq:3DCK:contactform} on $\S^{3}_{[\kappa_{1}],\kappa_{2}, \kappa_{3}}$ expressed in geodesic parallel coordinates $(x,y,z)$.}
\label{table:contactck:sp4}
\centering
 \begin{tabular}{ll}
\hline

\hline \\[-6pt]
   \qquad\qquad\qquad\qquad\qquad\qquad\qquad Hamiltonian functions \\[4pt]
\hline
 \\[-6pt]
$\begin{array}{ll}
  \displaystyle h_{\k, 1} = \frac{1}{2} \Sin_{\kappa_{01}}(2x) \Cos_{\kappa_{02}}^{2}(y) \Cos_{\kappa_{03}}^{2}(z) & \qquad  \displaystyle h_{\k, 2} = \frac{1}{2} \Cos_{\kappa_{01}}(x) \Cos_{\kappa_{02}}(y) \Sin_{\kappa_{03}}(2z)  \\[10pt]
\displaystyle h_{\k, 3} = \frac{1}{2} \Sin_{\kappa_{01}}(x) \Sin_{\kappa_{02}}(2y) \Cos_{\kappa_{03}}^{2}(z) & \qquad \displaystyle h_{\k,4} = \frac{1}{2} \Sin_{\kappa_{02}}(y) \Sin_{\kappa_{03}}(2z) \\[10pt]\displaystyle h_{\k, 5} = \frac{1}{2} \Cos_{\kappa_{01}}^{2}(x) \Cos_{\kappa_{02}}^{2}(y) \Cos_{\kappa_{03}}^{2}(z) & \qquad \displaystyle h_{\k, 6} = \frac{1}{2} \Cos_{\kappa_{01}}(x) \Sin_{\kappa_{02}}(2y) \Cos_{\kappa_{03}}^{2}(z)  \\[10pt]
  \displaystyle h_{\k,7} = \frac{1}{2} \Sin_{\kappa_{02}}^{2}(y) \Cos_{\kappa_{03}}^{2}(z) &\qquad\displaystyle h_{\k, 8} = \frac{1}{2} \Sin_{\kappa_{01}}^{2}(x) \Cos_{\kappa_{02}}^{2}(y) \Cos_{\kappa_{03}}^{2}(z)  \\[10pt]
  \displaystyle h_{\k,9} = \frac{1}{2} \Sin_{\kappa_{01}}(x) \Cos_{\kappa_{02}}(y) \Sin_{\kappa_{03}}(2z) &\qquad \displaystyle h_{\k, 10} = \frac{1}{2} \Sin_{\kappa_{03}}^{2}(z)
 \end{array}$ \vspace*{6pt}   \\
\hline \\[-6pt]

   \qquad\qquad\qquad\qquad\qquad\qquad\quad   Hamiltonian vector fields & \\[4pt]
\hline
 \\[-6pt]

 $\ \displaystyle \X_{\k,1} = - \Sin_{\kappa_{01}}(2x)\partial_{x} - \frac{1}{2} \Cos_{\kappa_{01}}(2x) \Sin_{\kappa_{02}}(2y) \partial_{y} - \frac{1}{2}\Cos_{\kappa_{01}}(2x)  \Cos_{\kappa_{02}}^{2}(y) \Sin_{\kappa_{03}}(2z) \partial_{z} $& \\ [10pt]  \rule{0pt}{\normalbaselineskip}

$\displaystyle \X_{\k,2} = - \frac{\Cos_{\kappa_{01}}(x) \Tan_{\kappa_{03}}(z)}{\Cos_{\kappa_{02}}(y)} \,\partial_{x} + \left(\!\Cos_{\kappa_{01}}(x) \Cos_{\kappa_{02}}^{2}(y) + \kappa_{01}\Sin_{\kappa_{01}}(x) \Sin_{\kappa_{02}}(y) \Tan_{\kappa_{03}}(z) \right)\! \partial_{y} $ \\ [12pt]  \rule{0pt}{\normalbaselineskip}
$\displaystyle \qquad\qquad \qquad + \left(\! \kappa_{01} \Sin_{\kappa_{01}}(x) \Cos_{\kappa_{02}}(y) \Sin_{\kappa_{03}}^{2}(z) - \frac{1}{4} \,\kappa_{02}\Cos_{\kappa_{01}}(x) \Sin_{\kappa_{02}}(2y) \Sin_{\kappa_{03}}(2z)\!\right)\! \partial_{z}$ \\ [8pt]  \rule{0pt}{\normalbaselineskip}

 $\displaystyle \X_{\k,3} = -\Sin_{\kappa_{01}}(x) \Tan_{\kappa_{02}}(y) \partial_{x} - \Cos_{\kappa_{01}}(x) \Sin_{\kappa_{02}}^{2}(y) \partial_{y}$ \\ [4pt]  \rule{0pt}{\normalbaselineskip}
 $\displaystyle \qquad\qquad \qquad
  - \Cos_{\kappa_{02}}(y) \Cos_{\kappa_{03}}^{2}(z) \bigl( \!\Sin_{\kappa_{01}}(x) + \Cos_{\kappa_{01}}(x) \Sin_{\kappa_{02}}(y) \Tan_{\kappa_{03}}(z)\bigr) \partial_{z}$ \\ [8pt]  \rule{0pt}{\normalbaselineskip}

  $\displaystyle \X_{\k, 4} = \frac{1}{2} \Sin_{\kappa_{02}}(2y) \partial_{y} + \frac{1}{4} \Sin_{\kappa_{03}}(2z) \bigl( \!\Cos_{\kappa_{02}}(2y)-3\bigr) \partial_{z}$\\[10pt]  \rule{0pt}{\normalbaselineskip}

  $\displaystyle \X_{\k,5} = - \Cos_{\kappa_{01}}^{2}(x) \partial_{x} + \frac{1}{4}\,\kappa_{01} \Sin_{\kappa_{01}} (2x) \Sin_{\kappa_{02}}(2y) \partial_{y} + \frac{1}{4}\,\kappa_{01} \Sin_{\kappa_{01}}(2x) \Cos_{\kappa_{02}}^{2}(y)\Sin_{\kappa_{03}}(2z) \partial_{z}$\\ [8pt]  \rule{0pt}{\normalbaselineskip}

  $\displaystyle \X_{\k,6} = - \Cos_{\kappa_{01}}(x) \Tan_{\kappa_{02}}(y) \partial_{x} + \kappa_{01} \Sin_{\kappa_{01}}(x) \Sin_{\kappa_{02}}^{2}(y) \partial_{y}$\\ [6pt]  \rule{0pt}{\normalbaselineskip}
  $\displaystyle   \qquad\qquad \qquad - \left( \! \!\Cos_{\kappa_{01}}(x) \Cos_{\kappa_{02}}(y) \Cos_{\kappa_{03}}^{2}(z) - \frac{1}{4}\,\kappa_{01} \Sin_{\kappa_{01}}(x) \Sin_{\kappa_{02}}(2y) \Sin_{\kappa_{03}}(2z) \! \right)\! \partial_{z}$ \\ [8pt]  \rule{0pt}{\normalbaselineskip}

 $\displaystyle \X_{\k,7} = - \Sin_{\kappa_{02}}(y)\Cos_{\kappa_{03}}^{2}(z)  \partial_{z}$ \\ [6pt]  \rule{0pt}{\normalbaselineskip}

 $ \displaystyle \X_{\k,8} = - \Sin_{\kappa_{01}}^{2}(x) \partial_{x} - \frac{1}{4} \Sin_{\kappa_{01}}(2x) \Sin_{\kappa_{02}}(2y) \partial_{y} - \frac{1}{4}  \Sin_{\kappa_{01}}(2x) \Cos_{\kappa_{02}}^{2}(y) \Sin_{\kappa_{03}}(2z) \partial_{z}$ \\ [10pt]  \rule{0pt}{\normalbaselineskip}

  $\displaystyle \X_{\k, 9} = - \frac{\Sin_{\kappa_{01}}(x) \Tan_{\kappa_{03}}(z)}{\Cos_{\kappa_{02}}(y)} \,\partial_{x} + \bigl(\!\Sin_{\kappa_{01}}(x)\Cos_{\kappa_{02}}^{2}(y) - \Cos_{\kappa_{01}}(x) \Sin_{\kappa_{02}}(y) \Tan_{\kappa_{03}}(z) \bigr)\partial_{y} $\\ [10pt]  \rule{0pt}{\normalbaselineskip}
    $\displaystyle   \qquad\qquad \qquad   - \frac{1}{2}\Cos_{\kappa_{02}}(y) \Sin_{\kappa_{03}}(2z) \bigl(\! \Cos_{\kappa_{01}}(x) \Tan_{\kappa_{03}}(z) + \kappa_{02} \Sin_{\kappa_{01}}(x) \Sin_{\kappa_{02}}(y) \bigr) \partial_{z} $ \\ [6pt]  \rule{0pt}{\normalbaselineskip}

  $\displaystyle \X_{\k, 10} = \Cos_{\kappa_{02}}(y) \Tan_{\kappa_{03}}(z) \partial_{y} - \kappa_{02} \Sin_{\kappa_{02}}(y)   \Sin^2_{\kappa_{03}}(z) \partial_{z}$
 \vspace*{6pt} \\  \hline

 \hline
\end{tabular}
\end{table}

%%%%%%%%%%%%%%%%%%%%%%

 \newpage

%%%%%%%%%%%%%%%%%%%%%%

\subsection{Contact Lie systems on $\mathbf{S}^{3}$ and $\mathbf{AdS}^{2+1}$} \label{subsection:contactck:liouville}

As pointed out in Section~\ref{subsection:contactmetric:sasakian}, by means of the Sasakian structure of $\S^{3}_{[+],\kappa_{2},+}$ with $\kappa_{2} = \pm1$, we can apply Theorem~\ref{th:contactmetric:firstintegrals} to construct contact Lie systems of Liouville type on
the sphere  $\mathbf{S}^{3}\equiv \S^{3}_{[+],+,+}$  and anti-de Sitter space  $\mathbf{AdS}^{2+1}\equiv \S^{3}_{[+],-,+}$; thus, in these cases, $\k=(+1,\pm 1,+1)$. Recall from \eqref{eq:3DCK:Killing}, that the Lie algebra of Killing vector fields of $\S^{3}_{[+],\kappa_{2},+}$ with respect to the metric $g_{\k}$ \eqref{eq:3DCK:coord} is spanned by the following six vector fields expressed in Weierstrass coordinates:
\begin{align}
&\mathbf{J}_{01} = x^{1} \pdv{x^{0}} - x^{0} \pdv{x^{1}},  &  \mathbf{J}_{02} = \kappa_{2}x^{2} \pdv{x^{0}} - x^{0} \pdv{x^{2}},  &\qquad\ \mathbf{J}_{03} = \kappa_{2} x^{3} \pdv{x^{0}} - x^{0} \pdv{x^{3}}, \nonumber \\[2pt]
&\mathbf{J}_{12} = \kappa_{2}x^{2} \pdv{x^{1}} - x^{1} \pdv{x^{2}},  &   \mathbf{J}_{13} =  \kappa_{2} x^{3} \pdv{x^{1}}- x^{1} \pdv{x^{3}},  &\qquad\ \mathbf{J}_{23}= x^{3} \pdv{x^{2}} - x^{2} \pdv{x^{3}} .
\nonumber
\end{align}
They close on the CK algebra $\so_{+,\kappa_{2},+}(4)$, isomorphic to $\so(4)$ for $\kappa_2=+1$, and to $\so(2,2)$ for $\kappa_2=-1$. With respect to the contact form $\eta_{\k}$ \eqref{eq:3DCK:contactform} on $\S^{3}_{[+],\kappa_{2},+}$ (see Table~\ref{table:3DCK:contactgen}), the procedure of Theorem~\ref{th:contactmetric:firstintegrals} leads to four independent first integrals of the Reeb vector field $\cR_{\k}$ \eqref{eq:3DCK:Reeb_desc}:
\begin{equation}
\begin{split}
&\eta_{\k} (\mathbf{J}_{01} ) = - \frac{1}{2} \bigl((x^{0})^{2} + (x^{1})^{2} \bigr), \qquad \eta_{\k}(\mathbf{J}_{13} ) = \kappa_{2} \eta_{\k} (\mathbf{J}_{02} ) = \frac{1}{2}\bigl( \kappa_{2} x^{0} x^{3} - x^{1} x^{2} \bigr), \\
&\eta_{\k} (\mathbf{J}_{03}  ) = -\kappa_{2}\eta_{\k}(\mathbf{J}_{12} ) = -\frac{1}{2}\bigl(x^{0}x^{2} + \kappa_{2} x^{1}x^{3} \bigr), \qquad \eta_{\k} (\mathbf{J}_{23}  ) = -\frac{1}{2}\bigl( (x^{2})^{2} + (x^{3})^{2} \bigr).
\end{split}
\nonumber
\end{equation}
We consider the change of basis defined by
\begin{equation}
\begin{split}
&h_{ 1}' := -\frac{1}{2} \bigl( \eta_{\k}(\mathbf{J}_{01}) - \kappa_{2} \eta_{\k}( \mathbf{J}_{23}) \bigr)=\frac{1}{4} \bigl( (x^{0})^{2} + (x^{1})^{2} -\kappa_2   (x^{2})^{2} -\kappa_2 (x^{3})^{2} \bigr) ,\\
& h_{ 2}':=  \eta_{\k} (\mathbf{J}_{13})=\frac{1}{2}\bigl( \kappa_{2} x^{0} x^{3} - x^{1} x^{2} \bigr), \qquad h_{ 3}':= -  \eta_{\k}(\mathbf{J}_{03}) =\frac{1}{2}\bigl(x^{0}x^{2} + \kappa_{2} x^{1}x^{3} \bigr) ,  \\[2pt]
&h_{ 4}':= -2 \bigl( \eta_{\k}(\mathbf{J}_{01}) + \kappa_{2} \eta_{\k}( \mathbf{J}_{23}) \bigr) =  (x^{0})^{2} + (x^{1})^{2} +\kappa_2   (x^{2})^{2} +\kappa_2 (x^{3})^{2}   =  1,
\end{split}
\label{eq:contactck:lh}
\end{equation}
thus finding the following identifications with respect to the Hamiltonian functions $h_{\k,i}$ on $\S^{3}_{[+], \kappa_{2},+}$ (see  \eqref{eq:s3:ham_sym}):
\begin{equation}
\begin{split}
&h_{ 1}' = \frac{1}{2} \bigl(h_{\k,5} + h_{\k,8} - \kappa_{2}h_{\k,7} - \kappa_{2} h_{\k,10}\bigr), \qquad h_{ 2}' =\frac 12 \bigl( \kappa_{2} h_{\k,2} - h_{\k,3} \bigr), \\[2pt]
&h_{ 3}' =\frac 12 \bigl( h_{\k,6} + \kappa_{2} h_{\k,9} \bigr),  \qquad
h_{ 4}' = -2 \bigl(h_{\k,5} + h_{\k,8} + \kappa_{2}h_{\k,7} + \kappa_{2} h_{\k,10} \bigr).
\end{split}
\nonumber
\end{equation}
Their commutation relations are obtained directly from Table~\ref{table:contactck:commutation}, namely
\begin{equation}
 \bigl\{h_{ 1}', h_{2}' \bigr\}_{\eta_{\k}} =  h_{ 3}', \qquad  \bigl\{h_{ 1}', h_{3}' \bigr\}_{\eta_{\k}} = -h_{ 2}', \qquad
   \bigl\{h_{ 2}', h_{ 3}' \bigr\}_{\eta_{\k}} =  \kappa_{2} h_{1}' ,\qquad\bigl\{h_{ 4}', \cdot \bigr\}_{\eta_{\k}} = 0.
\label{eq:contactk:commlh}
\end{equation}
Therefore, the Lie algebra spanned by the first integrals (\ref{eq:contactck:lh}) with respect to the Jacobi bracket $\set{\cdot, \cdot}_{\eta_{\k}}$ (\ref{eq:contactck:Jacobi})  is isomorphic to ${\so}(3) \oplus \R$ for  $\mathbf{S}^{3}$ and to ${\so}(2,1) \oplus \R$ for $\mathbf{AdS}^{2+1}$.

Let $\Y_{ i}$ be the Hamiltonian vector field associated with $h_{ i}'$ $(1 \leq i \leq 4)$, then we have that
\begin{equation}
\begin{split}
&\Y_{ 1} = \frac{1}{2}\bigl(\X_{\k,5} + \X_{\k,8} - \kappa_{2} \X_{\k,7} - \kappa_{2} \X_{\k,10}\bigr), \qquad  \Y_{ 2} = \frac12\bigl( \kappa_{2} \X_{\k,2} - \X_{\k,3} \bigr), \\[2pt]
& \Y_{ 3}= \frac12\bigl(  \X_{\k,6} + \kappa_{2} \X_{\k,9} \bigr), \qquad \Y_{ 4} = -2(\X_{\k,5} + \X_{\k,8} + \kappa_{2} \X_{\k,7} + \kappa_{2} \X_{\k,10}) \equiv  \cR_{\k},
\end{split}
\label{eq:contactck:hamvf}
\end{equation}
where $\X_{\k,i}$ is the Hamiltonian vector field associated with $h_{\k,i}$  $(1 \leq i \leq 4)$ and
$\cR_{\k}$ is the Reeb vector field (\ref{eq:3DCK:Reeb_desc}). From Table~\ref{table:contactck:sp4}, they can be expressed in geodesic parallel coordinates $(x, y,z)$ as follows:
\begin{equation}
\begin{split}
\Y_{ 1} &=-\frac{1}{2} \bigl(  \partial_{x} + \kappa_{2} \Cos_{\kappa_{2}}(y) \Tan_{\kappa_{2}}(z) \partial_{y} - \kappa_{2} \Sin_{\kappa_{2}}(y) \partial_{z} \bigr), \\[2pt]
\Y_{ 2} &= \frac{1}{2} \left(  \left( \sin x \Tan_{\kappa_{2}}(y) - \kappa_{2} \frac{\cos x}{\Cos_{\kappa_{2}}(y)} \Tan_{\kappa_{2}}(z) \right) \partial_{x} + \kappa_{2}\bigl(\cos x  + \sin x \Sin_{\kappa_{2}}(y) \Tan_{\kappa_{2}}(z) \bigr) \partial_{y}  \right)
 \\ & \qquad  + \frac{1}{2}\sin x \Cos_{\kappa_{2}}(y) \partial_{z},  \\[2pt]
 \Y_{ 3} &= - \frac{1}{2} \left( \left( \cos x \Tan_{\kappa_{2}}(y) + \kappa_{2}\frac{\sin x}{\Cos_{\kappa_{2}}(y)} \Tan_{\kappa_{2}}(z) \right) \partial_{x} - \kappa_{2}\bigl(\sin x - \cos x\Sin_{\kappa_{2}}(y) \Tan_{\kappa_{2}}(z) \bigr) \partial_{y} \right) \\
 &\qquad -\frac{1}{2} \cos x \Cos_{\kappa_{2}}(y) \partial_{z}, \\[2pt]
 \Y_{ 4} &= \cR_{\k} =  2 \bigl(  \partial_{x} - \kappa_{2} \Cos_{\kappa_{2}}(y) \Tan_{\kappa_{2}}(z) \partial_{y} + \kappa_{2} \Sin_{\kappa_{2}}(y) \partial_{z} \bigr),
\end{split}
\nonumber
\end{equation}
which formally satisfy the same commutation rules (\ref{eq:contactk:commlh}).

As a straightforward consequence, the $t$-dependent vector field on $\S^{3}_{[+],\kappa_{2},+}$ defined by
\begin{equation}
\Y_{\k}:= a_{1}(t) \Y_{1} + a_{2}(t) \Y_{ 2} + a_{3}(t) \Y_{ 3} + a_{4}(t) \cR_{\k}, \qquad a_{i} \in C^{\infty}(\R),
\label{eq:contactk:liouvilletype}
\end{equation}
is a contact Lie system of Liouville type on $\mathbf{S}^{3}$ or  $\mathbf{AdS}^{2+1}$ with respect to the contact form $\eta_{\k}$, {  with a  VG Lie algebra, spanned by the Hamiltonian vector fields \eqref{eq:contactck:hamvf}, isomorphic to ${\so}(3) \oplus \R$ or    to ${\so}(2,1) \oplus \R$, respectively.}

\begin{remark}
The contact Lie system of Liouville type $\Y_{\k}$ is the subsystem of the contact Lie system $\X_{\k}$ \eqref{eq:contactck:system} on $\S^{3}_{[\kappa_{1}],\kappa_{2},+}$ obtained after the following choices for the ten $t$-dependent functions $b_{i}(t)$:
\begin{equation}
\begin{split}
&b_{1}(t) = 0, \qquad b_{2}(t) = \tfrac 12\kappa_{2} a_{2}(t), \qquad b_{3}(t) = - \tfrac 12 a_{2}(t), \qquad b_{4}(t) = 0, \\[2pt]
&  b_{5}(t) = \tfrac{1}{2} a_{1}(t) - 2 a_{4}(t),\qquad b_{6}(t) = \tfrac{1}{2} a_{3}(t), \qquad\quad\ b_{7}(t) = - \tfrac{1}{2}  {\kappa_{2}} a_{1}(t) - 2 \kappa_{2} a_{4}(t), \\[2pt]
& b_{8}(t) = \tfrac{1}{2} a_{1}(t) - 2 a_{4}(t),\qquad  b_{9}(t) = \tfrac{1}{2}  \kappa_{2}a_{3}(t), \qquad b_{10}(t) = - \tfrac{1}{2}   {\kappa_{2}} a_{1}(t) - 2 \kappa_{2} a_{4}(t).
\end{split}
\nonumber
\end{equation}
Observe that this procedure can also be regarded as taking the restriction of the $t$-dependent vector field $\X_{\k}$ { to the VG Lie subalgebra spanned by  the vector fields \eqref{eq:contactck:hamvf}.}
\end{remark}

%%%%%%%%%%%%%%%%%%%%%%

\subsection{Contact Lie systems on $\mathbf{E}^{3}$, $\mathbf{G}^{2+1}$, $\mathbf{M}^{2+1}$, $\mathbf{NH}_{\pm}^{2+1}$ and $\mathbf{H}^{3}$ }
\label{subsection:contactck:liouville2}

In contrast to the sphere $\mathbf{S}^{3}$  and anti-de Sitter space  $\mathbf{AdS}^{2+1}$, where contact Lie systems of Liouville type have previously been obtained in a very geometrical way through Theorem~\ref{th:contactmetric:firstintegrals}, we now look for subsystems of Liouville type within the contact Lie system $\X_{\k}$ \eqref{eq:contactck:system} on the remaining six spaces covered by Theorem~\ref{th:3DCK:fibrebundles}. For this purpose, we consider those first integrals of the Reeb vector field $\cR_{\k}$ belonging to the $\sp(4, \R)$-Lie algebra spanned by the Hamiltonian functions $h_{\k,i}$ on the corresponding spaces.

\begin{itemize}
\item \textit{Euclidean $\E^{3}$, Galilean $\G^{2+1}$ and Minkowskian $\M^{2+1}$ spaces.}
Recall that these are the three flat spaces $\S^{3}_{[0],\kappa_{2},+}$, placed in the middle column in Table~\ref{table:3DCK:contactnine}, so that we take   $\k = (0, \kappa_{2}, +1)$. There are  six independent first integrals of the Reeb vector field, whose expressions in terms of geodesic parallel (Cartesian) coordinates $(x, y, z)$ follow from Table~\ref{table:contactck:sp4}:
\begin{equation}
   h_{\k,2}= z,      \qquad    h_{\k,4}= y z,  \qquad   h_{\k,5}= \tfrac 12, \qquad
    h_{\k,6}=y,   \qquad  h_{\k,7}= \tfrac 12 y^2,   \qquad   h_{\k,10}= \tfrac 12  z^2 ,
\nonumber
\end{equation}
 which span a Lie algebra isomorphic to the two-photon Lie algebra $\h_{6}$ with respect to the Jacobi bracket $\set{\cdot, \cdot}_{\eta_{\k}}$ (see Table~\ref{table:contactck:commutation}).  Their associated
 Hamiltonian vector fields, obtained from Table~\ref{table:contactck:sp4}, read  as
 \begin{equation}
\begin{split}
&\X_{\k,2} = - z \pdv{x} + \pdv{y}, \qquad \X_{\k,4} = y \pdv{y} - z \pdv{z}, \qquad  \X_{\k, 5} = - \pdv{x}\equiv -\frac 12\cR_{\k}  , \\[2pt]
&\X_{\k, 6} = - y \pdv{x} - \pdv{z}, \qquad \X_{\k,7} = - y \pdv{z}, \qquad\qquad\ \X_{\k,10} = z \pdv{y} .
\end{split}
\label{eq:contactck:realizationh6}
\end{equation}
  Thus, the $t$-dependent vector field
\begin{equation}
\Y_{\k} = b_{2}(t) \X_{\k,2} + b_{4}(t) \X_{\k,4} + b_{6}(t) \X_{\k,6} + b_{7}(t) \X_{\k,7} + b_{10}(t) \X_{\k,10} - \tfrac 12  {b_{5}(t)}  \cR_{\k} ,
\label{eq:contactck:Lflat}
\end{equation}
coming from $\X_{\k}$ \eqref{eq:contactck:system} after the choices $b_{1}(t) = b_{3}(t) = b_{8}(t) = b_{9}(t) = 0$, is a contact Lie system of Liouville type. Hence the VG Lie algebra is isomorphic to $\h_{6}$. Note that the signature of the metric, determined in this case by the contraction parameter $\kappa_{2}$, plays no role in the system \eqref{eq:contactck:Lflat}. Moreover, this is, precisely, the $\h_{6}$-class of Lie systems on $\R^{3} \simeq \S_{[0],\kappa_{2},+}$ recently constructed in \cite{Campoamor2024}.

\item  \textit{Oscillating and expanding NH spaces $\NH_{\pm}^{2+1}$.} They correspond to the two curved non-relativistic spacetimes $\S^{3}_{[\kappa_{1}],0,+} $ (see Table~\ref{table:3DCK:contactnine})
with curvature $\kappa_1 = \pm 1$. There are four first integrals of  $\cR_{\k}$ that in the geodesic parallel  coordinates $(x, y, z)$ shown in  Table~\ref{table:contactck:sp4} are given by
\begin{equation}
          h_{\k,4}= y z,           \qquad h_{\k,7}= \tfrac 12 y^2,    \qquad  h_{\k,10}= \tfrac 12  z^2 ,  \qquad  h_{\k,5}+\kappa_1h_{\k,8}= \tfrac 12.
\label{hamNH}
\end{equation}
These span a Lie algebra isomorphic to $\sl(2, \R) \oplus \R$ with respect to the Jacobi bracket $\set{\cdot, \cdot}_{\eta_{\k}}$ in Table~\ref{table:contactck:commutation}. The corresponding
 Hamiltonian vector fields turn out to be
 \begin{equation}
   \X_{\k,4} = y \pdv{y} - z \pdv{z}, \quad\    \X_{\k,7} = - y \pdv{z}, \quad\  \X_{\k,10} = z \pdv{y} , \quad\      \X_{\k, 5}+\kappa_1   \X_{\k, 8} = - \pdv{x}\equiv -\frac 12\cR_{\k}   .
   \nonumber
\end{equation}
 Therefore, the $t$-dependent vector field
\begin{equation}
\Y_{\k} =    b_{4}(t) \X_{\k,4} + b_{7}(t) \X_{\k,7} + b_{10}(t) \X_{\k,10} - \tfrac{1}{2} b_{5}(t)\cR_{\k} ,
\label{eq:contactck:LNH}
\end{equation}
provided by $\X_{\k}$ in \eqref{eq:contactck:system} with   $b_{8}(t) = \kappa_{1} b_{5}(t)$ and $b_{1}(t) =b_{2}(t) = b_{3}(t) = b_{6}(t) = b_{9}(t) = 0$, is a contact Lie system of Liouville type. { The} VG Lie algebra  is isomorphic to $\sl(2, \R) \oplus \R$ for the two possible values of the curvature $\kappa_1 = \pm 1$.

\item { \textit{Hyperbolic space $\H^{3}$.}} Finally, we set $\k = (-1,+1,+1)$, and conclude that there are only two first integrals of $\cR_{\k}$ given, in  geodesic parallel  coordinates $(x, y, z)$, by
\begin{equation}
          h_{\k,7}+ h_{\k,10} = \frac 12  \left( \sinh^2\! y \cosh^2\! z+\sinh^2 \!z \right)    ,     \qquad   h_{\k,5}-h_{\k,7}-h_{\k,8}-h_{\k,10}= \frac 12,
\nonumber
\end{equation}
 closing on a Lie algebra isomorphic to $\R^{2}$ with respect to the Jacobi bracket $\set{\cdot, \cdot}_{\eta_{\k}}$. From Table~\ref{table:contactck:sp4}, their associated
 Hamiltonian vector fields are found to be
 \begin{equation*}
\begin{aligned}
&  \X_{\k,7} + \X_{\k,10} = \cosh y \tanh z  \pdv{y} -  \sinh y \pdv{z} ,\\
&  \X_{\k,5} - \X_{\k,7} - \X_{\k,8} - \X_{\k,10}  = -\pdv{x}-\cosh y \tanh z \pdv{y}+\sinh y \pdv{z}\equiv -\frac 12  \cR_{\k} .
\end{aligned}
\end{equation*}
 In this way, the $t$-dependent vector field
\begin{equation}
\Y_{\k} =   \bigl(b_{5}(t) + b_{7}(t) \bigr) (\X_{\k,7} + \X_{\k,10})  - \tfrac 12  {b_{5}}(t)  \cR_{\k}\label{eq:contactck:LH3}
\end{equation}
arising as the subsystem of  $\X_{\k}$ \eqref{eq:contactck:system}  with $ b_{8}(t) =- b_{5}(t)$, $b_{10}(t) = b_{7}(t)$ and $b_{1}(t) = b_{2}(t) = b_{3}(t) = b_{4}(t) = b_{6}(t) = b_{9}(t) = 0$,   is a contact Lie system of Liouville type { with a VG Lie algebra isomorphic to $\R^{2}$}.

\end{itemize}

%%%%%%%%%%%%%%%%%%%%%%

\section{Curvature-dependent reductions of contact Lie systems of\\  Liouville type} \label{section:reduction}

We proceed to apply Theorem~\ref{th:s3:reduction} to the  eight contact Lie systems of Liouville type obtained in Sections~\ref{subsection:contactck:liouville} and \ref{subsection:contactck:liouville2} by means of the $\S^{1}_{[\kappa_1]}$-principal bundles given in Theorem~\ref{th:3DCK:fibrebundles} and studied in detail in Section~\ref{section:3DCK:contactstructure}.

\begin{remark}
The Reeb vector field $\cR_{\k}$ is a central generator of the VG Lie algebras $V^{Y_{\k}}$ of the contact Lie systems of Liouville type \eqref{eq:contactk:liouvilletype}, \eqref{eq:contactck:Lflat}, \eqref{eq:contactck:LNH} and \eqref{eq:contactck:LH3}, inducing in these cases the following exact sequence of Lie algebras:
\begin{equation}
\begin{tikzcd}
0  \arrow[hookrightarrow]{r} & \langle \cR_{\k} \rangle \arrow[hookrightarrow]{r}  & V^{Y_{\k}} \arrow["(\pi_{\cR_{\k}})_{*}"]{r}  & V^{(\pi_{\cR_{\k}})_{*}Y_{\k}} \arrow{r} & 0 .
\end{tikzcd}
\nonumber
\end{equation}
 Thus, the VG Lie algebras $V^{Y_{\k}}$ of these non-reduced systems are one-dimensional central extensions of the VG Lie algebras $V^{(\pi_{\cR_{\k}})_{*}Y_{\k}}$ of the reduced systems.
\end{remark}

 For the sake of simplicity,  the index `$\k$' is omitted hereafter.

%%%%%%%%%%%%%%%%%%%%%%

\subsection{Reductions on $\mathbf{S}^{3}$ and $\mathbf{AdS}^{2+1}$}

The contact Lie systems of Liouville type on $\S^{3}$ obtained in Section~\ref{subsection:contactck:liouville}, as pointed out in Section~\ref{section:s3}, can be  projected to LH systems on the sphere $\S^{2}$ with respect to the symplectic form \eqref{eq:s3:vols2} via the Hopf fibration $\pi_{\cR}: \S^{3} \to \S^{2}$ (\ref{eq:s3:Hopffibration}). First of all, let us recall that the unit $2$-sphere $\S^{2} \equiv \set{  (x^{0}, x^{1}, x^{2}) \in \R^{3}: (x^{0})^{2} + (x^{1})^{2} + (x^{2})^{2} = 1  }$  can be regarded as $\S^{3} \cap \set{x^{3} = 0}$ with respect to the ambient description \eqref{eq:3DCK:model} which, in geodesic parallel coordinates $(x, y, z)$ of $\S^{3}$  (see (\ref{coord})), is specified as the submanifold $\set{z = 0} \subset \S^{3}$. This leads to the geodesic parallel coordinates $(x, y)$ of $\S^{2}$, in terms of which the ambient coordinates $(x^{0}, x^{1}, x^{2})$ are described as:
\begin{equation}
x^{0} = \cos x \cos y, \qquad x^{1} = \sin x \cos y, \qquad x^{2} = \sin y.
\nonumber
\end{equation}
In these coordinates, the symplectic form \eqref{eq:s3:vols2} becomes
\begin{equation}
\omega =   -  \frac {\cos y}4 \, \dd x \wedge \dd y .
\label{eq:red:s2symp}
\end{equation}
Now, the contact Lie system of Liouville type $\Y$ \eqref{eq:contactk:liouvilletype} on $(\S^{3}, \eta)$ is projected onto the LH system $\pi_{\cR_{*}} \Y$ on $\S^{2}$ with respect to the symplectic form \eqref{eq:red:s2symp} given by
\begin{equation}
\pi_{\cR_{*}} \Y = \sum_{i = 1}^{3} a_{i}(t) \pi_{\cR_{*}} \Y_{i} ,
\label{xxa}
\end{equation}
where the vector fields $\pi_{\cR_{*}} \Y_{i}$ read, in geodesic parallel coordinates $(x, y)$ of $\S^{2}$, as
\begin{equation}
\begin{split}
 &  \pi_{\cR_{*}} \Y_{1} = \cos x \tan y \pdv{x} - \sin x \pdv{y}, \qquad \pi_{\cR_{*}} \Y_{2} =  \pdv{x},\\[2pt]
&  \pi_{\cR_{*}} \Y_{3} = \sin x \tan y \pdv{x} + \cos x \pdv{y}.
\end{split}
\label{xa}
\end{equation}
 Thus they  span a Lie algebra isomorphic to $\so(3)$ satisfying the usual commutation rules
\be
\bigl[  \pi_{\cR_{*}} \Y_{i} , \pi_{\cR_{*}} \Y_{j} \bigr]=\epsilon_{ijk}\, \pi_{\cR_{*}} \Y_{k}.
\nonumber
\ee
The Hamiltonian functions associated with the vector fields (\ref{xa}) with respect to the symplectic form \eqref{eq:red:s2symp} turn out to be
\begin{equation}
h_{1} =\frac 14 \cos x\cos y , \qquad h_{2} =-\frac 14 \sin y, \qquad h_{3} =\frac 14\sin x \cos y ,
\nonumber
\end{equation}
which { satisfy} the Poisson brackets $\{h_i,h_j\}_\omega=-\epsilon_{ijk}  h_k$.  Consequently, they span a LH algebra $\cH_{\omega} \simeq \so(3) $ with respect to the Poisson bracket induced by the symplectic form \eqref{eq:red:s2symp}.

It is worth remarking that the projected vector fields (\ref{xa}) are, precisely, the generators of the VG Lie algebra of the LH systems on $\S^{2}$ obtained in \cite{Herranz2017} and where the corresponding constants
of the motion and superposition rules can be found. In fact, they are Killing vector fields of the usual Riemannian metric on $\S^{2}$ given in geodesic parallel coordinates by $g=\cos^2\!y\,\dd x\otimes\dd x+ \dd y\otimes \dd y$. Hence   the $\S^{2}$-LH system (\ref{xxa}) is  the spherical counterpart of the   so-called  P$_1$-LH class on $\R^2$~\cite{Ballesteros2015,Blasco2015}.

 Similarly, the contact Lie systems of Liouville type on $\AdS^{2+1}$ obtained in Section~\ref{subsection:contactck:liouville} can be projected to LH systems on $\C\H^{1}$ with respect to the symplectic structure \eqref{eq:3DCK:symplectic_complexhyper} by means of the fibration $\pi_{\cR}: \AdS^{2+1} \to \C \H^{1} $ (\ref{fibAds}). More precisely, the projection of the contact Lie system of Liouville type $\Y$ \eqref{eq:contactk:liouvilletype} on $\AdS^{2+1}$ gives rise to a LH system $\pi_{\cR_{*}} \Y$ on $\C \H^{1}$ with respect to the symplectic form \eqref{eq:3DCK:symplectic_complexhyper} formally given by (\ref{xxa})  where the vector fields $\pi_{\cR_{*}} \Y_{i}$ are expressed in the global coordinates $(u, v)$ of $\C \H^{1}$ as
\begin{equation}
\begin{split}
&\pi_{\cR_{*}} \Y_{1} = v \pdv{u} - u \pdv{v}, \qquad \pi_{\cR_{*}} \Y_{2} = \frac{1}{2} \bigl(  u^{2}- v^{2}-1 \bigr) \pdv{u} + uv \pdv{v},\\[2pt]
& \pi_{\cR_{*}} \Y_{3} = uv \pdv{u} - \frac{1}{2} \bigl(  u^{2} -v^{2}+1\bigr) \pdv{v}.
\end{split}
\label{eq:red:ads:tdep}
\end{equation}
Hence, they close on a VG Lie algebra isomorphic to $\so(2,1)\simeq \sl(2, \R)$:
\be
\bigl[  \pi_{\cR_{*}} \Y_{1} , \pi_{\cR_{*}} \Y_{2} \bigr]=   \pi_{\cR_{*}} \Y_{3},\qquad
\bigl[  \pi_{\cR_{*}} \Y_{1} , \pi_{\cR_{*}} \Y_{3} \bigr]= -  \pi_{\cR_{*}} \Y_{2},\qquad
\bigl[  \pi_{\cR_{*}} \Y_{2} , \pi_{\cR_{*}} \Y_{3} \bigr]= -  \pi_{\cR_{*}} \Y_{1}.
\nonumber
\ee
The associated Hamiltonian functions turn out to be
\begin{equation}
h_{1} =  \frac{1+   u^{2}+ v^{2}}{4 \bigl(1-   u^{2} - v^{2} \bigr)}, \qquad h_{2} =-  \frac{v}{2 \bigl(1-   u^{2} - v^{2} \bigr)}, \qquad h_{3} = \frac{ u}{2 \bigl(1-   u^{2} - v^{2} \bigr)} ,
\label{eq:red:ads:hamfun}
\end{equation}
spanning a LH algebra $\cH_{\omega} \simeq \so(2,1)$ with respect to the Poisson bracket induced by the symplectic form \eqref{eq:3DCK:symplectic_complexhyper} as
\be
\{ h_1,h_2\}_\omega = - h_3,\qquad \{ h_1,h_3\}_\omega =   h_2,\qquad \{ h_2,h_3\}_\omega = h_1.  \nonumber
\ee

Recall that the above results are expressed in terms of the usual projective Poincar\'e coordinates
for the 2D hyperbolic space $\H^2$.  In order to establish a relationship with the (curved) LH systems studied in the literature, let us consider the change of coordinates from $(u,v)$ to the geodesic parallel $(x,y)$, which is given by
 \be
u=\frac{\sinh x\cosh y}{1+\cosh x\cosh y},\qquad v=\frac{\sinh y}{1+\cosh x\cosh y}.
\nonumber
 \ee
 The symplectic form \eqref{eq:3DCK:symplectic_complexhyper} transforms into
 \begin{equation}
\omega =      \frac {\cosh y}4 \, \dd x \wedge \dd y ,
\label{eq:red:h2symp}
\end{equation}
while the vector fields (\ref{eq:red:ads:tdep}) and their associated Hamiltonian functions  (\ref{eq:red:ads:hamfun}), now with respect to (\ref{eq:red:h2symp}),   become
\begin{equation}
\begin{split}
&\pi_{\cR_{*}} \Y_{1} = \cosh x\tanh y  \pdv{x} - \sinh x \pdv{y}, \qquad \pi_{\cR_{*}} \Y_{2} = - \pdv{x}  ,\\[2pt]
& \pi_{\cR_{*}} \Y_{3} =  \sinh x\tanh y  \pdv{x} - \cosh x  \pdv{y},
\end{split}
\label{xxcc}
\end{equation}
\begin{equation}
h_{1} =  \frac 14 \cosh x\cosh y, \qquad h_{2} =-  \frac 14  \sinh y, \qquad h_{3} =  \frac 14 \sinh x\cosh y .
\nonumber
\end{equation}
In this explicit form,  it is easily recognized that the projected vector fields (\ref{xxcc}) are just the generators of the VG Lie algebra of the LH systems on $\H^{2}$ constructed in \cite{Herranz2017} (their constants
of the motion and superposition principles can also be found there), so that they are  Killing vector fields of the   Riemannian metric on $\H^{2}$ reading as $g=\cosh^2\!y\,\dd x\otimes\dd x+ \dd y\otimes \dd y$. Therefore,   the  reduced LH system $\pi_{\cR_{*}} \Y$ on $\C \H^{1}$ is  the hyperbolic partner of the Euclidean class   P$_1$~\cite{Ballesteros2015,Blasco2015}.

%%%%%%%%%%%%%%%%%%%%%%

\subsection{Reductions on $\mathbf{E}^{3}$, $\mathbf{G}^{2+1}$, $\mathbf{M}^{2+1}$, $\mathbf{NH}_{\pm}^{2+1}$ and $\mathbf{H}^{3}$}

Through the trivial fibration $\pi_{\cR}: \S_{[0], \kappa_{2}, +} \to \R^{2}$ \eqref{eq:3DCK:fibrationflat}, covering $\mathbf{E}^{3}$, $\mathbf{G}^{2+1}$ and $\mathbf{M}^{2+1}$, the contact Lie system of Liouville type \eqref{eq:contactck:Lflat} is projected onto the LH system on $\R^{2}$ with respect to the canonical symplectic form \eqref{eq:3DCK:symplecticplane} given by
\begin{equation}
\pi_{\cR_{*}} \Y = b_{2}(t)  \pdv{q} + b_{4}(t) \left( q \pdv{q} - p \pdv{p} \right) - b_{6}(t)  \pdv{p} -  b_{7}(t) q \pdv{p}+ b_{10}(t) p \pdv{q}.
\nonumber
\end{equation}
This is, exactly, the vector field of the so-called P$_{5}$-LH class on $\R^{2}$ in the classification obtained in~\cite{Ballesteros2015} { with a VG Lie algebra} isomorphic to $\mathfrak{sl}(2,\R ) \ltimes  \mathbb{R}^2$. Observe that this reduction procedure is equivalent to the one recently used in \cite{Campoamor2024} by means of invariants of the two-photon $\h_{6}$-realization \eqref{eq:contactck:realizationh6}.

Following a similar ansatz for the NH spacetimes $\NH_{\pm}^{2+1}$, the trivial  fibration
$\pi_{\cR}: \NH_{\pm}^{2+1} \to \R^{2}$, shown in  (\ref{eq:3DCK:fibrationNHp})  and (\ref{eq:3DCK:fibrationNHm}), we project  the contact Lie system of Liouville type \eqref{eq:contactck:LNH} onto the LH system on $\R^{2}$ with respect to the canonical symplectic form \eqref{eq:3DCK:symplecticplane}
\begin{equation}
\pi_{\cR_{*}} \Y = b_{4}(t) \left(q \pdv{q} - p \pdv{p} \right) - b_{7}(t) q \pdv{p} + b_{10}(t) p \pdv{q}.
\nonumber
\end{equation}
The VG Lie algebra is thus isomorphic to $\mathfrak{sl}(2,\R)$, and as the value of the Casimir function $C$ of $\sl(2, \R)^{*}$ coming from the Killing--Cartan form of $\sl(2, \R)$ under the functional realization \eqref{hamNH} is
\begin{equation}
C = h_{4}h_{10} - h_{4}^{2}=  0 ,
\nonumber
\end{equation}
 such LH system corresponds to the  I$_{5}$-LH class on $\R^{2}$ in~\cite{Ballesteros2015,Blasco2015}.
Recall that it encompasses systems such as the harmonic oscillator or certain types of Ermakov, Kummer--Schwarz and Riccati equations.

Finally, the case of the system \eqref{eq:contactck:LH3} is rather trivial, as the reduced system provided by the fibration $\pi_{\cR}: \H^{3} \to \R^{2}$ is just the generator of the so-called I$_{1}$-LH class~\cite{Ballesteros2015} which, up to a local change of coordinates, is simply $\mathrm{I}_{1} = \bigl\langle \pdv{q} \bigr\rangle$.

 %%%%%%%%%%%%%%%%%%%%%%%%%%%%%%%%%%%%%%%%%%%%%%%%%%%%

\section{Concluding remarks}
\label{conclusions}
In this work, the notion of scaling symmetries for autonomous Hamiltonian systems \cite{Grabowska2022,Sloan2018,Bravetti2023, Bravetti2024} has been extended to LH systems on symplectic manifolds, enabling their subsequent reduction to contact Lie systems. This formalism has been applied to study the reduction of a time-dependent harmonic oscillator and also provides the first known example of thermodynamical systems analyzed through Lie system theory. Next, $\sp(4, \R)$-contact Lie systems on $\S^{3}$ are naturally derived from the $\sp(4, \R)$-LH systems constructed in \cite{Campoamor2024} coming from the fundamental representation of $\sp(4, \R)$. Furthermore, it is shown that these systems are a particular case within a larger family of contact Lie systems { in} three dimensions, thereby yielding the first known classes of contact Lie systems on homogeneous spaces. Additionally, it is shown that the contact structure of certain 3D CK spaces induces a principal bundle, whose associated 1D Lie group is determined by the sign of the curvature of the corresponding CK space. In the cases of the sphere $\S^{3}$ and the anti-de Sitter space $\AdS^{2+1}$, the contact structure is closely related with the associated (pseudo)-Riemannian structure, since they are Sasakian manifolds. This compatibility condition is later employed to construct contact Lie systems of Liouville type in a highly geometrical manner. Subsequently, Liouville-type subsystems contained within the $\sp(4, \R)$-contact Lie systems are reduced to 2D LH systems studied in previous works via the aforementioned curvature-dependent principal bundles. The significance of our approach lies in showing that, through a suitable combination of the algebraic techniques proposed in \cite{Campoamor2024} with those developed here, Problems~\ref{problemA} and \ref{problemB} posed in the Introduction regarding the study of Lie systems can be addressed in a very geometrical way.

In this context, several open problems emerge naturally. First, it would be interesting to obtain a classification of invariant contact structures on the $(2N +1)$D CK spaces through a { cohomology}-based approach \cite{Khakimdjanov2004,Ghosh2019}. Concerning applications, the theory of contact Lie systems can potentially be applied to the study of specific thermodynamic systems, particularly those related to the thermodynamics of black holes in anti-de Sitter spacetimes, where contact geometry has been shown to provide an elegant geometrical framework \cite{Alekseevskii1990}. In addition, it is worthy to develop new theoretical insights into the deformation of contact Lie systems, mainly starting from the formalism of Poisson--Hopf deformations of LH systems introduced in \cite{PH2018,PH2021}. Finally, we stress that { our novel results} have a natural physical application in 3D gravity, which deserves further investigation. It is worth recalling that 3D gravity has been formulated as a Chern--Simons gauge theory \cite{AT1986,Witten1988} and, notably, the three Riemannian and three Lorentzian spaces considered here are precisely those that appear in this framework, while the three Newtonian ones emerge as their non-relativistic counterparts (see \cite{BHM2010plb} and references therein). Therefore, the application of contact geometry in this framework could reveal valuable structural properties, as well as compatibility with additional geometric structures. Work in these directions is currently in progress.

 %%%%%%%%%%%%%%%%%%%%%%%%%%%%%%%%%%%%%%%%%%%%%%%%%%%%

\section*{Acknowledgements}

\phantomsection
\addcontentsline{toc}{section}{\\[-20pt]Acknowledgements}

\small

The authors thank E.~Padr{\'o}n for useful comments related with the contents of this paper. This work has been supported by Agencia Estatal de Investigaci\'on (Spain)   under  the grant PID2023-148373NB-I00 funded by MCIN/AEI/10.13039/501100011033/FEDER, UE.  F.J.H.~acknowledges support  by the  Q-CAYLE Project  funded by the Regional Government of Castilla y Le\'on (Junta de Castilla y Le\'on, Spain) and by the Spanish Ministry of Science and Innovation (MCIN) through the European Union funds NextGenerationEU (PRTR C17.I1).  O.C. acknowledges a fellowship (grant C15/23) supported by Universidad Complutense de Madrid and Banco de Santander.  The authors also acknowledge the contribution of RED2022-134301-T funded by MCIN/AEI/10.13039/501100011033 (Spain).

 %%%%%%%%%%%%%%%%%%%%%%%%%%%%%%%%%%%%%%%%%%%%%%%%%%%%


\small

 \begin{thebibliography}{99}

\phantomsection
\addcontentsline{toc}{section}{\\[-20pt]References}


 \bibitem{Lie1888} S.~Lie. Classification und Integration von gew{\"o}hnlichen Differentialgleichungen zwischen $x, y$, die eine Gruppe von Transformationen gestatten. {\it Math. Ann.} {\bf 32} (1888) 213--281. \doi{10.1007/BF01444068}

 \bibitem{Stephani1990} H.~Stephani. {\it Differential Equations: Their Solution Using Symmetries}. (Cambridge: Cambridge University Press) 1990. \doi{10.1017/CBO9780511599941}

\bibitem{AbrahamShrauner1995} B.~Abraham-Shrauner, P.~G.~Leach, K.~S.~Govinder and G.~Ratcliff. Hidden and contact symmetries of ordinary differential equations. {\it J. Phys. A: Math. Gen.} {\bf 28} (1995) 6707--6716. \doi{10.1088/0305-4470/28/23/020}.

\bibitem{CampoamorStursberg2016}R.~Campoamor-Stursberg. Generating functions and existence of contact symmetries of third order scalar ordinary differential equations. {\it Appl. Math. Comput.} {\bf 273} (2016) 1179--1189. \doi{10.1016/j.amc.2015.08.131}

\bibitem{Kushner2006} A.~Kushner, V.~Lychagin and V.~Rubtsov. {\it Contact Geometry and Nonlinear Differential Equations} (Cambridge: Cambridge University Press) 2006. \doi{10.1017/CBO9780511735141}

{
\bibitem{Kruglikov1998} B.~S.~Kruglikov. Symplectic and contact Lie Algebras with an application to the
Monge-Amp{\`e}re equations. {\it Tr. Mat. Inst. Steklova} {\bf 221} (1998), 232--246 (in Russian). \\
Ibid {\it Proc. Steklov Inst. Math.} {\bf 221} (1998) 221--235 (in English).
 }
\bibitem{Khakimdjanov2004} Yu.~Khakimdjanov, M.~Goze and A.~Medina. Symplectic or contact structures on Lie groups. {\it Differ. Geom. Appl.} {\bf 21} (2004) 41--54. \doi{10.1016/j.difgeo.2003.12.006}

\bibitem{Ancochea2006} J.~M.~Ancochea, R.~Campoamor-Stursberg and L.~Garcia Vergnolle. Solvable Lie algebras with naturally graded nilradicals and their invariants. {\it J. Phys. A: Math. Gen.} {\bf 39} (2006) 1339--1355. \doi{10.1088/0305-4470/39/6/008}

 \bibitem{Blaszak2017}
M. B{\l}aszak and A.~Sergyeyev. Dispersionless $(3+1)$-dimensional integrable hierarchies. {\it Proc. R. Soc. A} {\bf 473} (2017) 20160857. \doi{10.1098/rspa.2016.0857}

\bibitem{Sergyeyev2018}
A.~Sergyeyev. New integrable $(3+1)$-dimensional systems and contact geometry. {\it Lett. Math. Phys.} {\bf 108} (2018) 359--376. \doi{10.1007/s11005-017-1013-4}

\bibitem{Sergyeyev2025}
A.~Sergyeyev. Multidimensional integrable systems from contact geometry. {\it Bol. Soc. Mat. Mex.} {\bf 31} (2025) 26. \doi{10.1007/s40590-024-00703-7}

{
\bibitem{Reeb1952} G.~Reeb. Sur certaines propri{\'e}t{\'e}s topologiques des trajectoires des syst{\`e}mes dynamiques. {\it M\'em.  Acad. Roy. Belgique, Sci.} {\bf 27} (1952) 130--194. {\sc url:}
\href{https://academieroyale.be/academie/documents/XXVII_9_Reeb_Surcertainesproprietestopologiquestrajectoiressystemes_195219564.pdf}{https://academieroyale.be/academie/documents/XXVII{\_}9{\_}Reeb{\_}Surcertainesproprietestopologiques\\trajectoiressystemes{\_}195219564.pdf}
}
\bibitem{Blair2010} D.~E.~Blair. {\it Riemannian Geometry of Contact and Symplectic Manifolds} (Boston: Birkh{\"a}user) 2010. \doi{10.1007/978-0-8176-4959-3}

\bibitem{Geiges2008} H.~Geiges. {\it An Introduction to Contact Topology} (Cambridge: Cambridge University Press) 2008. \doi{10.1017/CBO9780511611438}

 \bibitem{Calvaruso2010}
 G.~Calvaruso and D.~Perrone. Contact pseudo-metric manifolds. {\it Differ. Geom. Appl.} {\bf 28} (2010) 615--634. \doi{10.1016/j.difgeo.2010.05.006}

  \bibitem{Takahashi1969}
T.~Takahashi. Sasakian manifold with pseudo-Riemannian metric. {\it Tohoku Math. J.} {\bf 21} (1969) 271--290. \doi{10.2748/tmj/1178242996}

\bibitem{Duggal1990}
K.~L.~Duggal. Space time manifolds and contact structures. {\it Int. J. Math. Math. Sci.} {\bf 13}  (1990) 545--554. \doi{10.1155/S0161171290000783}

\bibitem{Bravetti2017} A.~Bravetti, H.~Cruz and D.~Tapias. Contact Hamiltonian mechanics. {\it Ann. Phys.} {\bf 376} (2017) 17--39. \doi{10.1016/j.aop.2016.11.003}

\bibitem{Ciaglia2018} F.~M.~Ciaglia, H.~Cruz and G.~Marmo. Contact manifolds and dissipation, classical and quantum. {\it Ann. Phys.} {\bf 398} (2018) 159--179. \doi{10.1016/j.aop.2018.09.012}

\bibitem{deLeon2019}M.~de Le{\'on} and M.~Lainz Valc{\'a}zar. Contact Hamiltonian systems. {\it J. Math. Phys.} {\bf 60} (2019) 102902. \doi{10.1063/1.5096475}

\bibitem{Bravetti2020} A.~Bravetti, M.~de Le{\'o}n, J.~C.~Marrero and E.~Padr{\'o}n. Invariant measures for contact Hamiltonian systems: symplectic sandwiches with contact bread. {\it J. Phys. A: Math. Theor.} {\bf 53} (2020) 455205. \doi{10.1088/1751-8121/abbaaa}

 \bibitem{Grabowska2022}
K.~Grabowska and J.~Grabowski. A geometric approach to contact Hamiltonians and contact Hamilton--Jacobi theory. {\it J. Phys. A: Math. Theor.} {\bf 55} (2022) 435204. \doi{10.1088/1751-8121/ac9adb}

\bibitem{LopezGordon2024} A.~L{\'o}pez-Gord{\'o}n. {\it The Geometry of Dissipation}. Ph.D. Thesis. Universidad Aut{\'o}noma de Madrid, 2024. {\sc arXiv:}\href{https://arxiv.org/abs/2409.11947}{2409.11947} [math-ph].

\bibitem{Rivas2021} X.~Rivas. {\it Geometrical Aspects of Contact Mechanical Systems and Field Theories}. Ph.D. Thesis. Universitat Polit\'ecnica de Catalunya (UPC), 2021. {\sc arXiv:}\href{https://arxiv.org/abs/2204.11537}{2204.11537} [math-ph].

\bibitem{deLeon2023} M.~de Le{\'o}n, J.~Gaset, M.~C.~Muñoz-Lecanda, X.~Rivas and N.~Rom{\'a}n-Roy. Multicontact formulation for non-conservative field theories. {\it J. Phys. A: Math. Theor.} {\bf 56} (2023) 025201. \doi{10.1088/1751-8121/acb575}

  \bibitem{deLucas2023}
J.~de Lucas and X.~Rivas. Contact Lie systems: theory and applications.
{\it J. Phys. A: Math. Theor.} {\bf 56} (2023) 335203. \doi{10.1088/1751-8121/ace0e7}

 \bibitem{Jacobi2015}
F.~J.~Herranz, J.~de Lucas and C.~Sard\'on.
Jacobi--Lie systems: Fundamentals and low-dimensional classification.  {\it Dyn. Sys.  Diff. Equ.  Appl. } AIMS Proceedings (2015) 605--614. {\sc doi:}\href{https://www.aimsciences.org/article/doi/10.3934/proc.2015.0605}{10.3934/proc.2015.0605}

 \bibitem{deLucas2020}
J.~de Lucas and  C.~Sard{\'o}n.
{\it A Guide to Lie Systems with Compatible Geometric Structures}. (Singapore:
World Scientific)  2020.
\doi{10.1142/q0208}

  \bibitem{Ballesteros2013}
A.~Ballesteros, J.~F.~Cari{\~n}ena, F.~J.~Herranz, J.~de Lucas and C.~Sard{\'o}n. From constants of motion to superposition rules for Lie--{H}amilton systems.  {\it J. Phys. A: Math. Theor.} {\bf 46} (2013) 285203.
\doi{10.1088/1751-8113/46/28/285203}

\bibitem{Lie1893} S.~Lie. {\it Theorie der Transformationsgruppen}. (Leipzig: Teubner) 1893.

\bibitem{Mostow1950} G.~D.~Mostow. The Extensibility of Local Lie Groups of Transformations and Groups on Surfaces. {\it Ann. Math.} {\bf 52} (1950) 606--636. \doi{10.2307/1969437}

\bibitem{Gorbatsevich1977} V.~V.~Gorbatsevich. Three-dimensional homogeneous spaces. {\it Sib. Math. J.} {\bf 18} (1977) 200--210. \doi{10.1007/BF00967152}

\bibitem{Shnider1984} S.~Shnider and P.~Winternitz. Classification of systems of nonlinear ordinary differential equations with superposition principles. {\it J. Math. Phys.} {\bf 25} (1984) 3155--3165. \doi{10.1063/1.526085}

  \bibitem{Shnider1984a} S.~Shnider and P.~Winternitz. Nonlinear equations with superposition principles and the theory of transitive primitive Lie algebras. {\it Lett. Math. Phys.} {\bf 8} (1984)   69--78. \doi{10.1007/BF00420043}

\bibitem{GonzalezLopez1992} A.~Gonz{\'a}lez-L{\'o}pez, N.~Kamran and P.~J.~Olver. Lie algebras of vector fields in the real plane. {\it Proc. London Math. Soc.} {\bf 64} (1992) 339--368. \doi{10.1112/plms/s3-64.2.339}

\bibitem{Ballesteros2015}
A.~Ballesteros, A.~Blasco, F.~J.~Herranz, J.~de Lucas and  C.~Sard{\'o}n. Lie--Hamilton systems on the plane: Properties, classification and applications.
 {\it J. Diff. Equ.} {\bf 258} (2015) 2873--2907. \doi{10.1016/j.jde.2014.12.031}

\bibitem{Doubrov2017} B.~Doubrov. Three-dimensional homogeneous spaces with non-solvable transformation groups. (2017) {\it Preprint}  {\sc arXiv:}\href{https://arxiv.org/abs/1704.04393}{1704.04393} [math.DG].

\bibitem{Gorbatsevich2024} V.~V.~Gorbatsevich. On decompositions and transitive actions of nilpotent Lie groups. {\it Russ. Math.} {\bf 68} (2024) 1--11. \doi{10.3103/S1066369X24700221}
{
\bibitem{Bodarenko2019} V.~M.~Bondarenko and  A.~P.~Petravchuk. Wildness of the problem of classifying nilpotent Lie algebras of vector fields in four variables. {\it Linear Algebra Appl.} {\bf 568} (2019) 165--172. \doi{10.1016/j.laa.2018.07.031}
}
  \bibitem{Campoamor2024}
R.~Campoamor-Stursberg, O.~Carballal and F.~J.~Herranz. A representation-theoretical approach to higher-dimensional Lie--Hamilton systems: The symplectic Lie algebra $\mathfrak{sp}(4, \mathbb{R})$. {\it Commun. Nonlinear Sci. Numer. Simulat.} {\bf 141}   (2025)  108452.
\doi{10.1016/j.cnsns.2024.108452}

\bibitem{Carballal2024sp6}
O.~Carballal, R.~Campoamor-Stursberg and F.~J.~Herranz. Lie--Hamilton systems associated with the symplectic Lie algebra $\mathfrak{sp}(6, \mathbb{R})$.  {\it J. Geom. Symmetry Phys.}  {\bf 69} (2024)  37--57.
 \doi{10.7546/jgsp-69-2024-37-57}

   \bibitem{Herranz2017}
F.~J.~Herranz, J.~de Lucas and M.~Tobolski. Lie--Hamilton systems on curved spaces: a geometrical approach. {\it J. Phys. A: Math. Theor.} {\bf 50} (2017) 495201. \doi{10.1088/1751-8121/aa918f}

  \bibitem{Campoamor2024conformes}
R.~Campoamor-Stursberg, O.~Carballal and F.~J.~Herranz. Lie--Hamilton systems on Riemannian and Lorentzian spaces from conformal transformations and some of their applications.
{\it J. Phys. A: Math. Theor.} {\bf 57} (2024) 485203. \doi{10.1088/1751-8121/ad8e1d}

 \bibitem{Ballesteros1994}
A.~Ballesteros, F.~J.~Herranz, M.~del~Olmo and M.~Santander. Quantum (2+1) kinematical algebras: a global approach. {\it J. Phys. A: Math. Gen.} {\bf 27} (1994) 1283--1297. \doi{10.1088/0305-4470/27/4/021}

 \bibitem{Herranz2006}
 F.~J.~Herranz and A.~Ballesteros. Superintegrability on three-dimensional Riemannian and relativistic spaces of constant curvature. {\em Symmetry Integrability Geom.: Methods Appl.}    {\bf 2} (2006) 010. \doi{10.3842/SIGMA.2006.010}
 {
\bibitem{Gromov1990}
N.~A.~Gromov and V.~I.~Man'ko.
\newblock The Jordan--Schwinger representations of
 Cayley--Klein groups. I. The orthogonal groups.
 \newblock{\em J. Math. Phys.} {\bf 31} (1990) 1047--1053.
\doi{10.1063/1.528781}
}
\bibitem{Gromov1992}
N.~A.~Gromov.
\newblock The Gel'fand--Tsetlin representations of the orthogonal Cayley--Klein algebras.
 \newblock{\em J. Math. Phys.} {\bf 33} (1992) 1363--1373.
\doi{10.1063/1.529711}

  \bibitem{Herranz1997}
  F.~J.~Herranz and  M.~Santander.
\newblock Casimir invariants for the complete family of quasisimple orthogonal algebras.
\newblock{\em  J. Phys. A: Math. Gen.}  {\bf 30} (1997) 5411--5426.
\doi{10.1088/0305-4470/30/15/026}

\bibitem{Azcarraga1998}
J.~A.~de~Azc{\'a}rraga, F.~J.~Herranz, J.~C.~P{\'e}rez~Bueno and M.~Santander. Central extensions of the quasi-orthogonal Lie algebras. {\it J. Phys. A: Math. Gen.} {\bf 31} (1998) 1373--1394. \doi{10.1088/0305-4470/31/5/008}

\bibitem{GH2021symmetry}
 I.~Gutierrez-Sagredo and F.~J.~Herranz. Cayley--Klein Lie bialgebras: Noncommutative spaces, Drinfel'd doubles and kinematical applications. {\it Symmetry} {\bf 13} (2021) 1249. \doi{10.3390/sym13071249}

 \bibitem{Sloan2018}
D.~Sloan. Dynamical similarity. {\it Phys. Rev. D} {\bf 97} (2018) 123541.
\doi{10.1103/PhysRevD.97.123541}

\bibitem{Bravetti2023}
A.~Bravetti, C.~Jackman and D.~Sloan. Scaling symmetries, contact reduction and Poincar{\'e}'s dream. {\it J. Phys. A: Math. Theor.} {\bf 56} (2023) 435203. \doi{10.1088/1751-8121/acfddd}

   \bibitem{Bravetti2024}A.~Bravetti, S.~Grillo, J.~C.~Marrero and E.~Padr{\'o}n.
  Kirillov structures and reduction of Hamiltonian systems by scaling and standard symmetries. {\it Stud. Appl. Math.} {\bf 153} (2024) e12681.  \doi{10.1111/sapm.12681}

  \bibitem{Blasco2015} A.~Blasco, F.~J.~Herranz, J.~de Lucas and C.~Sard\'on.
{Lie--Hamilton systems on the plane: Applications and superposition rules.}
 {\it J. Phys. A: Math. Theor.}  {\bf 48} (2015)  345202.
 \doi{10.1088/1751-8113/48/34/345202}

 \bibitem{Poincare1908}
H.~Poincar{\'e}. {\it Science et M{\'e}thode}. (Paris: Flammarion) 1908.

\bibitem{Gryb2021}
S.~Gryb and D.~Sloan. When scale is surplus. {\it Synthese} {\bf 199} (2021) 14769--14820. \doi{10.1007/s11229-021-03443-7}

\bibitem{Libermann1987}
P.~Libermann and C.-M.~Marle. {\it Symplectic Geometry and Analytical Mechanics}. (Dordrecht: D. Reidel Publishing Co.) 1987. \doi{10.1007/978-94-009-3807-6}

\bibitem{Arnold1978}
V.~I.~Arnold. {\it Mathematical Methods of Classical Mechanics}. (Berlin: Springer) 1978. \doi{10.1007/978-1-4757-1693-1}

  \bibitem{Balian2001}A.~Balian and P.~Valentin.
  Hamiltonian structure of thermodynamics with gauge.
  {\it Eur. Phys. J. B} {\bf 21} (2001) 269--282. \doi{10.1007/s100510170202}

  \bibitem{Bazarov1964}I.~P.~Bazarov.
  {\it Thermodynamics}.
  (New York: Pergamon Press) 1964.

  \bibitem{Mrugala1991} R.~Mruga{\l{}}a, J.~D.~Nulton, J.~C.~Sch{\"o}n and P.~Salamon.
  Contact structure in thermodynamic theory.
  {\it Rep. Math. Phys.} {\bf 29} (1991) 109--121. \doi{10.1016/0034-4877(91)90017-H}

  \bibitem{Mrugala2000} R.~Mruga{\l{}}a.
  On a special family of thermodynamic processes and their invariants.
 {\it Rep. Math. Phys.} {\bf 46} (2000) 461--468.
 \doi{10.1016/S0034-4877(00)90012-0}

  \bibitem{Eberard2007} D.~Eberard, B.~M.~Maschke and A.~J.~van der Schaft.
An extension of Hamiltonian systems to the thermodynamic phase space: Towards a geometry of nonreversible processes.
 {\it Rep. Math. Phys.} {\bf 60} (2007) 175--198.
 \doi{10.1016/S0034-4877(07)00024-9}

\bibitem{Olver} P.~J.~Olver. {\em Applications of Lie Groups to Differential Equations}. (New York: Springer) 1993.
\doi{10.1007/978-1-4612-4350-2}

{
\bibitem{Wei1963} J.~Wei and E.~Norman. Lie algebraic solution of linear differential equations. {\it J. Math. Phys.} {\bf 4} (1963) 575--581. \doi{10.1063/1.1703993}

\bibitem{Wei1964} J.~Wei and E.~Norman. On global representations of the solutions of linear differential equations as a product of exponentials. {\it Proc. Amer. Math. Soc.} {\bf 15} (1964) 327--334. \doi{10.2307/2034065}
}
\bibitem{Ryder1980}
L.~H.~Ryder. Dirac monopoles and the Hopf map $S^{3} \to S^{2}$.
{\it J. Phys. A: Math. Gen.} (1980) {\bf 13} 437--447. \doi{10.1088/0305-4470/13/2/012}
{
\bibitem{Kegel2021} M.~Kegel and C.~Lange. A Boothby--Wang theorem for Besse contact manifolds. {\it Arnold Math. J.} {\bf 7} (2021) 225--241. \doi{10.1007/s40598-020-00165-5}
}

{
\bibitem{Grabowska2024}
K.~Grabowska and J.~Grabowski. The regularity and products in contact geometry. {\it Ann. Mat. Pura Appl.} (2025). \doi{10.1007/s10231-025-01631-7}
}
\bibitem{Montigny1991} M.~de Montigny and J.~Patera.
\newblock Discrete and continuous graded contractions of Lie algebras and superalgebras.
 \newblock {\it J. Phys. A: Math. Gen.}  {\bf 24} (1991)  525--547.
 \doi{10.1088/0305-4470/24/3/012}

    \bibitem{Moody1991}
 R.~V.~Moody and J.~Patera.
\newblock Discrete and continuous graded contractions of representations of Lie algebras.
\newblock{\em  J. Phys. A: Math. Gen.}  {\bf 24} (1991) 2227--2257.
\doi{10.1088/0305-4470/24/10/014}

   \bibitem{Herranz1994}
 F.~J.~Herranz, M.~de Montigny, M.~del Olmo and  M.~Santander.
\newblock Cayley--Klein algebras as graded contractions of $\so(N+1)$.
\newblock{\em  J. Phys. A: Math. Gen.}  {\bf 27} (1994) 2515--2526.
\doi{10.1088/0305-4470/27/7/027}

\bibitem{Inonu1953}
 E.~In{\"o}n\"u and E.~P.~Wigner.
 On the contraction of groups and their representations.
 {\em Proc. Natl.
Acad. Sci. USA.} {\bf 39} (1953) 510--524.
\doi{10.1073/PNAS.39.6.510}

\bibitem{Herranz2000}
F.~J.~Herranz, R.~Ortega and M.~Santander.
\newblock Trigonometry of spacetimes: a new self-dual approach to a curvature/signature (in)dependent trigonometry.
\newblock{\em J. Phys. A: Math. Gen.} {\bf 33} (2000) 4525--4551.
\doi{10.1088/0305-4470/33/24/309}

   \bibitem{Herranz2002}
 F.~J.~Herranz  and  M.~Santander.
\newblock Conformal symmetries of spacetimes.
\newblock{\em  J. Phys. A: Math. Gen.}  {\bf 35} (2002) 6601--6618.
\doi{10.1088/0305-4470/35/31/306}

  \bibitem{Ballesteros2003}
 A.~Ballesteros, F.~J.~Herranz, M.~Santander and T.~Sanz-Gil. Maximal superintegrability on $N$-dimensional curved spaces. {\it J. Phys. A: Math. Gen.} {\bf 36} (2003) L93--L99. \doi{10.1088/0305-4470/36/7/101}

\bibitem{Bacry1968} H.~Bacry and J.-M.~L\'evy-Leblond. Possible Kinematics.  {\it J. Math. Phys.} {\bf 9} (1968) 1605--1614. \doi{10.1063/1.1664490}

   \bibitem{Goldman1999}
W.~M.~Goldman. {\it Complex Hyperbolic Geometry}. (Oxford: Oxford University Press)  1999. \doi{10.1093/oso/9780198537939.001.0001}

  \bibitem{Yaglom1979}
  I.~M.~Yaglom.
  \newblock {\em A Simple Non-Euclidean Geometry and Its Physical Basis}.
  \newblock (New York: Springer) 1979.
  \doi{10.1007/978-1-4612-6135-3}

\bibitem{Kisil2012}
 V.~V.~Kisil.
\emph{Geometry of M\"obius Transformations: Elliptic, Parabolic and Hyperbolic Actions of $SL_2(\mathbb R)$}.
(Singapore: World Scientific)  2012.
\doi{10.1142/p835}

\bibitem{Bejancu1993}
A.~Bejancu and K.~L.~Duggal. Real hypersurfaces of indefinite Kaehler manifolds. {\it Int. J. Math. Math. Sci.} {\bf 16} (1993) 545--556. \doi{10.1155/S0161171293000675}

 \bibitem{Sasaki1962}
 S.~Sasaki and Y.~Hatakeyama. On differentiable manifolds with contact metric structures. {\it J. Math. Soc. Japan} {\bf 14} (1962) 249--271. \doi{10.2969/jmsj/01430249}

 \bibitem{Tashiro1963}
 Y.~Tashiro. On contact structure of hypersurfaces in complex manifolds, I. {\it Tohoku Math. J.} {\bf 15} (1963) 62--78. \doi{10.2748/tmj/1178243870}

 \bibitem{Ghosh2019} A.~Ghosh and C.~Bhamidipati. Contact geometry and thermodynamics of black holes in AdS spacetimes. {\it Phys. Rev. D} {\bf 100} (2019) 126020. \doi{10.1103/PhysRevD.100.126020}

 \bibitem{Alekseevskii1990} D.~V.~Alekseevskii. Contact homogeneous spaces. {\it Funct. Anal. Its Appl.} {\bf 24} (1990) 324--325. \doi{10.1007/BF01077337}

  \bibitem{PH2018}
A.~Ballesteros, R.~Campoamor-Stursberg, E.~Fern\'andez-Saiz, F.~J.~Herranz and J.~de Lucas. Poisson--Hopf algebra deformations of Lie--Hamilton systems.  {\it J. Phys. A: Math. Theor.} {\bf 51} (2018) 065202.
\doi{10.1088/1751-8121/aaa090}

   \bibitem{PH2021}
A.~Ballesteros, R.~Campoamor-Stursberg, E.~Fern\'andez-Saiz, F.~J.~Herranz and J.~de Lucas. Poisson--Hopf deformations of Lie--Hamilton systems revisited: deformed superposition rules and applications to the oscillator algebra.  {\it J. Phys. A: Math. Theor.} {\bf 54} (2021) 205202.
\doi{10.1088/1751-8121/abf1db}

 \bibitem{AT1986}
A.~Ach{\'{u}}carro and P.~K.~Townsend.
\newblock {A Chern-Simons action for three-dimensional anti-de Sitter
  supergravity theories}.
   {\it Phys. Lett. B}  {\bf 180} (1986)  89--92.
\doi{10.1016/0370-2693(86)90140-1}

 \bibitem{Witten1988}
E.~Witten.
\newblock 2+1 dimensional gravity as an exactly soluble system.
\newblock {\em Nucl. Phys. B} {\bf 311} (1988) 46--78.
\doi{10.1016/0550-3213(88)90143-5}

\bibitem{BHM2010plb}
A.~Ballesteros, F.~J.~Herranz and C.~Meusburger.
\newblock {Three-dimensional gravity and Drinfel'd doubles: Spacetimes and
  symmetries from quantum deformations}.
\newblock {\em Phys. Lett. B} {\bf 687} (2010) 375--381.
\doi{10.1016/j.physletb.2010.03.043}

 \end{thebibliography}
\end{document}